\documentclass[11pt]{emulateapj}

\usepackage{epsfig}
\usepackage{graphicx}
\usepackage{float}
\usepackage[T1]{fontenc}
\usepackage[utf8]{inputenc}
\usepackage{lmodern}
\usepackage{amsmath}
\usepackage{placeins}

\newcommand{\fmrsfr}{FMR$_{\text{SFR}}$}
\newcommand{\fmrhi}{FMR$_{\text{HI}}$}
\newcommand{\fmlsfr}{FML$_{\text{SFR}}$}
\newcommand{\fmlhi}{FML$_{\text{HI}}$}

\begin{document}
\title{The Gas Phase Mass Metallicity Relation for Dwarf Galaxies: Dependence on Star Formation Rate and HI gas mass}
\author{Jimmy\footnotemark[1], Kim-Vy Tran\footnotemark[1], Am\'elie Saintonge\footnotemark[2], Gioacchino Accurso\footnotemark[2], Sarah Brough\footnotemark[3], Paola Oliva-Altamirano\footnotemark[3]\textsuperscript{,}\footnotemark[4]}
\footnotetext[*]{Based on VLT service mode observations (Programs 081.B-0649 and 083.B-0662) gathered at the European Southern Observatory, Chile.}
\footnotetext[1]{ George P. and Cynthia W. Mitchell Institute for Fundamental Physics and Astronomy, Department of Physics and Astronomy, Texas A\&M University, College Station, TX 77843, USA }
\footnotetext[2]{Department of Physics \& Astronomy, University College London, Gower Place, London WC1E 6BT, UK }
\footnotetext[3]{Australian Astronomical Observatory, PO Box 915, North Ryde, NSW 1670, Australia}
\footnotetext[4]{Centre for Astrophysics \& Supercomputing, Swinburne University of Technology, Hawthorn, VIC 3122, Australia}

\begin{abstract}

Using a sample of dwarf galaxies observed using the VIMOS IFU on the VLT, we investigate the mass-metallicity relation (MZR) as a function of star formation rate (\fmrsfr ) as well as HI-gas mass (\fmrhi ).  We combine our IFU data with a subsample of galaxies from the ALFALFA HI survey crossmatched to the Sloan Digital Sky Survey to study the \fmrsfr\ and \fmrhi\ across the stellar mass range 10$^{6.6}$ to 10$^{8.8}$ M$_\sun$, with metallicities as low as 12+log(O/H) = 7.67.  We find the 1$\sigma$ mean scatter in the MZR to be 0.05 dex.  The 1$\sigma$ mean scatter in the \fmrsfr\ (0.02 dex) is significantly lower than that of the MZR.  The \fmrsfr\ is not consistent between the IFU observed galaxies and the ALFALFA/SDSS galaxies for SFRs lower than 10$^{-2.4}$ M$_\sun$ yr$^{-1}$, however this could be the result of limitations of our measurements in that regime.  The lowest mean scatter (0.01 dex) is found in the \fmrhi .  We also find that the \fmrhi\ is consistent between the IFU observed dwarf galaxies and the ALFALFA/SDSS crossmatched sample.  We introduce the fundamental metallicity luminosity counterpart to the FMR, again characterized in terms of SFR (\fmlsfr ) and HI-gas mass (\fmlhi ).  We find that the \fmlhi\ relation is consistent between the IFU observed dwarf galaxy sample and the larger ALFALFA/SDSS sample.  However the 1$\sigma$ scatter for the \fmlhi\ relation is not improved over the \fmrhi\ scenario.  This leads us to conclude that the \fmrhi\ is the best candidate for a physically motivated fundamental metallicity relation.

{\bf Key words: } galaxies: abundances - galaxies: formation  - galaxies: starburst - galaxies: dwarf - galaxies:evolution
\end{abstract}

\maketitle

\section{Introduction}

The mass-metallicity relation (MZR) is an important tool for understanding the underlying processes of galaxy formation.  A galaxy's metal content provides insights into the history of star formation within a galaxy.  \citet{Lequeux:79} and \citet{Kinman:81} first reported on the correlation between galaxy stellar-mass and gas-phase metallicity.  Later studies confirmed the relation using the Sloan Digital Sky Survey (SDSS) for large number statistics \citep{Tremonti:04, Kewley:08, Michel-Dansac:08, Salim:14}. Other studies have focused on the redshift evolution of the MZR \citep{Savaglio:05, Erb:06, Cowie:08, Liu:08, Maiolino:08, Panter:08, Rodrigues:08, Hayashi:09, Mannucci:09, Perez-Montero:09, Cresci:12}.

The luminosity-metallicity relation (LZR) has also been proposed as a relation of merit \citep{Skillman:89, Garnett:02, Perez-Gonzalez:03, Pilyugin:04, Lee:06b}.  The advantage of the LZR is that reliable luminosity measurements are generally easier to obtain than reliable stellar mass estimations.  The disadvantage of the LZR is that it is dependent upon the bandpass being used, and could be affected by dust extinction \citep{Salzer:05}.  The LZR is observed to hold over a range of 10 magnitudes in galaxy optical luminosity \citep{Zaritsky:94, Tremonti:04, Lee:06b}.  The MZR and the LZR reveal that some physical mechanism must be driving a correlation between stars, metals, and gas flows across a vast range of scales.  However, the exact physical mechanism(s) responsible for forming this relation are still uncertain.

Recent studies have investigated the mass-metallicity relation with a particular focus on low stellar-mass or low-luminosity systems \citep{Berg:12, Nicholls:14, Haurberg:15, McQuinn:15, James:15}, which is the population of galaxies of interest to this work.  The MZR is known to exhibit scatter greater than would be expected by the uncertainties on the individual data points \citep{Tremonti:04, LaraLopez:10, Mannucci:10}.

To explain this apparently enhanced scatter, it has been suggested that the MZR has a dependence on Star Formation Rate \citep[SFR; ][]{LaraLopez:10, Mannucci:10, Andrews:13}, which could alternatively be explained as a dependence on HI-gas content \citep{Bothwell:13, LaraLopez:13}.  This 3 dimensional extension of the mass-metallicity relation is known as the fundamental metallicity relation (FMR).  The FMR has been shown to be consistent, with little evolution, up to a redshift of 2.5 \citep{LaraLopez:10, Mannucci:10}.  It is currently uncertain as to which observable (HI mass or SFR) better explains the scatter in the MZR, or if some other unknown parameter is responsible.

Several hypotheses have been proposed for the mechanisms that guide the FMR, such as pristine gas inflows diluting metal abundances \citep{Finlator:08, Dave:10}.  Inflows of pristine gas would cause a galaxy to exhibit lower gas-phase metallicity than would be predicted by the FMR.  Recent evidence suggests that infall of pristine gas is the more dominant component at high redshift \citep{Agertz:09, Bournaud:09, Brooks:09, Dekel:09}

As argued in \citet{Mannucci:10} infalling pristine gas must eventually ignite star formation, which will create strong supernova winds that will expel metals.  Gas outflows expelling metal-rich gas into the interstellar medium provide an alternative explanation for the observed gas-phase metallicity being lower than the MZR prediction \citep{Edmunds:90, Lehnert:96, Garnett:02, Tremonti:04, Kobayashi:07, Scannapieco:08, Spitoni:10}.

Studies on the FMR thus far, either as a function of SFR (\fmrsfr ) or HI mass (\fmrhi ), have focused on the SDSS sample of spectroscopically observed galaxies \citep{LaraLopez:10, Mannucci:10, Bothwell:13}.  As such these galaxies are biased towards the intermediate to high stellar-mass galaxies that dominate the SDSS spectroscopic sample.  However it is important to understand the low-mass galaxy population as well and to test whether or not they fall onto the same FMR.  

Dwarf galaxies are the most common type of galaxies in the Universe, comprising $\sim$ 85\% of the objects in the volume within 10Mpc \citep{Karachentsev:04}.  Even though they are not as widely studied as their more massive counterparts, they may hold the answers to several fundamental questions concerning the processes of galaxy formation and evolution.  

The data presented here consists of integral field unit (IFU) observations of a sample of 11 dwarf galaxies selected from the ALFALFA survey.  ALFALFA is a blind HI survey of the local universe ($cz<$180,000 km s$^{-1}$; \citealt{Haynes:11}).  The IFU observations were taken using the VIMOS IFU spectrograph on the Very Large Telescope (VLT).  

To compliment the IFU observed dwarf galaxy dataset, we utilize a larger ALFALFA/SDSS cross-matched sample of spectroscopically observed galaxies.  
In addition we also include samples from the literature of low-mass/low-luminosity galaxies such as the \citet{Berg:12}, long-slit spectroscopic observations of low-luminosity galaxies, the SHIELD survey of ALFALFA dwarf galaxies \citep{Haurberg:15,McQuinn:15} and the \citet{James:15} long-slit morphologically selected survey of dwarf irregular galaxies.  

Throughout this paper, we assume a Hubble constant of H$_0$ = 67 km s$^{-1}$ Mpc$^{-1}$, $\Omega_{M}$ = 0.32, and $\Omega_{\Lambda}$ = 0.68. 

\vspace{10 mm}

\section{Sample}
\label{Observations}

\begin{figure*}[!htbp]
\epsfig{ file=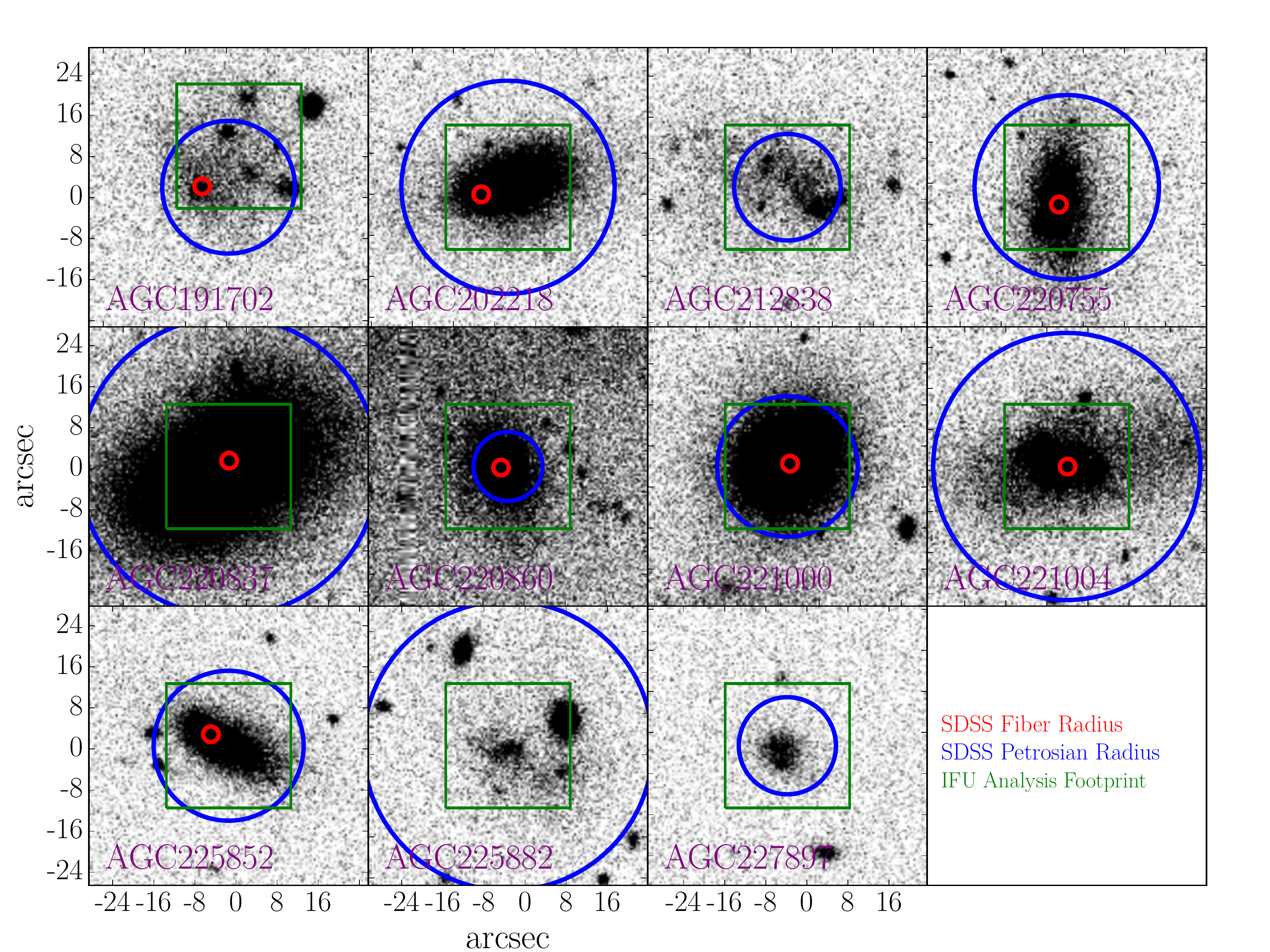, scale=0.59}
\caption{SDSS {\it r}-band images of the 11 galaxies presented in this study, demonstrating the need for IFU spectroscopy.  The red circle indicates the location of the 3$^{\prime \prime}$ SDSS fiber which would be used to obtain spectral information without IFU observations.  The SDSS fiber often misses a considerable amount of light from the clumpy star-forming regions of these nearby dwarf galaxies.  Shown in blue is the Petrosian radius as determined by the SDSS automated data reduction procedure.  The SDSS photometry pipeline visibly overestimates or underestimates the flux in galaxies from this sample.  Each thumbnail image is the size of the full VIMOS low resolution multiplexed field-of-view (54$^{\prime \prime}$x54$^{\prime \prime}$ ).  The green boxes indicate the subset of this field-of-view that our IFU analysis focuses on.}
\label{field_of_views}
\end{figure*}

The IFU observed dwarf galaxy sample includes nearby (D $<$ 20Mpc), low HI mass (M$_{\text{HI}} < 10^{8.2}$), low surface brightness dwarf galaxies, selected to study the mass-metallicity relation on the low stellar-mass end.  Due to the nature of observing nearby dwarf galaxies, IFU spectroscopy is vital to capture the properties of the entire galaxy.  As can be seen in Figure \ref{field_of_views}, for the galaxies in our sample that have Sloan Digital Sky Survey (SDSS) fiber observations, the single SDSS fiber only captures a small portion of the galaxy's flux, and the fiber may miss bright star forming regions (e.g. \citealt{Richards:14}), this would cause star formation rate estimations of nearby dwarf galaxies such as these to be inaccurate.   

IFU observations also have the advantage of not requiring additional photometric observations to identify bright HII regions to target.  Performing such observations using long-slit spectroscopy would require {\it a-priori} knowledge of the positions of HII regions which can be obtained from narrow-band H$\alpha$ images, introducing an optical bias in the otherwise HI-selected sample.  The most metal poor galaxies are also likely to have the faintest HII regions, making them barely detectable in narrow-band images.   Integrating over a larger surface area allows us to target fainter galaxies.  Because the dwarf galaxies that we are trying to observe are so faint, only 6 galaxies in this sample have reliable SDSS spectroscopy.  A purely SDSS based study, or a long-slit spectroscopy study are both likely to preferentially exclude the faintest dwarf galaxies.  

\subsection{Spectroscopic Observations}

Spectroscopic data of 28 dwarf galaxies were initially taken using the VIMOS \citep{LeFevre:03} IFU spectrograph on the Very Large Telescope (VLT) located at Paranal Observatory.  Of the 28 observed galaxies, 11 have H$\alpha$ emission lines greater than our amplitude over noise (AoN) cut of 5 for more than 20 spaxels and therefore are included in this study.  Integration times for the remaining 17 galaxies were insufficient to reach our target depth.  The data was obtained starting on April 11, 2008 and ending on May 19, 2010 under program IDs 081.B-0649(A) and 083.B-0662(A).  Data was obtained using the VIMOS Low Resolution Blue Grism which has a wavelength range of 4000-6700 \AA\ and a spectral resolution of 5.3 \AA\ pixel$^{-1}$ (R $\sim$ 1000).

Using the LR Blue grism provides the full 54$^{\prime \prime}$x54$^{\prime \prime}$ field of view possible with VIMOS which allows us to obtain in a single pointing spectra across the complete stellar disk of each galaxy.  Each object was observed using a 3 dither pattern, with each dither being integrated for 20 minutes.  Average seeing across all observations is 1.05" FWHM.

The LR blue grism provides both a wider spectral range, and a wider field of view when compared to the HR blue grism.  The LR grism wavelength range allows for simultaneous observations of the emission lines from H$\beta$ ($\lambda$ = 4861) to [NII] ($\lambda$ = 6583).  The major drawback to using the low-resolution spectra ($\sim$5.3 \AA\ pixel$^{-1}$) is that we are unable to measure gas kinematics from the emission lines and the instrumental dispersion causes the H$\alpha$ and [NII] $\lambda \lambda$ = 6549,6583 \AA\ emission lines to blend together.  

The spectroscopic data obtained with the VIMOS IFU is reduced from its raw form using the Reflex environment for ESO pipelines \citep{Freudling:13}.  The standard VIMOS template is used within the Reflex environment to produce the master bias and calibration frames containing the fiber traces and wavelength solution.  Many of the raw data and calibration frames contain a bright artifact across the surface of the chip, identified to be an internal reflection within the instrument.  This contamination interferes with the wavelength calibration routine within Reflex because it is often misidentified as a skyline, causing the spectrum to be shifted incorrectly.  In order to compensate, we disable the skyline shift in the calibration steps.  The wavelength solution provided by the ESO pipeline proved to be accurate without using the skyline shift.  We also use the flux standardization routine within the Reflex VIMOS pipeline.  We then apply the calibrations frames to the science frames to produce the fully reduced Row-Stacked Spectra (RSS).

The final output of the Reflex pipeline is four quadrants per observation.  We input these individual RSS quadrants into routines written in IDL and Python\footnotemark[1].  These routines fit Gaussian curves to the skylines and subtract the sky background while retaining the flux information from each skyline to normalize the transmission of each spaxel across the entire field of view.  After the normalization and sky subtraction steps are completed, the two dimensional RSS are converted into three dimensional data cubes.  These data cubes consist of two spatial dimensions with the wavelength axis being the third dimension.  Further details of the process these scripts follow can be found in \citet{Jimmy:13}.

\footnotetext[1]{Available publically: http://jimmy.pink/\#code}

Once the data cubes have been built, the individual dithers (typically 3 per galaxy) are stacked using a 5$\sigma$ clipped mean.  We use the AoN to select for spaxels containing sufficient emission line flux for oxygen abundance estimations.  AoN is defined as the amplitude of the emission line divided by the noise after the linear offset is subtracted.  An AoN threshold of 5 is used to identify spaxels which contain emission lines for gas-phase metallicity analysis.  Only spaxels which pass the AoN cut in all of the three following lines are included in the integrated spectrum for oxygen abundance estimations: H$\beta$ $\lambda = 4861$, [OIII], $\lambda = $5007, H$\alpha\ \lambda = 6563$.  We do not require [NII] $\lambda = 6583$ to be separately detected because it is blended with H$\alpha$ (Figure \ref{line_fit_output}).

All spaxels which pass our AoN cut are then fed into Python based Gaussian fitting routines to measure the emission line flux ratios for gas-phase metallicity estimations.  To obtain the integrated spectrum used to derive oxygen abundances, we sum the spectra from each spaxel that passes the AoN cut to produce a single spectrum that we perform this analysis upon.  We chose to write our own routines instead of using a publicly available program such as GANDALF \citep{Sarzi:06}.  Because GANDALF fits the stellar continuum and emission lines simultaneously, GANDALF is unable to fit robustly the weak stellar continuum in the dwarf galaxies.

Using an AoN cut is necessary because a significant fraction of the spaxels in the data cubes are excessively noisy, and including them in the integrated spectrum would significantly reduce the signal-to-noise ratio.  Unfortunately the AoN cut causes us to discard a number of spaxels containing H$\alpha$ flux, resulting in an underestimate of the SFR.  In order to ensure that we capture all of the H$\alpha$ flux, we use the segmentation images described in Section \ref{PhotometricObservations} to select spaxels belonging to a galaxy.  We rescale and rotate the segmentation map to match the VIMOS data cube using the IDL routine HASTROM.PRO\footnotemark[3].  The integrated spectra obtained using the segmentation map selection is used to measure H$\alpha$ flux for the purposes of determining the SFR for each galaxy.

\footnotetext[3]{Available as part of the IDL Astronomy User's Library}

\subsubsection{Emission Line Fitting}

\begin{figure}
\epsfig{ file=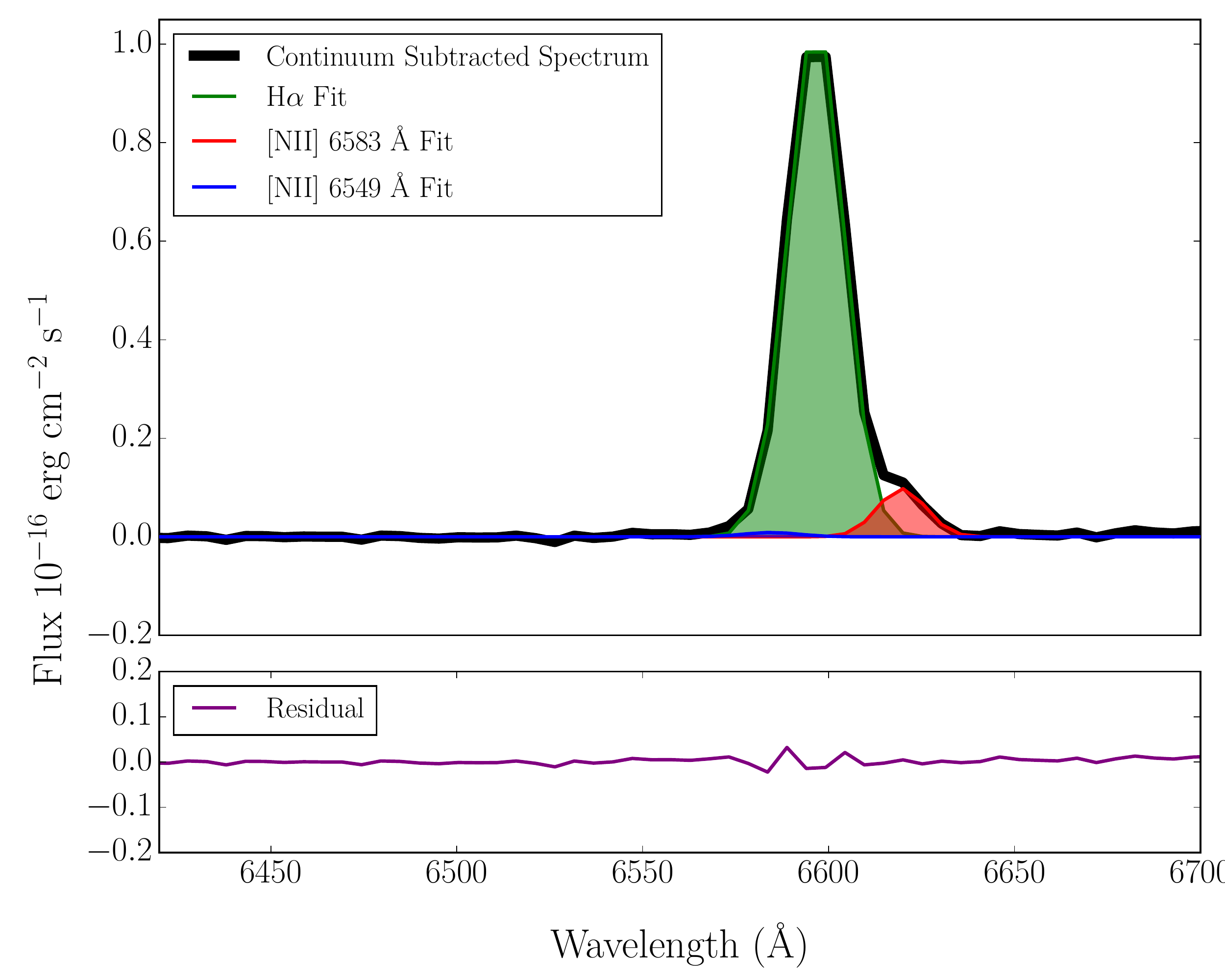, scale=0.35}
\caption{Example spectra showing the blending within the [NII] and H$\alpha$ emission lines due to the instrumental resolution (7.5 \AA ).  Our Python based Gaussian fitting routines are able to deconvolve the three emission lines into their constituent parts, as can be seen by the three Gaussian curves plotted in red, green, and blue.  The flux in the blue [NII] 6549 \AA\ emission line is negligible. Details of our ability to recover the three emission line fluxes can be found in Appendix \ref{AppendixA}.  Our oxygen abundance estimations rely upon measurements of the H$\alpha$ and [NII] emission lines.} 
\label{line_fit_output}
\end{figure}

The basic procedure that our Python based Gaussian fitting routines follows is as such: First a linear fit to the spectrum continuum is performed using the optimize\_curve routine to fit a line using a least squares optimization.  The fitting window around each emission line is approximately 400 \AA\ ($\sim$75 pixels) with nearby emission, sky lines, and internal reflections regions being masked out.  After the continuum has been subtracted, the [NII] and H$\alpha$ emission lines are fit using gaussfitter\footnotemark[2] to fit 3 Gaussian curves simultaneously.  This allows us to separate the individual components of the 3 blended emission lines.

The output measurements of gaussfitter are not heavily dependent upon initial input parameters assumed, therefore we provide it an initial input intensity of 0.5 erg cm$^{-2}$s$^{-1}$, an initial input peak position of the redshift corrected expected peak for each line, and an initial input dispersion of 7.5 \AA , equal to the instrumental dispersion of the observations.  No line flux ratios are assumed within our fits.  The gaussfitter bounds are set up such that each Gaussian fit is limited to a maximum 10\% variation from its expected peak position and FWHM.  Because the instrumental dispersion dominates the width of the emission lines, we do not expect line widths to vary widely across spaxels.

We do not include a fit to the Balmer absorption in our procedures.  Our observations are insufficient to reliably estimate the stellar continuum for accurate absorption measurements.  However it is likely that Balmer absorption effects are negligible and well within our uncertainties \citep{Rosa-Gonzalez:02}.  Further details about the emission line flux fitting code and our ability to recover the blended emission lines can be found in Appendix \ref{AppendixA}.

Uncertainties on our flux measurements are estimated using a Monte Carlo Markov chain (MCMC) technique with 1000 iterations.  For each iteration, random noise is added to the spectrum on the order of the residuals to the linear fit.  Then the emission line fluxes are remeasured.  The 1$\sigma$ width of the Gaussian distribution of flux values provided by 1000 iterations provides the uncertainties in our line fluxes.  These uncertainties are then propagated using standard Taylor series error propagation methods in order to determine the uncertainties in our derived quantities.

We perform the linear and Gaussian fits on the blue and the red end of the spectrum separately, with the blue end containing the H$\beta$ and [OIII] lines, and the red end containing the H$\alpha$ and [NII] lines.  The results of our fitting routine on the red end of an integrated spectrum can be seen in Figure \ref{line_fit_output}.  

\footnotetext[2]{written by Adam Ginsburg http://www.adamgginsburg.com/pygaussfit.htm}

\subsection{Photometric Observations}
\label{PhotometricObservations}

To obtain stellar mass estimations for our galaxies, we utilize the stellar mass estimation procedure outlined in \citet{West:10}.  Doing so requires accurate photometric observations from the Sloan Digital Sky Survey (SDSS; \citealt{Eisenstein:11}) Data Release 12 \citep{Alam:15} which is the final data release of SDSS-III.  The automated SDSS photometric pipeline \citep{Lupton:02} is optimized for small, well resolved objects and is known to be unreliable for clumpy and diffuse objects such as dwarf galaxies, mostly due to the SDSS algorithm's aggressive parent-child splitting routines. As can be seen in Figure \ref{field_of_views}, Petrosian flux measurements could possibly be an inaccurate representation of the flux of our galaxies.  \citet{West:10} found that roughly 25\% of SDSS galaxies have less than 90\% of their flux contained in
the brightest child.  Therefore, to obtain more accurate stellar mass estimations, we have chosen to run Source Extractor \citep{Bertin:96} manually on each of the SDSS bandpass image frames for each of the 11 galaxies within our IFU sample.  

To ensure that the SDSS images are aligned properly, we first use HASTROM.PRO to align each individual bandpass fits image to the r-band image coordinates.  We stack all 5 photometric bands to create a higher signal/noise detection image.  We run Source Extractor in dual image mode with the detection image and each of the individual frames.   The default values and keyword searches are used as inputs to Source Extractor except the following: MAG\_ZEROPOINT = 22.5, BACK\_SIZE = 256, and PIXEL\_SCALE = 0.396.  Additionally for every galaxy we begin with the following parameters: DEBLEND\_NTHRESH, DETECT\_THRESH, ANALYSIS\_THRESH set to 1.0, and then adjust them accordingly to ensure that each dwarf galaxy is detected as a single source in the segmentation map, and that the full flux is being captured when viewing the residuals map.  This segmentation map is also used to select the spaxels to be summed for SFR estimation.

\subsection{Stellar Masses}
\label{StellarMasses}

Stellar masses are estimated according to the color-derived M/L ratio of \citet{West:10}.  The \citet{West:10} stellar mass estimation is a modification of the work done by \citet{Bell:03}.  The \citet{Bell:03} mass estimation assumes a ``Diet Salpeter'' IMF, however the \citet{West:10} estimation modifies the \citet{Bell:03} estimation to a Kroupa IMF.  This stellar mass estimation technique has been shown to be accurate within 20\% of the actual stellar mass \citep{Bell:03}.  Photometric data for the IFU observed dwarf galaxies are derived from the Source Extractor measurements of the SDSS {\it g}, {\it r}, and {\it i} filters as detailed in Section \ref{PhotometricObservations}.  With proper SDSS photometric measurements obtained, we utilize the \citet{West:10} stellar mass to light estimation:  \begin{equation}\text{log}(M_*/L_i) = -0.222+0.864(g-r) + \text{log}(0.71)\label{stellar_mass_to_light} \end{equation}

Assuming a solar {\it i}-band absolute magnitude of 4.57 \citep*{Sparke:00}, we then convert the {\it i}-band photometry into stellar mass estimations using the conversion outlined in \citet{West:10} along with the M/L ratio defined in Equation \ref{stellar_mass_to_light}.  \begin{equation}M_* = (M_*/L_i) \times 10^{-((M_i-4.57)/2.51)}\label{stellar_mass} \end{equation}

Because these stellar mass estimations rely upon {\it i}-band absolute magnitudes, we require accurate distance estimations.  For the local universe dwarf IFU observed galaxies, we cannot rely upon the redshift to provide accurate distances due to gravitational interactions within the local universe.  Therefore, we utilize the distance estimations from the ALFALFA catalog for all galaxies below a redshift of 6,000 km s$^{-1}$ ($z<$0.02) where the errors on the distance estimates become comparable to those of Hubble flow estimations \citep{Haynes:11}.  The ALFALFA catalog distances are calculated using the local universe peculiar velocity model of \citet{Masters:04} derived from the SFI++ catalog of galaxies \citep{Springob:07} and published literature values are adopted as appropriate.  See \citet{Haynes:11} for a full explanation of the model used.  Above a redshift of 6,000 km s$^{-1}$  we calculate the luminosity distance (D$_L$) from the SDSS redshift assuming the motions of galaxies are dominated by the Hubble flow.  

Using this method, stellar masses for each of the 11 galaxies presented here are calculated to be in the range 6.58 $\leq$ log(M$_*$) $\leq$ 8.78.  The full list of stellar masses can be found in Table \ref{results_table}.  Uncertainties in our stellar masses are derived from propagating the Source Extractor provided photometric uncertainties through the standard Taylor series technique.  

\subsection{ALFALFA/SDSS Sample Selection}
\label{SDSS_Section}
To compliment the IFU observed dwarf galaxy results, we compare this sample to galaxies selected from the ALFALFA HI survey.  The ALFALFA team has crossmatched their HI detections to the SDSS database in order to associate the radio HI-gas detection to optical photometric galaxy detections.  We cross reference the ALFALFA $\alpha$.40 database (15,855 objects) with the SDSS DR12 spectroscopic database and find 7,773 ALFALFA galaxies with SDSS spectroscopic observations.  We first remove the 6 galaxies from the IFU observed dwarf galaxy sample that overlap with the ALFALFA/SDSS sample in order to ensure that they are not counted twice within our combined sample.  We remove AGN contamination from our sample with the \citet{Kauffmann:03b} selection criteria, leaving us with 6,151 galaxies.  We remove galaxies for which the aperture correction is exceedingly large by cutting those galaxies for which the difference between the Petrosian {\it r}-band magnitude and the fiber {\it r}-band magnitude exceeds 5 mag, bringing our sample down to 6,137 galaxies.  We also select for galaxies with a redshift $z<$0.05, bringing our sample to 5,759 galaxies. We then perform a cut to ensure that the SDSS spectra contain H$\alpha$ emission at a S/N greater than 3 leaving 5,751 galaxies.  After performing a final cut to exclude galaxies for which the Petrosian stellar mass estimations disagree with the MPA-JHU catalog stellar mass estimations by more than 0.5 dex, we have 5,436 galaxies in our final crossmatched ALFALFA/SDSS sample.  In comparison, \citet{Bothwell:13} perform similar cuts, although they specify H$\alpha$ S/N $>$ 25 and remove galaxies for which their two oxygen abundance estimations disagree, leaving them with a sample containing 4,253 galaxies.

\subsection{Comparison Data}
To provide additional galaxies in the low stellar-mass/ low luminosity regime that is poorly constrained by the ALFALFA/SDSS sample, we also include in our MZR/LZR analysis data from similar surveys found in the literature of low stellar-mass/low luminosity galaxies.  We will utilize data from the following surveys:

\begin{itemize}

\item \citet{Saintonge:PhD} - Observations of 25 low HI-mass ALFALFA dwarf galaxies observed using long-slit spectroscopy from the Double Spectrograph (DBSP; Oke \& Gunn 1982) at Mount Palomar's Hale 5m telescope.  Stellar masses are derived from B and V band photometry using the simple stellar population models from \citet{Bell:03}.  These are galaxies very similar in nature to those of the IFU observed dwarf galaxy sample, except that narrow-band H$\alpha$ imaging had revealed the presence of at least one bright HII region per galaxy making long-slit spectroscopy possible.

\item \citet{Berg:12} - A sample of low-luminosity galaxies selected from the Spitzer Local Volume Legacy survey and observed using MMT long-slit spectroscopy.  Stellar masses are derived from B and K band photometry using models from \citet{Bell:01}.

\item \citet{James:15} - A morphologically selected survey of 12 SDSS dwarf irregular galaxies, observed using MMT long-slit spectroscopy.  Stellar masses are derived from \citet{Bell:03} {\it g} and {\it r} band photometry.

\item Survey of HI in Extremely Low-mass Dwarfs \citep[SHIELD; ][]{Haurberg:15} - An ALFALFA derived survey of 8 galaxies observed using long-slit spectroscopy from the Mayall 4m telescope at KPNO.  Stellar mass estimations come from IR (3.6 and 4.5 $\mu$m) photometric colors using the method from \citet{Eskew:12}.  B-band magnitudes come from WIYN 3.5m observations.
\end{itemize}

Due to offsets between oxygen abundance calibrations \citep{Kewley:08}, it is necessary to perform oxygen abundance estimations consistently in order to obtain meaningful comparisons between samples.  Therefore, for each of the comparison samples from other studies, we have recalculated oxygen abundance estimations (as described in Section \ref{OxygenAbundance}) using the emission line fluxes reported within each survey.  

Similarly stellar mass estimations from all of the above surveys have been rescaled onto a Kroupa initial mass function (IMF) to match the stellar mass estimations that we perform on both the dwarf IFU observed sample and the larger crossmatched ALFALFA/SDSS sample in Section \ref{StellarMasses}.

\vspace{10 mm}

\section{Analysis}

\subsection{Oxygen Abundance}
\label{OxygenAbundance}

As a proxy for the gas-phase metallicity of a galaxy, we estimate the oxygen abundance using strong emission line based measurements.  Due to the low surface brightness of the IFU observed dwarf galaxy sample, we are unable to detect any of the faint auroral lines required for direct (T$_e$ based) oxygen abundance measurements.  We are also limited by our wavelength range to using the [OIII], [NII], and Balmer hydrogen emission lines to perform our oxygen abundance estimations.  We attempted to edit the VIMOS pipeline source code to extend the wavelength range which would allow us to capture the [OII] feature required for R23 (R23$ = ([\text{OII}]\lambda 3727 + [\text{OIII}]\lambda 4958, 5007)/\text{H}\beta $) based estimations, however issues with interference between overlapping pseudo-slits on the CCD proved to be a hinderance to accurate [OII] emission line measurements.

Ideally we would use the full range of emission lines within our spectrum.  The O3N2 based oxygen abundance estimation technique of \citet{Pettini:04} (hereafter PP04) is calculated using the flux of the emission lines in [OIII] $\lambda$5007, H$\beta$ $\lambda$4861, H$\alpha$ $\lambda$6563, and [NII] $\lambda$6583.  Unfortunately, the calibration of the PP04 O3N2 based estimations becomes unreliable near or below 12+log(O/H) = 8.09 \citep*{Pettini:04}.  For completeness, we include in Appendix \ref{AppendixB} a parallel analysis utilizing the O3N2 oxygen abundances and find that our conclusions would not change.  

N2 based estimations rely only on the H$\alpha$ $\lambda$6563 and [NII] $\lambda$6583 features and are defined as \begin{equation}N2 = \text{log} \left\{  \frac {[NII] \lambda 6583 }{\text{H}\alpha} \right\}. \label{N2} \end{equation}  N2 based oxygen abundance estimations have the benefit of being less sensitive to the ionization parameter (although not completely independent; see \citealt{Morales-Luis:14}).  The PP04 N2 estimation was calibrated down to metallicities as low as 12+log(O/H) = 7.48 \citep*{Pettini:04}.

Throughout this paper we will utilize the N2 oxygen abundance estimation as calibrated by (\citealt{Denicolo:02}; hereafter D02) which was targeted towards low gas-phase metallicity galaxies, calibrated over a range 7.2 $\leq$ 12+log(O/H) $\leq$ 9.1, and is well behaved at the low metallicity regime.  In the D02 system the oxygen abundance is defined as: \begin{equation} 12+\text{log(O/H)} = 9.12 + 0.73 \times \text{N2.} \label{D02}\end{equation}

The blending of the emission lines due to instrumental resolution in the dwarf IFU observed galaxy sample causes N2 based oxygen abundance estimations to become less accurate in low flux spaxels.  However, when we use the integrated AoN selected spaxels for the entire galaxy, the integrated flux is of sufficient quality to reliably deblend the three Gaussian components that compromise [NII] and H$\alpha$ as demonstrated in Figure \ref{line_fit_output}.  For our dwarf galaxy IFU observed sample, we calculate oxygen abundance estimations in the range 7.67 $<$ 12+log(OH) $<$ 8.56, indicating that our low stellar mass sample does indeed exhibit low gas-phase metallicity.  Uncertainties in our oxygen abundance estimations are derived from Taylor series propagation of the uncertainties in the emission line fluxes measured in the IFU observed dwarf galaxy sample.  The full list of oxygen abundances calculated for each IFU observed dwarf galaxy can be found in Table \ref{results_table}.

\subsection{Luminosity}
\label{luminosity}

The B-band Luminosity-Metallicity Relation (LZR) has been claimed to be as strong of a gas-phase metallicity indicator as the MZR when reliable distance measurements and appropriate photometry are used \citep{Berg:12}.  Throughout this work, we will perform a parallel analysis to the FMR using luminosity in place of stellar mass.  We calculate the B-band absolute magnitudes (M$_B$) from SDSS photometry using the relation from \citet{Lupton:05}.  \begin{equation}m_B = g + 0.3130*(g - r) + 0.2271.\label{m_B} \end{equation}  
 
We calculate B-band luminosities for the IFU observed dwarf galaxy population in the range -16.11 $<$ M$_B$ $<$ -12.08.  As detailed in Section \ref{StellarMasses},  this requires accurate luminosity distances, therefore we follow the same procedure and utilize ALFALFA catalog distances for $cz<$6,000 and assume a Hubble flow for $cz>$6,000.  Uncertainties in our B-band luminosities are derived from propagating the Source Extractor provided photometric uncertainties through the standard Taylor series technique.  The full list of luminosities and their uncertainties can be found in Table \ref{results_table}.  

\subsection{Star Formation Rates}

\begin{figure*}
\epsfig{ file=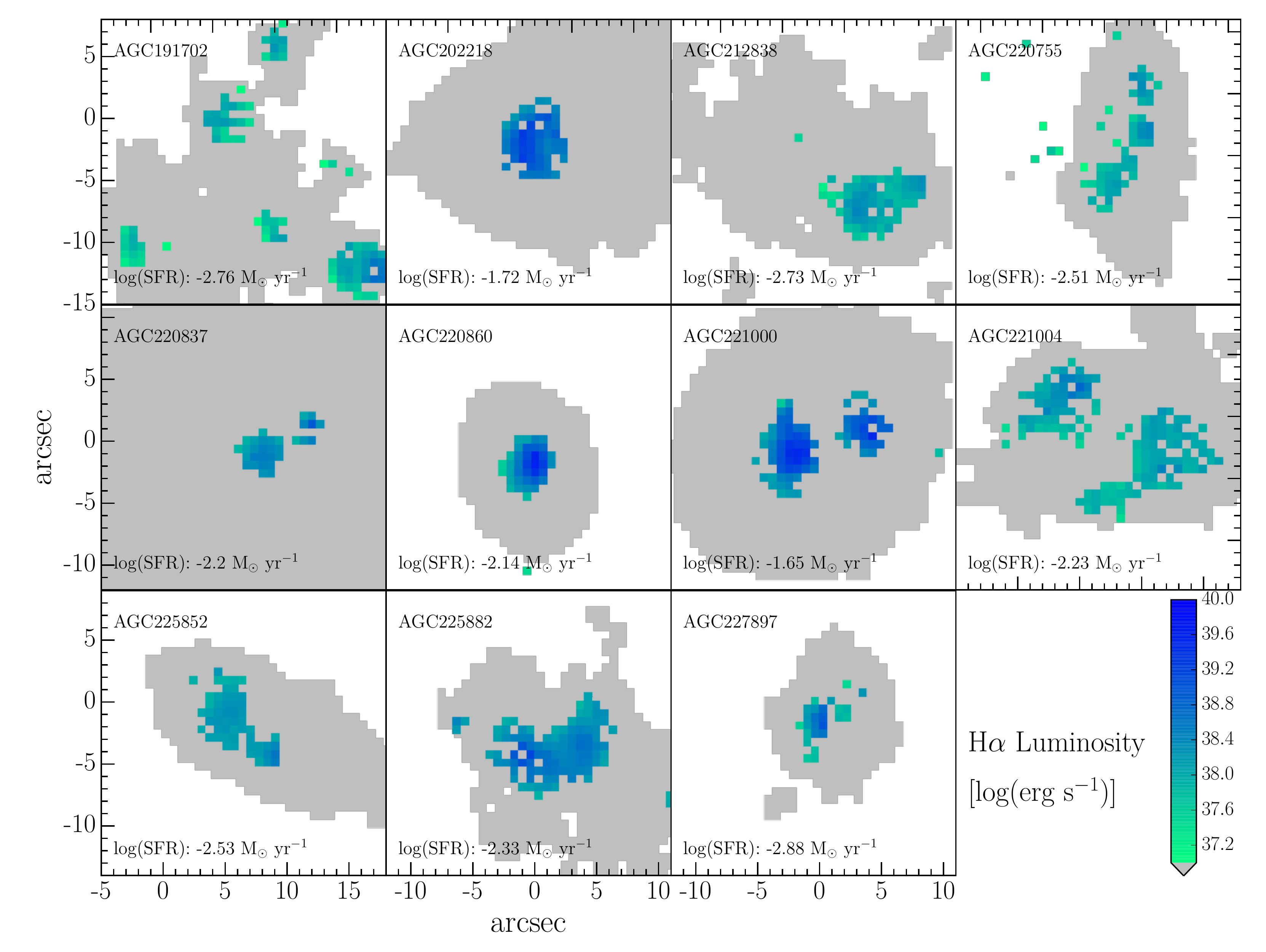, scale=0.5}
\caption{Spatial mapping of the H$\alpha$ emission line flux within a subset of the IFU field-of-view (Figure \ref{field_of_views}).  The gray regions indicate the stellar continuum as selected by the segmentation maps.  The blue-green pixels indicate the spaxels which pass the AoN of 5 cut.  The intensity of the flux varies widely between galaxies, and even within galaxies the flux intensity appears to be quite patchy.  Also shown for each galaxy is the SFR for each galaxy calculated using the flux integrated over the entire gray region showing that our galaxies cover a range of approximately 1.5 dex in SFR.}
\label{h_alpha}
\end{figure*}

Star Formation Rates (SFRs) for each galaxy are derived from the integrated flux of H$\alpha$ emission within the segmentation map selected spaxels of each galaxy.  Spatial maps of the stellar continuum and H$\alpha$ emission for each galaxy can be seen in Figure \ref{h_alpha}.  The SFR for each galaxy is determined by first integrating the flux of every spaxel selected by the segmentation map selection criteria (the gray shaded regions in Figure \ref{h_alpha}).  We use the integrated flux because, at the average distance of our galaxies, each spaxel subtends 25 pc on a side and H$\alpha$ based SFR estimations have been shown to be unreliable on scales less than 100-1000 pc \citep{Kennicutt:12}, or when the luminosity drops below 10$^{38}$-10$^{39}$ erg s$^{-1}$ (which corresponds to an SFR of $\sim$ 0.001-0.01 M$_\sun$ yr$^{-1}$).  This would mean that several galaxies (AGC220837, AGC221004, AGC225852, and AGC228882) have unreliable SFR estimates on a spaxel by spaxel basis as can be seen in Figure \ref{h_alpha}.

To convert the H$\alpha$ emission line measurements into SFR estimations, we convert our H$\alpha$ emission line flux (H$\alpha$) into luminosity (H$\alpha _L$) using the luminosity distance (D$_L$) as \begin{equation}(\text{H}\alpha _L)_{\text{obs}} =  4\pi \text{H}\alpha D_L^2.\end{equation} where the subscript ``obs'' denotes the observed H$\alpha$ luminosity.  

We need to correct for reddening to obtain the intrinsic H$\alpha$ luminosity (H$\alpha _L$)$_{\text{int}}$ where \begin{equation}(\text{H}\alpha _L)_{\text{int}} = (\text{H}\alpha_L)_{\text{obs}}10^{0.4A_{\text{H}\alpha}}.\end{equation}  The attenuation at a specific wavelength ($A_\lambda$) is defined as \begin{equation} A_\lambda = \kappa (\lambda )E(B-V) \label{generala}\end{equation} where $E(B-V)$ is the broadband color excess and $\kappa$($\lambda$) is the value of an attenuation curve at wavelength $\lambda$.  In order to obtain the reddening for each galaxy based on the intrinsic value of the Balmer decrement, we will utilize the color excess between the H$\alpha$ and H$\beta$ emission lines.  This can be found by substituting the wavelengths of H$\alpha$ and H$\beta$ into Equation \ref{generala} to find \begin{equation}A(\text{H}\beta ) - A(\text{H}\alpha ) = \kappa (\text{H}\beta )E(B-V) - \kappa (\text{H}\alpha )E(B-V).\label{betaminusalpha}\end{equation}  Equation \ref{betaminusalpha} could alternatively be defined as $E(\text{H}\beta - \text{H}\alpha)$, which is analogous to $E(B-V)$ such that \begin{equation}A(\text{H}\beta ) - A(\text{H}\alpha ) = E(\text{H}\beta -\text{H}\alpha ) = 2.5\times \text{log}_{10} \left\{ \frac{\text{H}\alpha/\text{H}\beta}{2.86} \right\} \label{ebetaminusalpha} \end{equation} in which we have used the fact that the intrinsic H$\alpha$ to H$\beta$ ratio is 2.86 for Case B recombination with a temperature $T = 10^4 $K and an electron density $n_e = 10^2$ cm$^{-3}$ \citep{Osterbrock:89}

By setting the right hand sides of Equations \ref{betaminusalpha} \& \ref{ebetaminusalpha} equal to each other and rearranging, we find that \begin{equation}E(B-V) = \frac {2.5}{\kappa (\text{H}\beta )- \kappa (\text{H}\alpha )} \text{log}_{10} \left\{ \frac{\text{H}\alpha/\text{H}\beta}{2.86} \right\}.\label{bminusv}\end{equation}  Which can then be used in Equation \ref{generala} to find the color excess at the wavelength for H$\alpha$.  We use the reddening curve from \citet{Calzetti:00} to obtain $\kappa$(H$\alpha$) = 3.33 and $\kappa$(H$\beta$) = 4.60 for the attenuation correction: \begin{equation}\begin{split} A(\text{H}\alpha )= \frac {2.5 \times \kappa (\text{H}\alpha )}{\kappa (\text{H}\beta )- \kappa (\text{H}\alpha )} \text{log}_{10} \left\{ \frac{\text{H}\alpha/\text{H}\beta}{2.86} \right\} \\ = 6.56\times \text{log}_{10} \left\{ \frac{\text{H}\alpha/\text{H}\beta}{2.86} \right\} .\label{attenuation}\end{split}\end{equation} 

Once we have our dust corrected H$\alpha$ luminosity values, we apply the \citet{Hao:11} conversion \begin{equation}SFR = \text{log} ((\text{H}\alpha _L)_\text{int}) - 41.27 \label{kennicutt12} \end{equation}  We calculate star formation rates for the IFU observed dwarf galaxy population in the range -2.88 $<$ log(M$_\sun$ yr$^{-1}$)$<$ - 1.65.  The full list of star formation rates can be found in Table \ref{results_table}.  Uncertainties in our star formation rates are propagated using standard Taylor series error propagation techniques from the H$\alpha$ flux measurement uncertainties obtained via 1000 MCMC iterations.

\begin{deluxetable*}{ l r r r r r r r r r r}
\tablecolumns{9}
\tablecaption{Observed Properties of Dwarf Galaxies  }
\startdata
\hline
\hline
	Galaxy & RA & Dec & Distance$^a$  & HI Mass$^a$ & Stellar Mass & Luminosity & Metallicity & SFR \\
	AGC\# & hh:mm:ss.s & $\pm$hh:mm:ss & Mpc ($\pm$ 2.43) & log(M$_\sun$) & log(M$_\sun$) & B Band mag & 12+log(O/H) & log(M$_\sun$ yr$^{-1}$)\\
	 
\hline
\\

AGC191702 & 09:08:36.5 & +05:17:32 & 8.7 $\pm$ 2.43 & 7.74 $\pm$ 0.18 & 6.67 $\pm$ 0.61 & -12.09 $\pm$ 0.63 & 7.94 $\pm$ 0.13 & -2.76 $\pm$ 0.34 \\
AGC202218 & 10:28:55.8 & +09:51:47 & 19.6 $\pm$ 2.43 & 7.75 $\pm$ 0.5 & 8.12 $\pm$ 0.38 & -14.95 $\pm$ 0.28 & 8.22 $\pm$ 0.11 & -1.72 $\pm$ 0.29 \\
AGC212838 & 11:34:53.4 & +11:01:10 & 10.3 $\pm$ 2.43 & 7.6 $\pm$ 0.19 & 6.94 $\pm$ 0.56 & -12.68 $\pm$ 0.53 & 8.05 $\pm$ 0.14 & -2.73 $\pm$ 0.32 \\
AGC220755 & 12:32:47.0 & +07:47:58 & 16.4 $\pm$ 2.43 & 7.18 $\pm$ 1.2 & 7.76 $\pm$ 0.43 & -13.94 $\pm$ 0.33 & 8.49 $\pm$ 0.16 & -2.51 $\pm$ 0.64 \\
AGC220837 & 12:36:34.9 & +08:03:17 & 16.4 $\pm$ 2.43 & 7.41 $\pm$ 0.54 & 8.78 $\pm$ 0.46 & -16.11 $\pm$ 0.33 & 8.56 $\pm$ 0.13 & -2.2 $\pm$ 1.28 \\
AGC220860 & 12:38:15.5 & +06:59:40 & 16.4 $\pm$ 2.43 & 7.22 $\pm$ 1.39 & 7.57 $\pm$ 0.42 & -13.92 $\pm$ 0.33 & 7.82 $\pm$ 0.13 & -2.14 $\pm$ 0.14 \\
AGC221000 & 12:46:04.4 & +08:28:34 & 16.5 $\pm$ 2.43 & 7.46 $\pm$ 0.83 & 8.35 $\pm$ 0.44 & -15.44 $\pm$ 0.33 & 8.35 $\pm$ 0.05 & -1.65 $\pm$ 0.08 \\
AGC221004 & 12:46:15.3 & +10:12:20 & 16.7 $\pm$ 2.43 & 7.66 $\pm$ 0.55 & 7.98 $\pm$ 0.3 & -14.61 $\pm$ 0.22 & 8.35 $\pm$ 0.16 & -2.23 $\pm$ 0.22 \\
AGC225852 & 12:59:41.9 & +10:43:40 & 16.6 $\pm$ 2.43 & 7.68 $\pm$ 0.53 & 7.57 $\pm$ 0.42 & -13.89 $\pm$ 0.33 & 8.27 $\pm$ 0.22 & -2.53 $\pm$ 0.38 \\
AGC225882 & 12:03:26.3 & +13:27:34 & 23.6 $\pm$ 2.43 & 8.15 $\pm$ 0.3 & 7.06 $\pm$ 0.35 & -13.66 $\pm$ 0.25 & 7.95 $\pm$ 0.18 & -2.33 $\pm$ 0.05 \\
AGC227897 & 12:50:04.2 & +06:50:51 & 16.6 $\pm$ 2.43 & 7.44 $\pm$ 0.89 & 6.58 $\pm$ 0.45 & -12.08 $\pm$ 0.35 & 7.67 $\pm$ 0.42 & -2.88 $\pm$ 0.24 \\

\enddata
\tablecomments{$^a$ Values obtained from the ALFALFA $\alpha$.40 catalog \citep{Haynes:11}.  Uncertainties in the distance are dominated by the local velocity
dispersion measured by \citep{Masters:05}}.
\label{results_table}
\end{deluxetable*}

\subsection{ALFALFA/SDSS Analysis}

\begin{figure}
\epsfig{ file=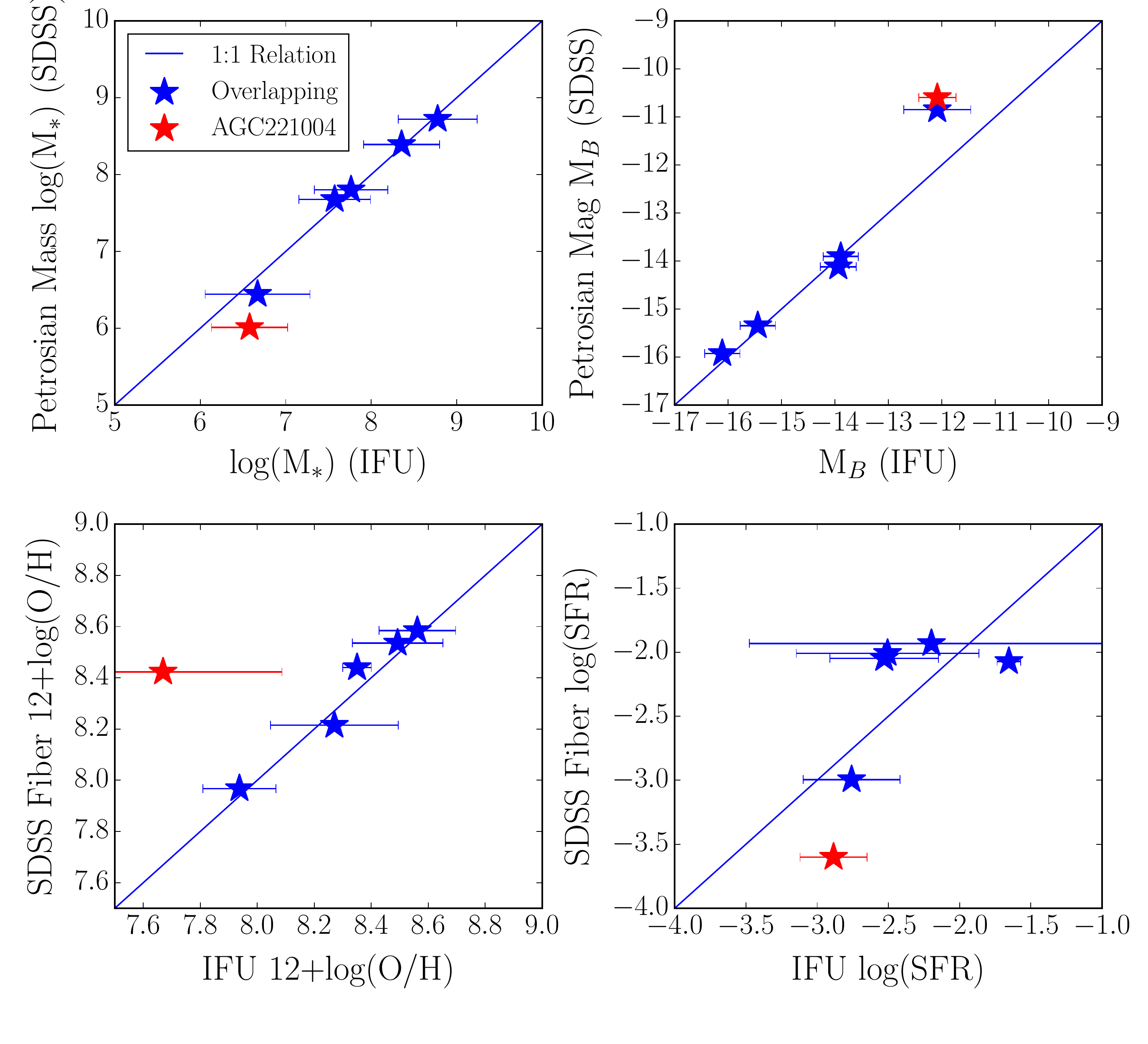, scale=0.30}
\caption{Comparison between the measured values in galaxies that overlap between the IFU observed dwarf galaxy sample and the larger ALFALFA/SDSS sample, showing the consistency between the data used to calculate observables for each of the two data sets.  Discrepancies in the stellar-mass and luminosity values can be explained by the issues with SDSS Petrosian magnitude measurements.  The poor correlation between SFR estimations can be explained by the patchy dwarf galaxy star formation being observed with a single SDSS fiber.  Galaxies in blue have passed the selection criteria show reasonable agreement between both samples, galaxies in red would be rejected from the ALFALFA/SDSS sample using the selection criteria outlined in Section \ref{SDSS_Section}.}
\label{consistency}
\end{figure}

In order to compare results derived for the IFU observed dwarf galaxy sample with the larger ALFALFA/SDSS sample, we must ensure that the ALFALFA/SDSS data is reduced in a manner as similar as possible to the IFU observed dwarf galaxy data.  HI masses between samples are obtained from the same ALFALFA database, therefore we are confident that all HI masses are derived similarly between the two samples.  For the stellar-mass, luminosity, gas-phase metallicity, and SFR calculations, we detail the similarities and differences between analysis of the two datasets in the following subsections.  Figure \ref{consistency} displays a comparison of the observables for 6 galaxies that overlap between the IFU observed dwarf galaxy sample and the larger ALFALFA/SDSS sample.  In general we find that they agree within 1 standard deviation, and outlying data points can be explained.  This leads us to conclude that comparisons between the samples are valid, however the outliers noted highlight the need for IFU observations and careful analysis in the low stellar-mass/luminosity regime.  The sample selection criteria outlined in Section \ref{SDSS_Section} is used to remove these outlier galaxies from our ALFALFA/SDSS sample.

\subsubsection{Stellar Mass}
\label{compare_stellar_mass}
Following the procedure in Section \ref{StellarMasses}, we perform the same calculations as in Equations \ref{stellar_mass_to_light} \& \ref{stellar_mass} to obtain stellar-mass estimations.  However, we use Petrosian magnitudes from the SDSS public database in place of Source Extractor derived photometry.  Although it would be ideal to perform the same Source Extractor procedure, doing so would be prohibitively time consuming for the entire ALFALFA/SDSS sample.  For the crossmatched ALFALFA/SDSS sample we utilize ALFALFA catalog model distances for galaxies where $cz <$ 6,000 km s$^{-1}$ and cosmological distances otherwise.  In Figure \ref{consistency} we find that 5 of the 6 galaxies that exist in both samples are within 1 standard deviation of the 1:1 correlation, with one significant outlier (AGC221004).

To relieve this tension, we examine the MPA-JHU value added catalog \citep{Kauffmann:03b, Brinchmann:04, Tremonti:04} which has stellar mass estimations based on spectral energy distributions (SED) fits to the SDSS fiber spectroscopy.  These stellar-mass estimations utilize \citet*{Bruzual:03} stellar templates and assume a Kroupa IMF \citep{Kauffmann:03b}.  We find that for galaxies which overlap between the IFU observed dwarf galaxy sample and the ALFALFA/SDSS sample, the Source Extractor based stellar-mass estimations agree within 7\% with the MPA-JHU SDSS spectroscopic fiber derived total stellar mass estimates, suggesting that the SDSS Petrosian photometry is inaccurate for AGC221004. 

As outlined in \citet{West:10}, roughly 25\% of Petrosian fluxes for the SDSS population are inaccurate due to issues with parent-child splitting in the SDSS algorithms, this is consistent with what we have observed in Figure \ref{consistency}.  The MPA-JHU catalog also contains a significant number of outliers, so would not be a complete replacement for the SDSS petrosian magnitudes, instead we have implemented a quality control cut on the ALFALFA/SDSS data in which we exclude galaxies from our sample for which the stellar mass estimates disagree by more than 0.5 dex between the Petrosian and MPA-JHU stellar mass estimations.  With this quality control check in place, we continue to use the \citet{West:10} based mass estimations for the IFU observed dwarf galaxies and the ALFALFA/SDSS population in order to keep the analysis as consistent as possible between the two samples.

\subsubsection{Luminosity}
Following the procedure in Section \ref{luminosity}, we once again utilize the Petrosian magnitudes provided by the SDSS catalog in order to derive B-band luminosities using Equation \ref{m_B}.  When we compare the results of the Petrosian magnitude derived luminosities to the Source Extractor derived luminosities in Figure \ref{consistency}, once again we find that AGC221004 exists considerably offset from the expected 1:1 correlation.  As observed in Section \ref{compare_stellar_mass}, this is likely due to the known issues with the SDSS pipeline aggressively splitting up the photometry of irregular HII regions.  This demonstrates the need for careful analysis of dwarf galaxies in the low-luminosity regime.   Errant galaxies such as AGC221004 are removed from the ALFALFA/SDSS sampled by the quality control cuts outlined in Section \ref{SDSS_Section}.

\subsubsection{Gas-phase Metallicity}
To derive the gas-phase metallicity estimations for the larger ALFALFA/SDSS sample, we rely upon emission line fluxes from SDSS single fiber spectroscopy.  We apply the same D02 oxygen abundance estimation (Equation \ref{D02}) as was performed for the IFU observed dwarf sample.  In Figure \ref{consistency} we find good agreement between both samples.  This is somewhat expected considering that the SDSS fibers target the brightest part of a galaxy, and the integrated IFU fluxes are dominated by the same bright areas.  It is worth noting that this lends credence to our emission line de-blending because the SDSS spectroscopy is higher spectral resolution and not significantly blended.  

\subsubsection{Star Formation Rate}
To determine SFRs for the larger ALFALFA/SDSS sample, we use the H$\alpha$ flux reported in the SDSS 3$^{\prime \prime }$ single fiber spectroscopy database.  We correct for dust using the same procedure as with the IFU observed dwarf galaxy sample.  In addition, we perform an aperture correction to our SFR calculations.  As can be seen in Figure \ref{field_of_views}, the SDSS fiber often misses a significant amount of flux in the emission regions of nearby galaxies.  Therefore, to more accurately compare our IFU observations which cover the entire star forming region to the SDSS observations, we apply the aperture correction technique outlined in \citet{Hopkins:03} \begin{equation} A = 10^{-0.4(r_{\text{petro}} - r_{\text{fiber}})} \end{equation} where $r_{\text{petro}}$ denotes the SDSS {\it r}-band Petrosian flux, and $r_{\text{fiber}}$ denotes the SDSS {\it r}-band flux within the fiber.

As observed in Figure \ref{consistency}, the calculated star formation rates may vary by as much as 0.5 dex in SFR between the two samples, emphasizing the need for IFU observations to collect all of the H$\alpha$ flux.  This is to be expected considering the patchy star forming nature of the IFU observed dwarf galaxy sample.  The patchy nature of star formations also means that this aperture correction is imperfect, as seen is Figure \ref{consistency}.  These facts indicate that ALFALFA/SDSS SFRs are likely to be unreliable for galaxies requiring a large aperture correction.  This necessitates the sample selection cut where we exclude galaxies for which the difference between the Petrosian {\it r}-band magnitude and the fiber {\it r}-band magnitude differ by more than 5 mag as described in Section \ref{SDSS_Section}.

\vspace{10 mm}

\section{Results}

\subsection{Mass-Metallicity Relation and Luminosity-Metallicity Relation}
\label{mzr_lzr_section}

\begin{figure}
\epsfig{ file=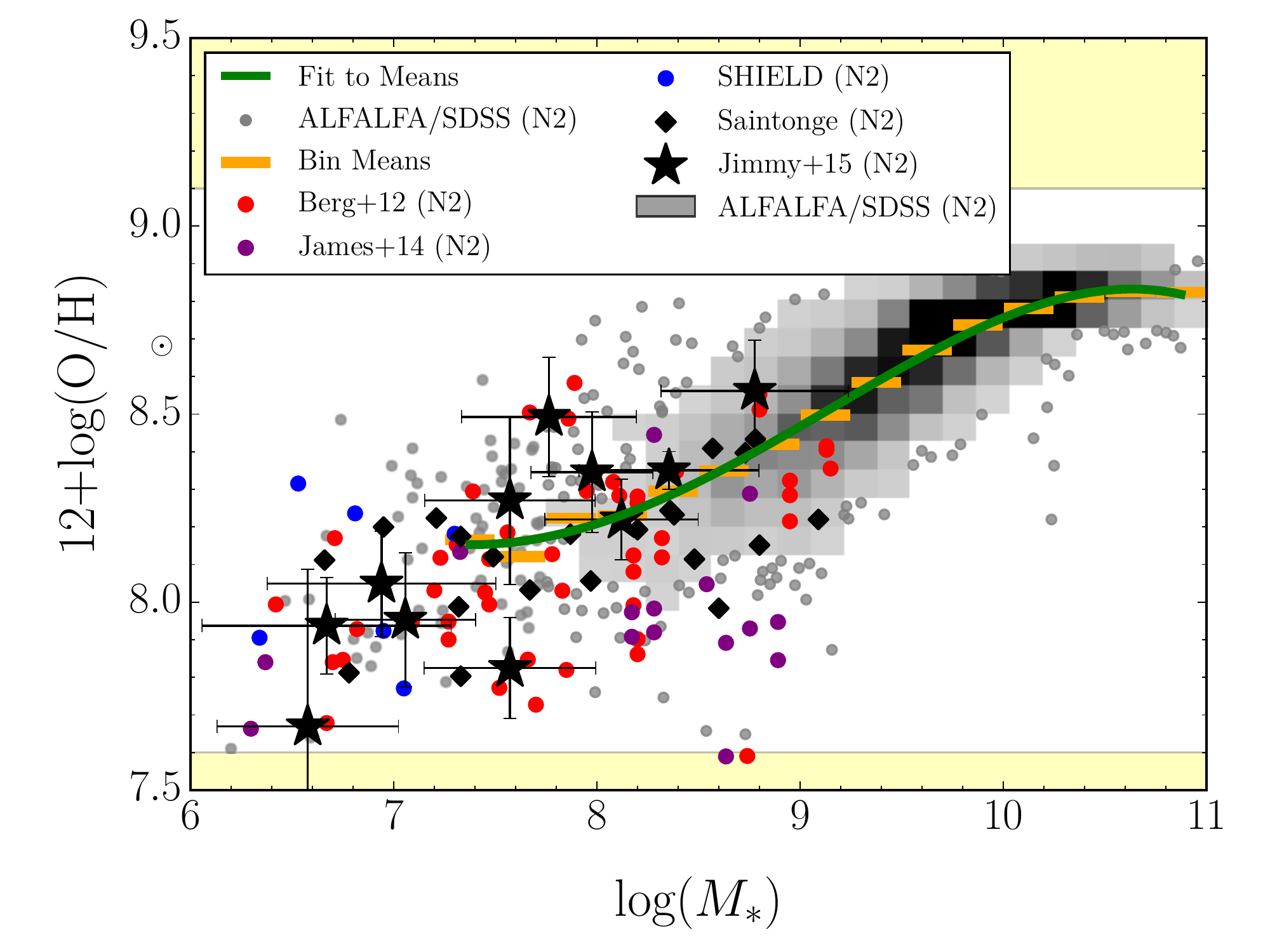, scale=0.45}
\epsfig{ file=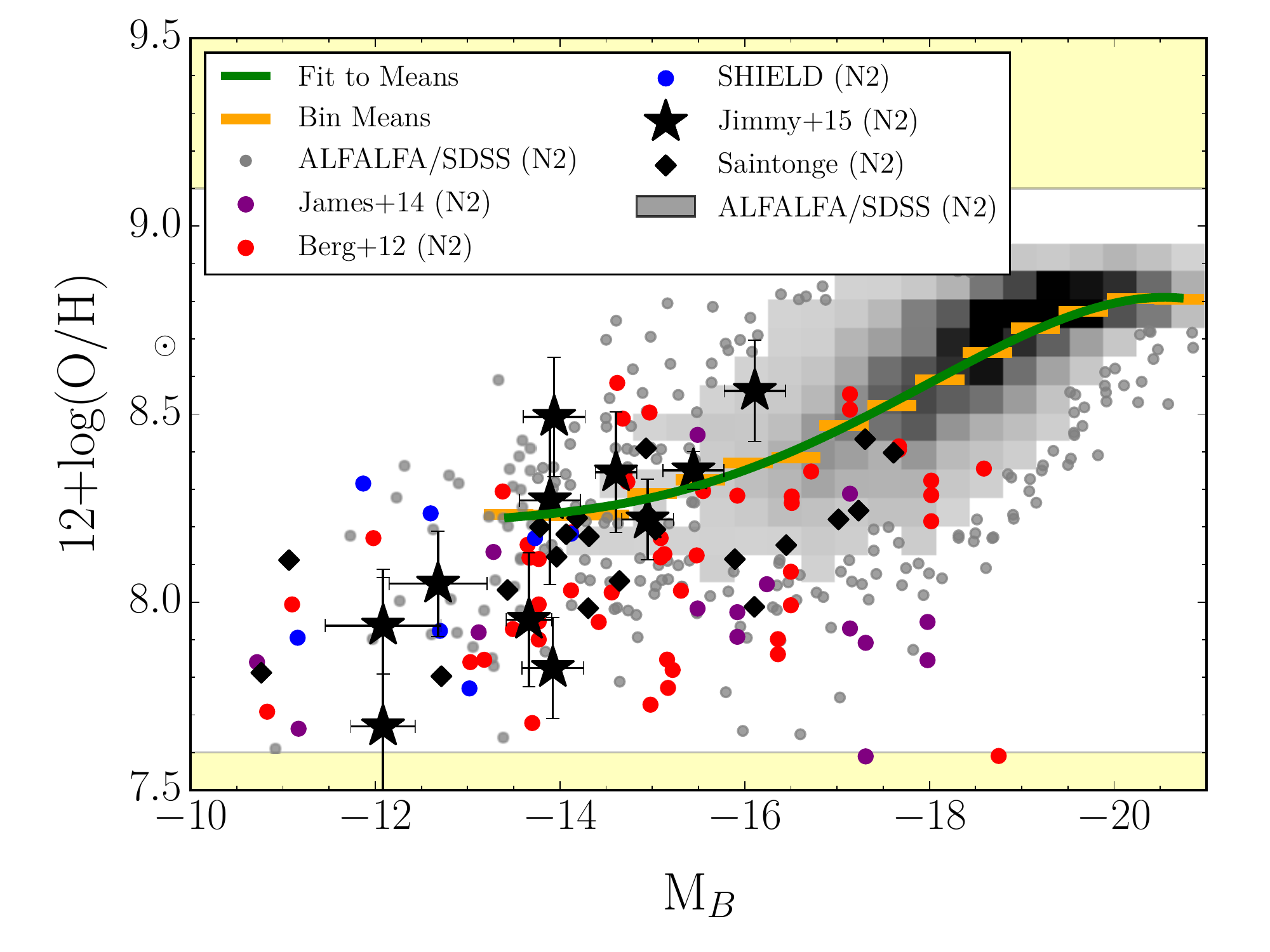, scale=0.45}
\caption{The mass-metallicity relation ({top}) and the luminosity-metallicity relation ({bottom}) for the IFU observed dwarf galaxy sample, along with several low stellar mass galaxy observations and the ALFALFA/SDSS sample.  The scatter in both the MZR (LZR) fit around the bin means is 0.05 dex (0.03 dex).  For reference we plot a sun symbol at solar metallicity, showing that all of the IFU observed dwarf galaxy oxygen abundances are below solar.  The yellow regions indicate the calibration limits of the D02 N2 oxygen abundance estimation.  We have used the published emission line fluxes from other authors to estimate the oxygen abundance in the same D02 system.  The MZR appears to hold down to the lowest stellar masses observed (log(M$_*$) = 7.25).  
}
\label{MZRLZR}
\end{figure}

The mass-metallicity relation (MZR) can be seen in Figure \ref{MZRLZR} along with homogenized observations from earlier dwarf/low luminosity galaxy surveys.  Similar to the process outlined in \citet{Mannucci:10}, we bin the ALFALFA/SDSS crossmatched sample by stellar mass, into bins 0.25 dex wide.  We then fit a 4 degree polynomial to the mean values of the bins and find the following relationship: 
\begin{equation}\begin{split}12+log(O/H) = 8.756 + 0.219 \times m -0.119 \times m^2 \\ -0.052 \times m^3 - 0.003 \times m^4\end{split} \label{mzr_fit} \end{equation} where $m$ = log(M$_*$)-10.  The MZR is fit to data from the literature collected samples, the ALFALFA/SDSS crossmatched sample, and the IFU observed dwarf galaxy sample.  In order to ensure robust results, we require that there are at least 21 galaxies per stellar mass bin.  We use the limit of 21 galaxies per bin in order to be consistent with the analysis performed in \citet{Bothwell:13}.  In total there are 15 stellar-mass bins covering the stellar-mass range 7.25 $<$ log(M$_*$) $<$ 11.0.  The residual 1$\sigma$ scatter of every galaxy to the MZR fit in Equation \ref{mzr_fit} is 0.11 dex, similar to the 0.1 dex scatter reported for the MZR of \citet{Tremonti:04}.  The 1$\sigma$ scatter of the mean values in each stellar-mass bin is 0.05 dex.  The 1$\sigma$ scatter of the means is the metric that we will use to compare scatter between various metallicity relations.

The LZR can also be seen in Figure \ref{MZRLZR} along with homogenized observations from similar dwarf/low luminosity galaxy surveys.  We bin the combined literature, ALFALFA/SDSS, and IFU observed sample into bins 0.52 dex wide.  We then fit a 4 degree polynomial to these mean values and find the following relationship: \begin{equation}\begin{split}12+log(O/H) = -18.6 - 7.23 \times \text{M}_B - 0.724 \times {\text{M}_B}^2 \\ - 0.032 \times {\text{M}_B}^3 - 0.0005 \times {\text{M}_B}^4.\end{split} \label{lzr_fit} \end{equation}  The LZR is fit to 15 luminosity bins over a range $-13.7 < \text{M}_B) < -21.0$ in order to ensure that there are at least 21 galaxies per luminosity bin.  The residual 1$\sigma$ scatter of every galaxy to the LZR fit in Equation \ref{lzr_fit} is 0.14 dex.  The residual 1$\sigma$ scatter of the mean values to the LZR in each luminosity bin is 0.03 dex.  Again, this is the metric that we will use to compare scatter between various metallicity relations.  

\subsection{MZR/LZR with SFR and HI Mass Dependance}
\label{colorful}

\begin{figure}
\epsfig{ file=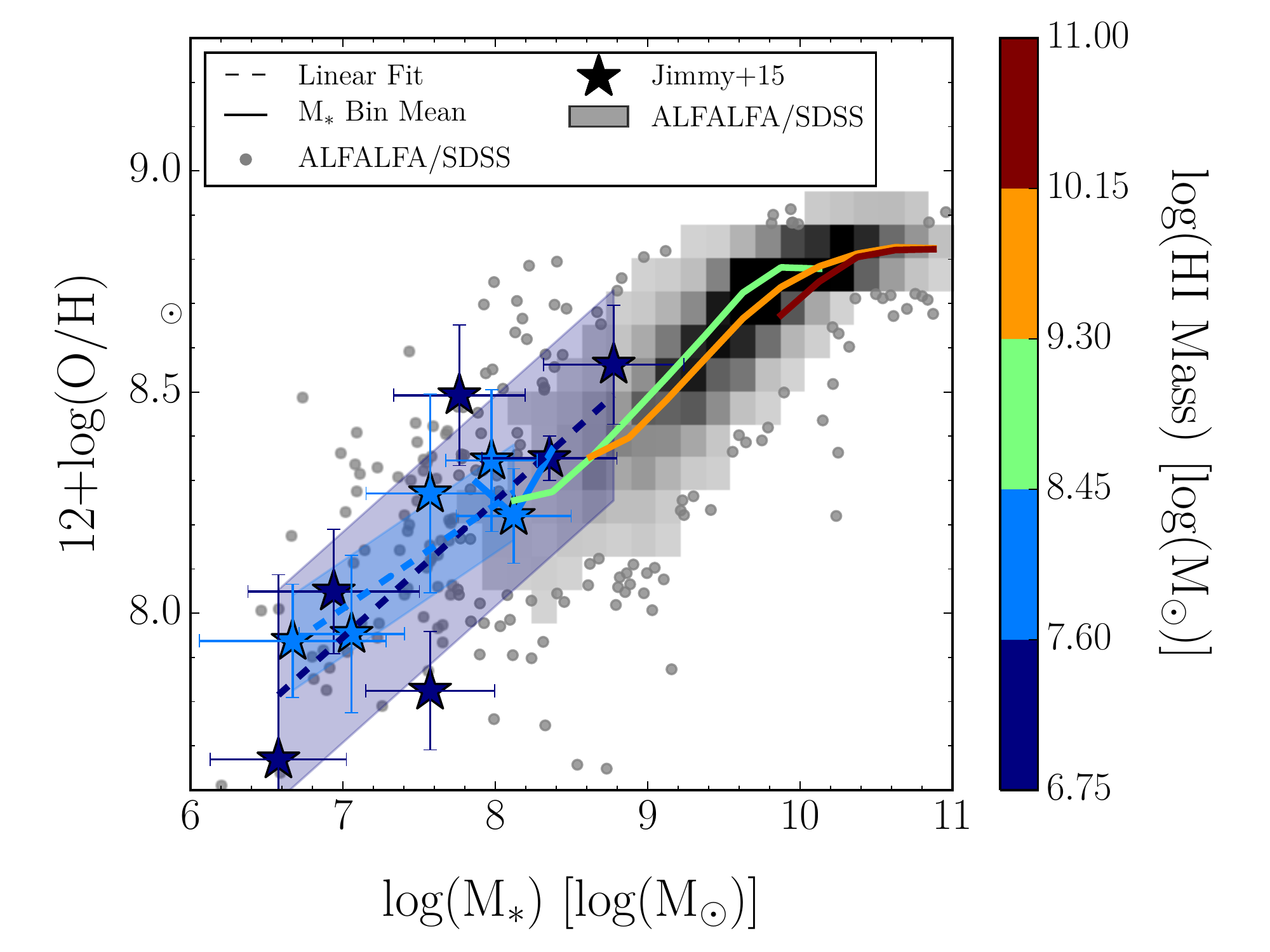, scale=0.45}
\epsfig{ file=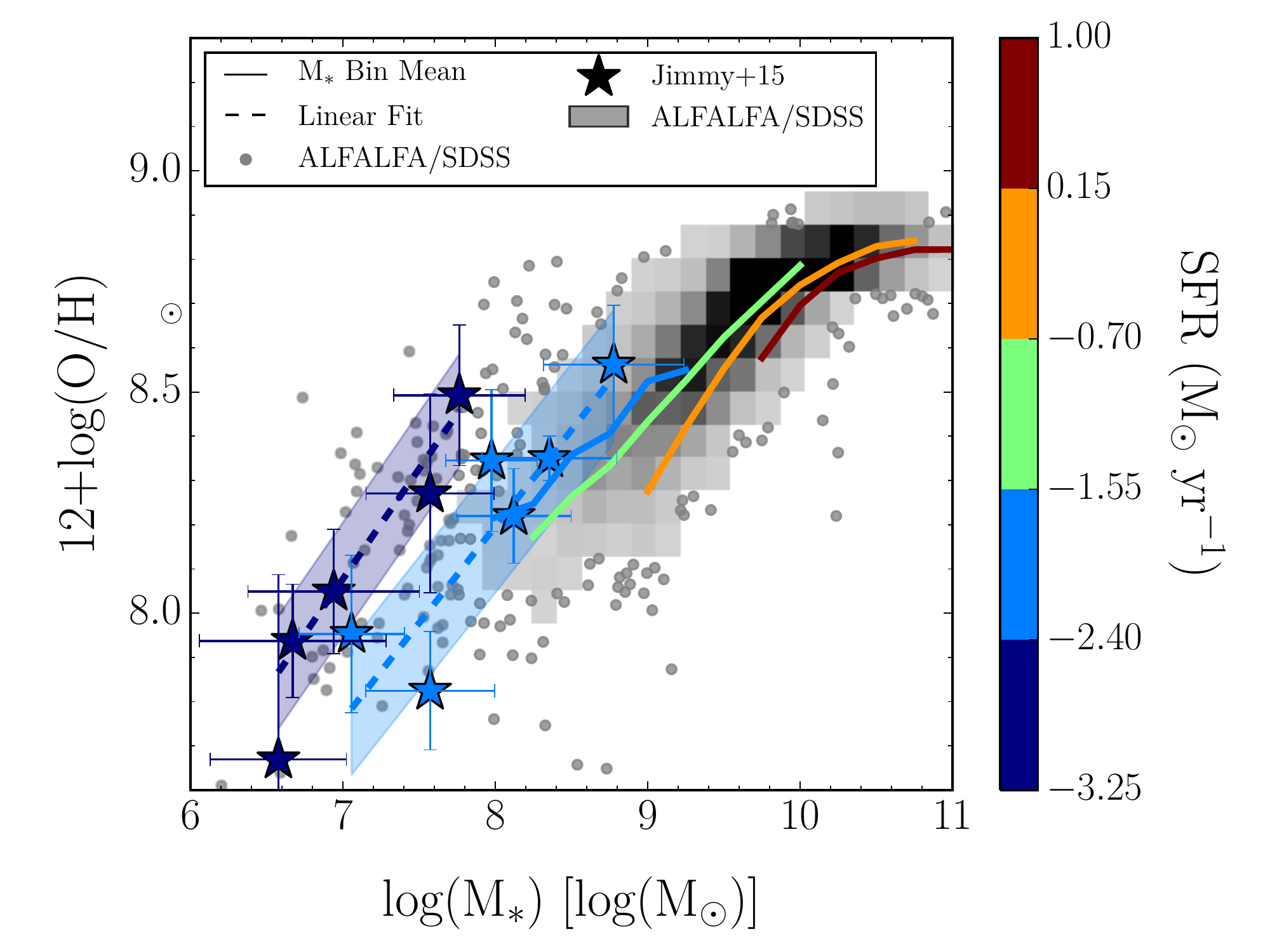, scale=0.45}
\caption{The mass-metallicity relation as seen in Figure \ref{MZRLZR} color-coded by HI mass {(top)} and SFR {(bottom)}.  The stars indicate individual observations within our IFU observed sample of dwarf galaxies.  The dashed lines indicate linear least squares fits to these points, and the shaded regions indicate the 1$\sigma$ standard deviations to these fits.  The solid curves indicate the mean values of the ALFALFA/SDSS sample, separated into HI mass bins (top) and SFR bins (bottom).  Shown in the background are the ALFALFA/SDSS points which are binned to produce the color-coded means.  We find little overlap between the HI mass bins (top) and SFR bins (bottom).}
\label{colorful_fmr}
\end{figure}

\begin{figure}
\epsfig{ file=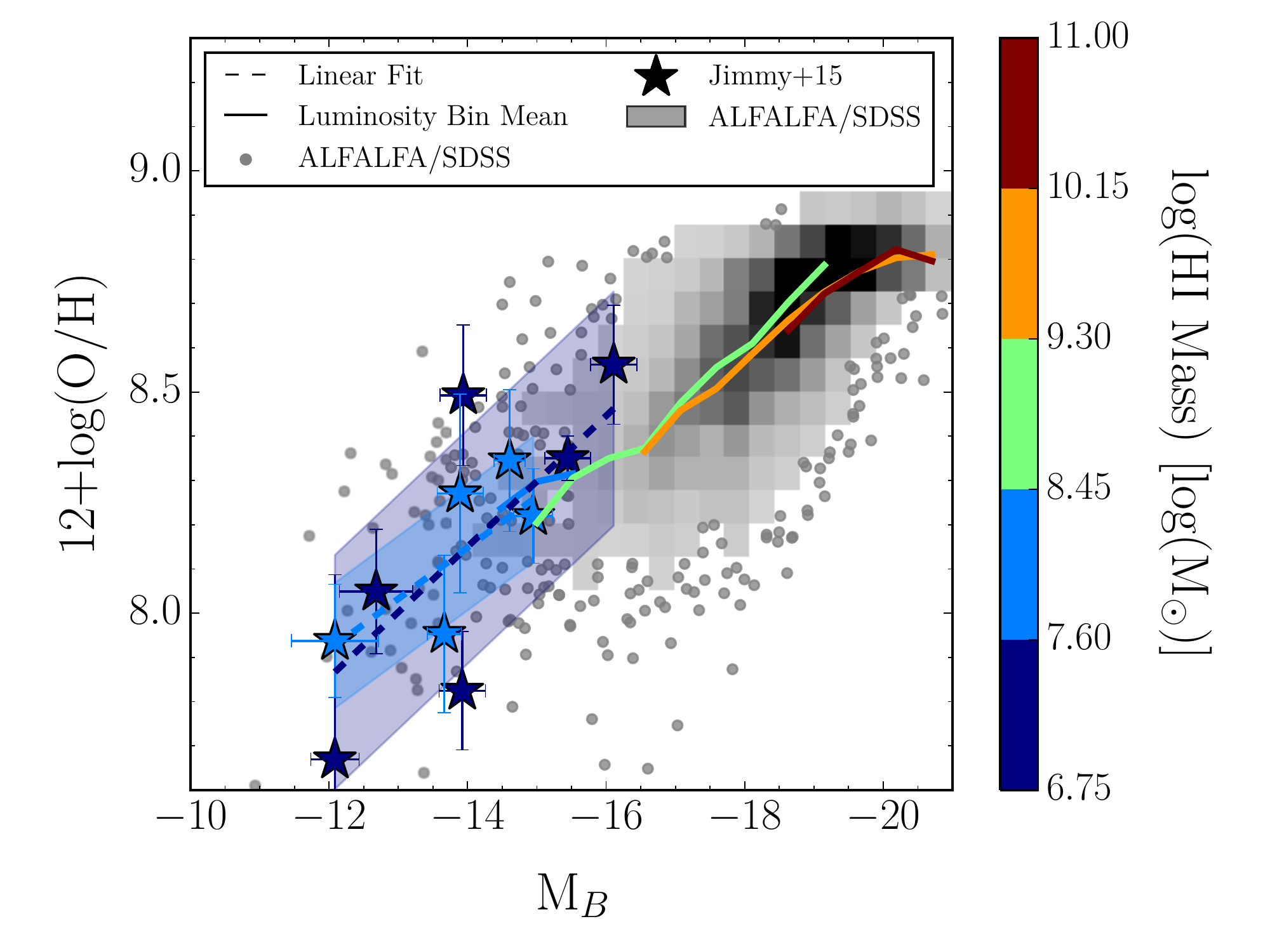, scale=0.45}
\epsfig{ file=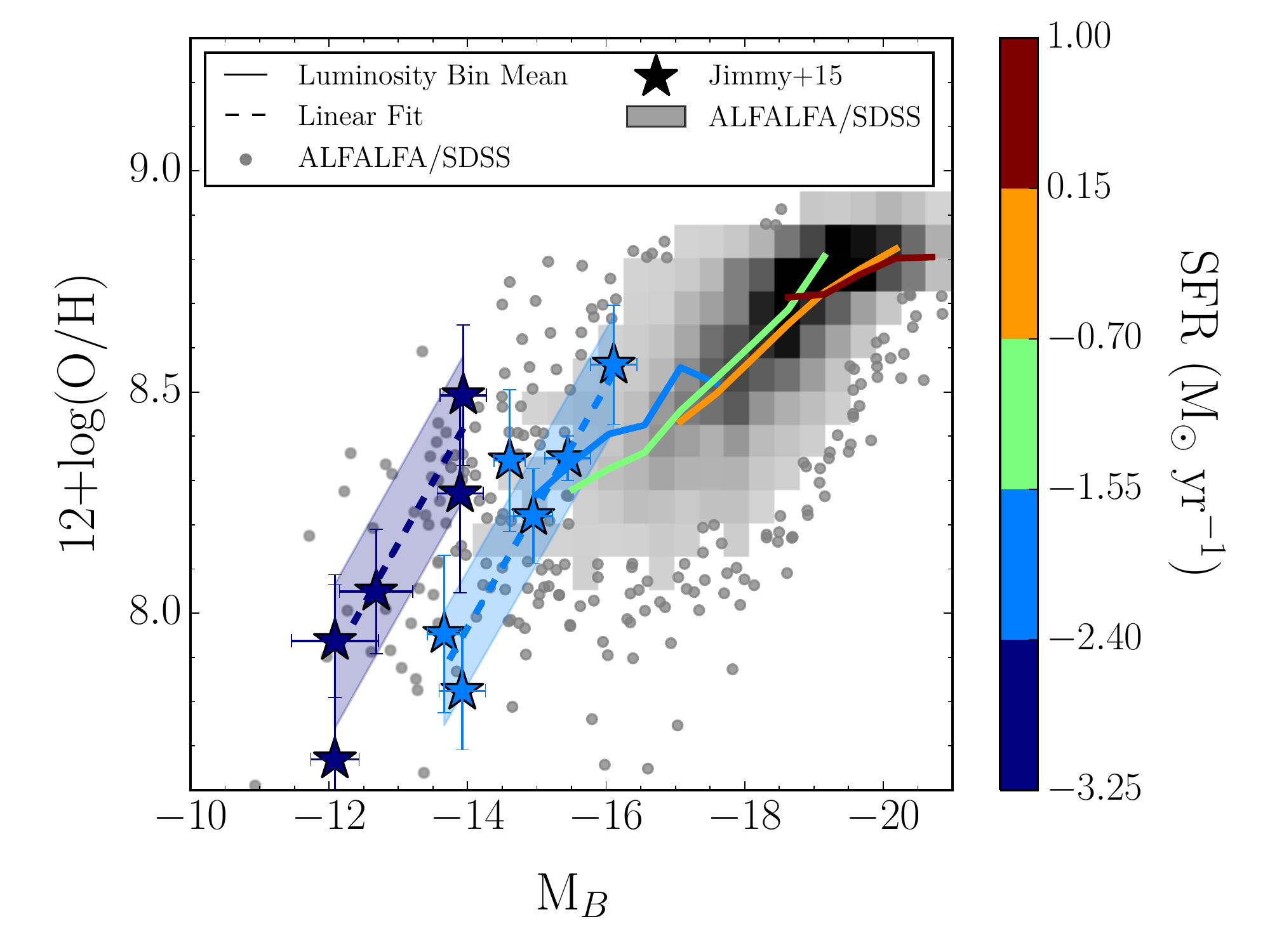, scale=0.45}
\caption{The luminosity-metallicity relation as seen in Figure \ref{MZRLZR} color-coded by HI mass {(top)} and SFR {(bottom)}.  The stars indicate individual observations within our IFU observed sample of dwarf galaxies.  The solid curves indicate the mean values of the ALFALFA/SDSS sample, separated into HI mass bins (top) and SFR bins (bottom).  Shown in the background are the ALFALFA/SDSS points which are binned to produce the color-coded means.  We find little overlap between the HI mass bins (top) and SFR bins (bottom).}
\label{colorful_lzr_fmr}
\end{figure}

We test whether or not a fundamental metallicity relation explains the enhanced scatter of the MZR and LZR by following a similar procedure as \citet{Mannucci:10} and \citet{Bothwell:13}.  We bin the crossmatched ALFALFA/SDSS data into SFR bins of 0.85 log(M$_\sun$ yr$^{-1}$) and HI mass bins of 0.85 log(M$_\sun$).  We use larger HI mass and SFR bins than \citet{Bothwell:13} to ensure that we have at least 21 objects per bin in our lower stellar mass/luminosity regimes.  Figures \ref{colorful_fmr} \& \ref{colorful_lzr_fmr} show the result of binning the MZR and the LZR in terms of both SFR and HI mass.  A clear separation exists between the HI mass bins shown in the top of Figures \ref{colorful_fmr} \& \ref{colorful_lzr_fmr}, as well as the SFR bins in the bottom of Figures \ref{colorful_fmr} \& \ref{colorful_lzr_fmr}.  The numerical values of the means used to produce the solid lines in Figures \ref{colorful_fmr} \& \ref{colorful_lzr_fmr} can be found in Appendix \ref{AppendixC}.  Using these larger HI mass and SFR bins, we are able to extend the ALFALFA/SDSS means relation down to stellar mass bins as low as log(M$_*$) = 7.75.

Also shown in Figures \ref{colorful_fmr} \& \ref{colorful_lzr_fmr} are the linear least squares fits to the IFU observed dwarf galaxies.  The dwarf galaxy sample of this study is not large enough to produce similarly binned mean relationships for dwarf IFU observed data alone, therefore we instead perform least squares optimized linear fits to the dwarf IFU data to demonstrate that it is consistent with the FMR.  The numerical values for the linear fits and their 1$\sigma$ standard deviations can be seen in Table \ref{linear_fits_table}.

The IFU observed dwarf galaxy sample overlaps with the larger ALFALFA/SDSS sample by $\sim$ 1 dex, providing a valuable consistency check.  The added benefit of the IFU observed dwarf galaxy sample is that it is able to test the fundamental metallicity relation down to stellar masses 2 orders of magnitude lower than observed in \citet{Mannucci:10} and \citet{Bothwell:13}.

\begin{deluxetable}{ l r r r }
\tablecolumns{4}
\tablecaption{Linear Fits to Dwarf IFU SFR \& HI Mass Bins  }
\startdata

\hline
\hline
	& {Stellar Mass} & & \\
	SFR Bin & Slope & Y-Intercept & 1$\sigma$ Scatter \\
\hline
\\
-2.4 to -1.55 & 0.44 $\pm$ 0.01 & 4.7 $\pm$ 0.7 & 0.15\\
-3.25 to -2.4 & 0.5 $\pm$ 0.0 & 4.59 $\pm$ 0.22 & 0.13\\

& & & \\
	HI Mass Bin & Slope & Y-Intercept & 1$\sigma$ Scatter \\
\hline
\\
6.75 to 7.6 & 0.3 $\pm$ 0.01 & 5.85 $\pm$ 0.85 & 0.22\\
7.6 to 8.45 & 0.24 $\pm$ 0.0 & 6.34 $\pm$ 0.26 & 0.1\\
& & & \\

\hline
\hline
	& Luminosity & & \\
	SFR Bin & Slope & Y-Intercept & 1$\sigma$ Scatter \\
\hline
\\
-2.4 to -1.55 & -0.27 $\pm$ 0.0 & 4.15 $\pm$ 0.64 & 0.13\\
-3.25 to -2.4 & -0.28 $\pm$ 0.0 & 4.54 $\pm$ 0.5 & 0.16\\
& & & \\
	HI Mass Bin & Slope & Y-Intercept & 1$\sigma$ Scatter \\
\hline
\\
6.75 to 7.6 & -0.15 $\pm$ 0.0 & 6.07 $\pm$ 0.85 & 0.24\\
7.6 to 8.45 & -0.13 $\pm$ 0.0 & 6.27 $\pm$ 0.32 & 0.11\\

\enddata
\tablecomments{ Linear fit results for the dotted lines as seen in Figures \ref{colorful_fmr} \& \ref{colorful_lzr_fmr}.  The 1$\sigma$ scatter is the width of the shaded regions in Figures \ref{colorful_fmr} \& \ref{colorful_lzr_fmr}.}
\label{linear_fits_table}
\end{deluxetable}

\subsection{Fundamental Metallicity Relations}

In order to test whether the IFU observed dwarf galaxy sample is constant with the offsets in SFR and HI Mass observed for the larger ALFALFA/SDSS crossmatched sample, we examine the fundamental metallicity relationships.  We also introduce nomenclature for the luminosity equivalent of the FMR as the fundamental metallicity luminosity (FML) relation, which can be dependent upon SFR (\fmlsfr ) or HI-gas mass (\fmlhi ).  The FMR and FML are a projection of three-dimensional parameter space into two dimensions to minimize the scatter observed. 

\subsubsection{MZR SFR Dependence (\fmrsfr )}
\label{fmrsfr_sec}
\begin{figure*}
\epsfig{ file=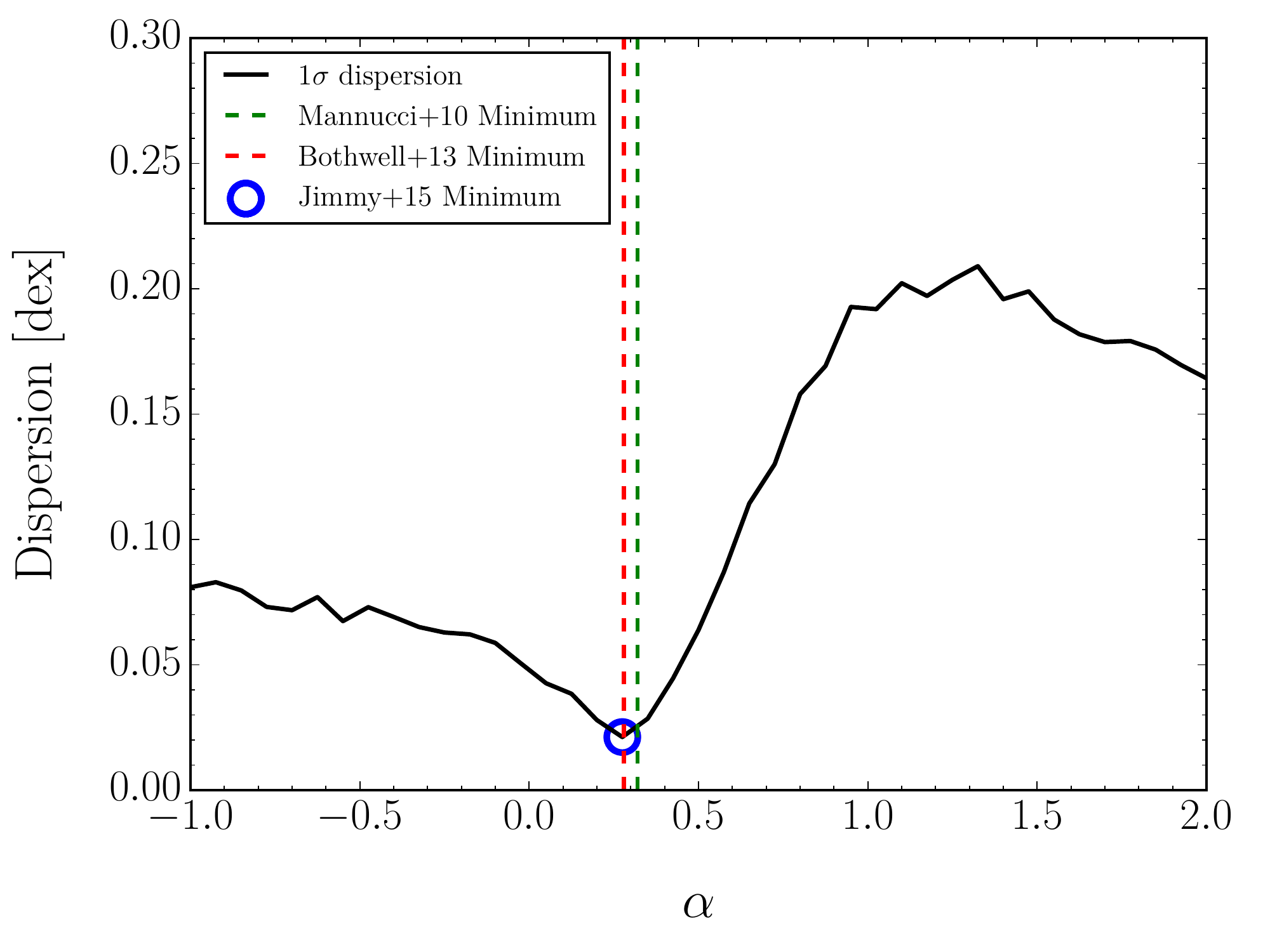, scale=0.45}
\epsfig{ file=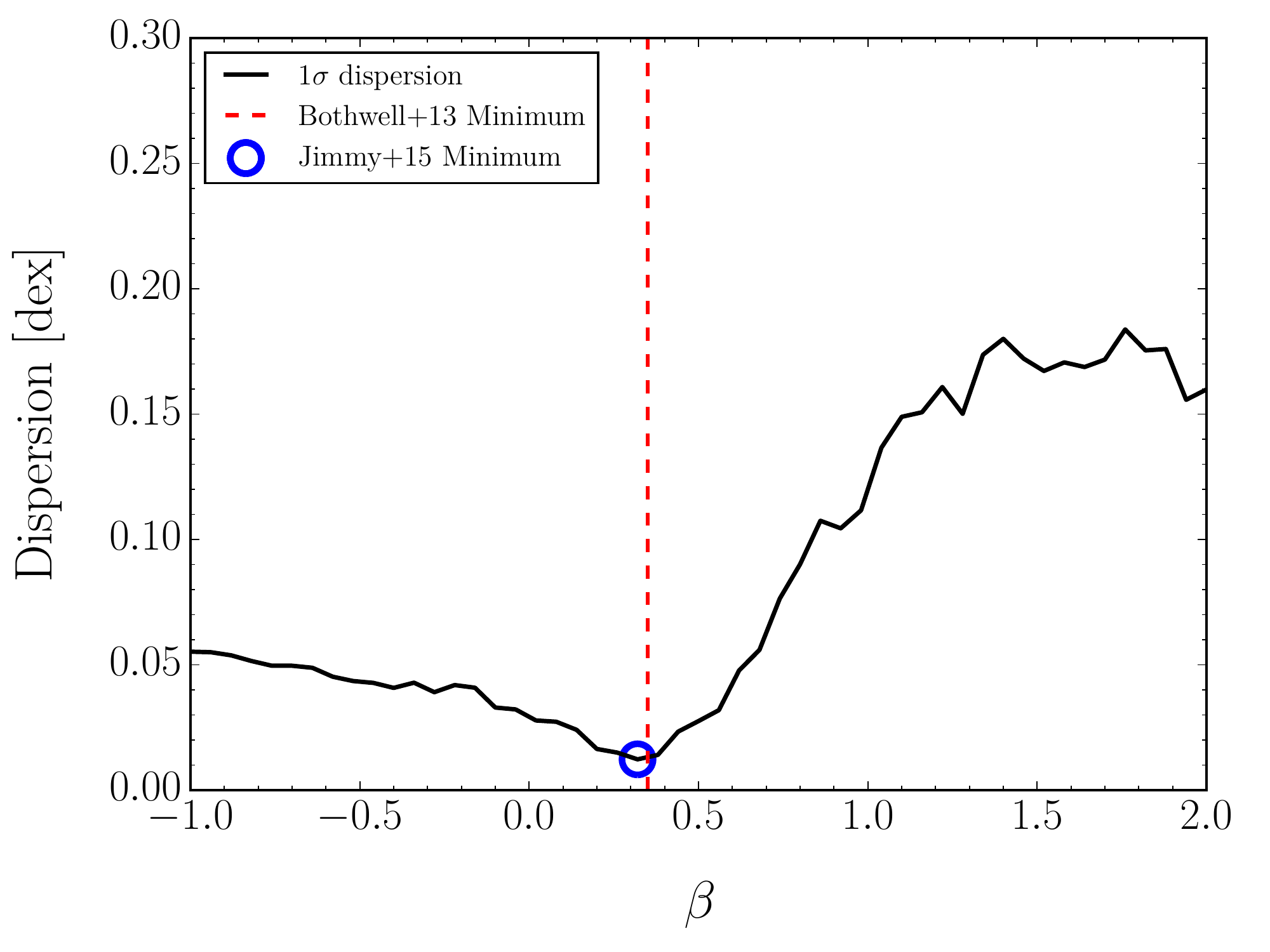, scale=0.45}
\caption{ Left: 1$\sigma$ scatter in the means of the \fmrsfr\ as a function of $\alpha$.  The optimal scatter found by \citet{Mannucci:10} is indicated by the dashed green vertical line, and the optimal alpha value found by \citet{Bothwell:13} is indicated by the dashed red vertical line.  We find the lowest scatter in the \fmrsfr\ when $\alpha = 0.28$  Right: 1$\sigma$ scatter in the means of the \fmrhi\ as a function of $\beta$.  The dashed red vertical line indicates the optimal $\beta$ value from \citet{Bothwell:13} (No $\beta$ minimum exists from \citet{Mannucci:10} as they did not study the \fmrhi\ ).   We find the lowest scatter in the \fmrhi\ when $\beta$ = 0.33. }
\label{scatter}
\end{figure*}
\citet{Mannucci:10} define a variable $\mu_\alpha$ which combines SFR with stellar mass such that \begin{equation} \mu_\alpha = \text{log}(\text{M}_*)-\alpha \text{log}(\text{SFR}). \label{fmrsfr_eq} \end{equation}  They find that $\alpha = 0.32$ minimizes the scatter in the \fmrsfr\ relation, whereas \citet{Bothwell:13} find $\alpha = 0.28$ is the optimal value.  

We perform the same tests on our larger crossmatched ALFALFA/SDSS sample in order to find the value of $\alpha$ minimizes the \fmrsfr\ scatter.  To find the optimal value of $\alpha$, we project the ALFALFA/SDSS data onto the $\mu$-metallicity plane and bin our data using the same SFR bins used for Figure \ref{colorful_fmr}.  Each SFR binned dataset is binned a 2nd time in $\mu$ space similar to the stellar-mass binning performed in Section \ref{colorful} and we once again perform the same cut of 21 galaxies per $\mu$-SFR bin.  We take the mean metallicity of each $\mu$-SFR bin and fit a 4th degree polynomial to the binned  mean metallicity values.  This process is repeated for values of $\alpha$ ranging from -1 to 2.  

The dataset used for this procedure includes only the larger ALFALFA/SDSS dataset.  We tested using the combined ALFALFA/SDSS and IFU observed dwarf data and found the same scatter per alpha value tested.  This is due to the fact that there are only 11 data points in the IFU observed data set being added to the considerably larger ALFALFA/SDSS sample.  Also producing the \fmrsfr\ using only the larger ALFALFA/SDSS data means that the test of whether or not the IFU observed dwarf galaxy data fits the same \fmrsfr\ is completely independent.

\begin{figure}
\epsfig{ file=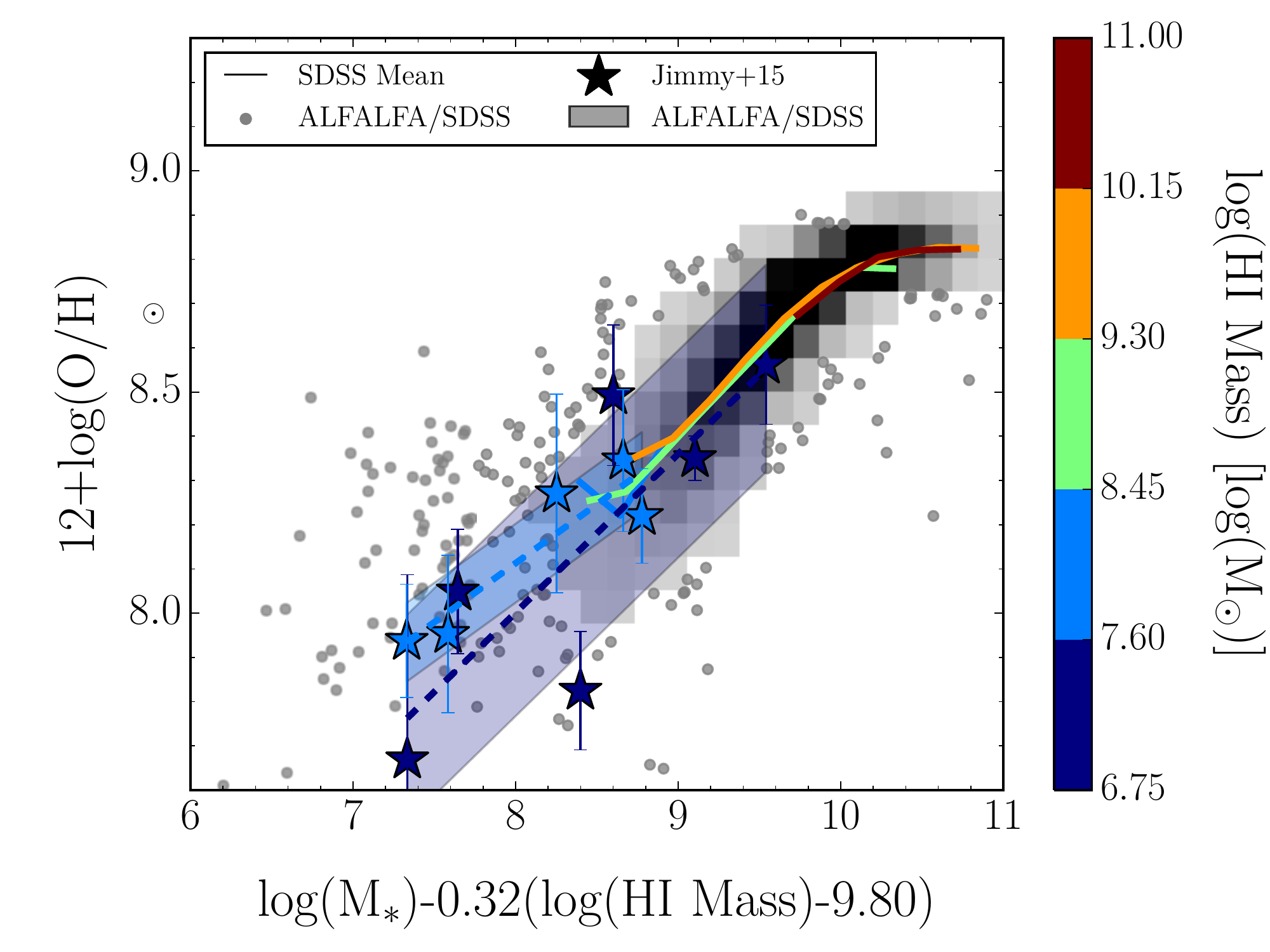, scale=0.45}
\epsfig{ file=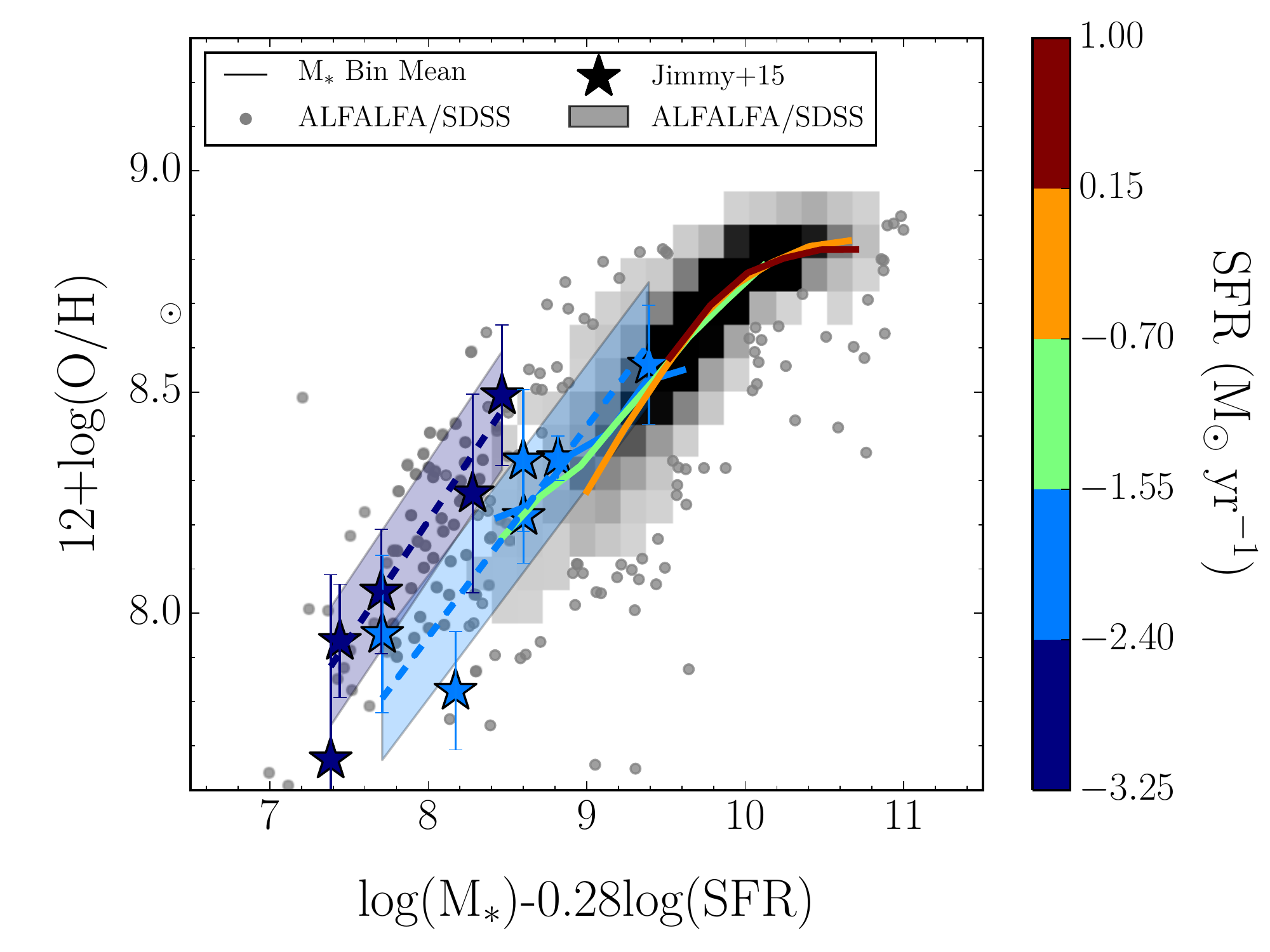, scale=0.45}
\caption{\fmrhi\ (top) and \fmrsfr\ (bottom) as calculated using the $\alpha$ (top) and $\beta$ (bottom) values found to minimize scatter (Figure \ref{scatter}).  We find that the \fmrhi\ relation is consistent down to the lowest HI-mass bin, where as the lowest SFR bin in the \fmrsfr\ is offset 1$\sigma$ above the \fmrsfr .}
\label{fmr}
\end{figure}

If $\alpha=0$ provides the optimal value, that would indicate that the MZR has the lowest scatter obtainable and that taking into account the SFR is unnecessary.  Also of note, if $\alpha = 1$ provides the optimal value, that would indicate that the specific star formation rate (M$_*$/SFR) would be the best projection to use.  The scatter obtained for each value of $\alpha$ can be seen in the left hand side of Figure \ref{scatter}.  We find that a value of $\alpha$ = 0.28 provides the lowest scatter for the \fmrsfr\ (Table \ref{fmr_params}).  Using this $\alpha$ value we then reproduce the \fmrsfr\ plot using our optimal projections for Figure \ref{fmr}.

For the \fmrsfr\ relation seen in Figure \ref{fmr}, we find that although the 2nd lowest (light blue) SFR bin overlaps between the IFU observed dwarf galaxy sample and the larger crossmatched ALFALFA/SDSS sample, the lowest (dark blue) SFR bin is not consistent between the IFU observed dwarf galaxy sample and the crossmatched ALFALFA/SDSS sample. In fact the lowest SFR bin is offset 1$\sigma$ above the \fmrsfr .  This is suggestive of a breakdown of the \fmrsfr\ for the lowest SFR galaxies.  It is possible that our SFRs are slightly inaccurate in this range considering they are near the limit discussed in \citet{Kennicutt:12}.  As star formation rates approach this region, they should be treated as upper limits of the true SFR.  

\subsubsection{MZR HI-Mass Dependence (\fmrhi )}
\citet{Bothwell:13} defines an equation similar to Equation \ref{fmrsfr_eq} for the \fmrhi\ relation: \begin{equation}\eta_\beta = \text{log}(\text{M}_*) - \beta (\text{log}(M_{HI})-9.80). \end{equation}  Wherein they find that a value of $\beta = 0.35$ minimizes the scatter in the gas-phase metallicity-$\eta$ plane.

To find the optimal value of $\beta$, we project the ALFALFA/SDSS data onto the $\eta$-metallicity plane and bin our data using the same HI mass bins used for Figure \ref{colorful_fmr}.  Each HI mass binned dataset is binned a 2nd time in $\eta$ space similar to the stellar-mass binning performed in Section \ref{colorful} and we once again perform the same cut of 21 galaxies per $\eta$-HI mass bin.  We take the mean metallicity of each $\eta$-HI mass bin and fit a 4th degree polynomial to the binned  mean metallicity values.  This process is repeated for values of $\beta$ ranging from -1 to 2.  

If $\beta=0$ provides the optimal value, that would indicate that taking into account the HI mass is unnecessary.  Also of note, if $\beta = 1$ provides the optimal value, then the ratio of HI-gas mass to stellar mass would be the best projection to use.  The scatter obtained for each value of $\beta$ can be seen in the right hand side of Figure \ref{scatter}.  We find that a value of $\beta$ = 0.32 provides the lowest scatter for the \fmrhi\ (Table \ref{fmr_params}).  Using this $\beta$ value we then produce the \fmrhi\ plot using our optimal projections for Figure \ref{fmr}.

For the \fmrhi\ relation, we find that our linear fits to the IFU observed dwarf galaxy sample are, within the 1$\sigma$ uncertainties, consistent with the binned means of the larger ALFALFA/SDSS crossmatched survey.  This suggests that the \fmrhi\ relation is likely to continue down to stellar masses as low as 10$^{6.6}$ M$_*$.  

\subsubsection{LZR SFR Dependence (\fmlsfr )}
\begin{figure*}
\epsfig{ file=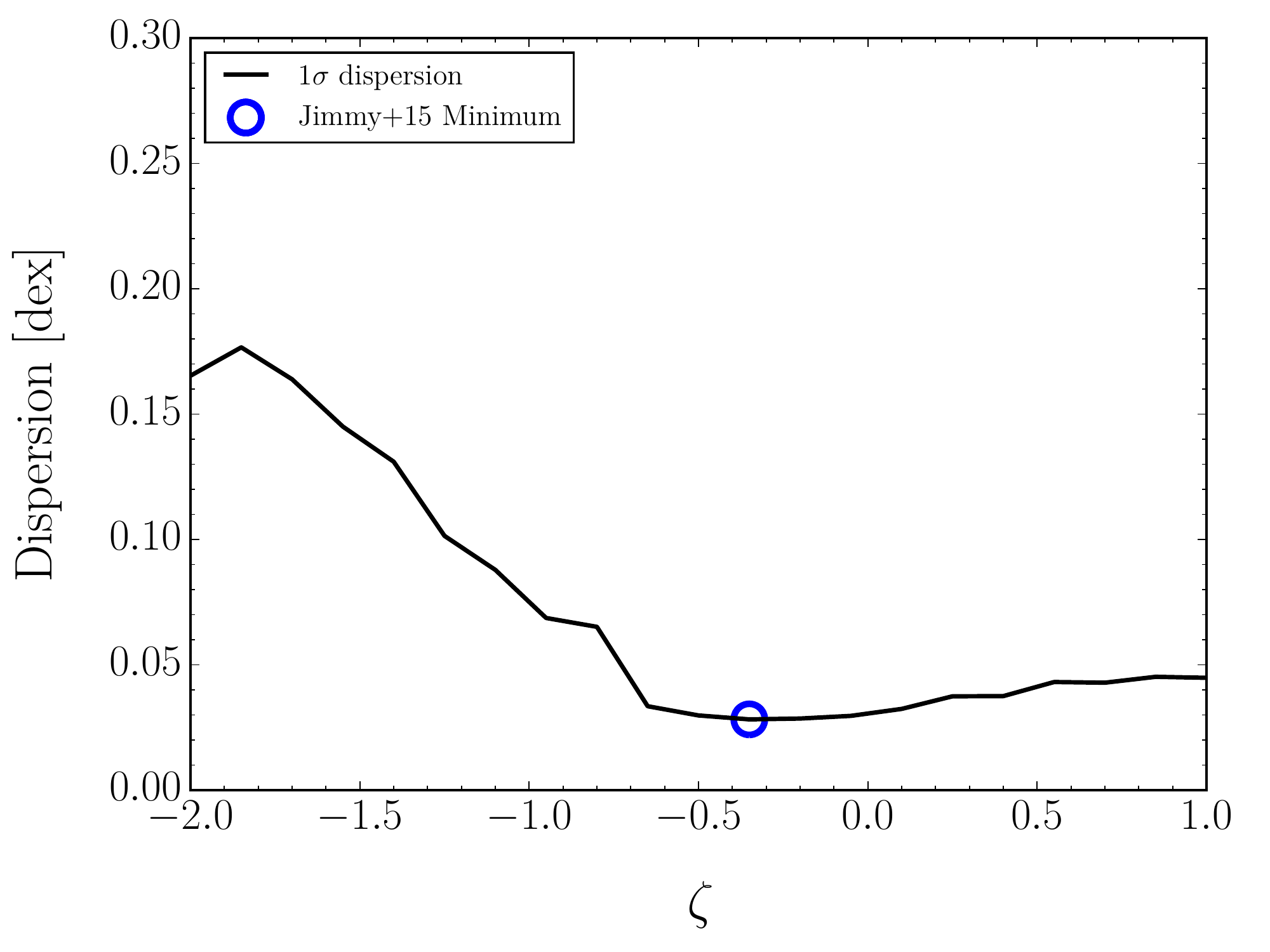, scale=0.45}
\epsfig{ file=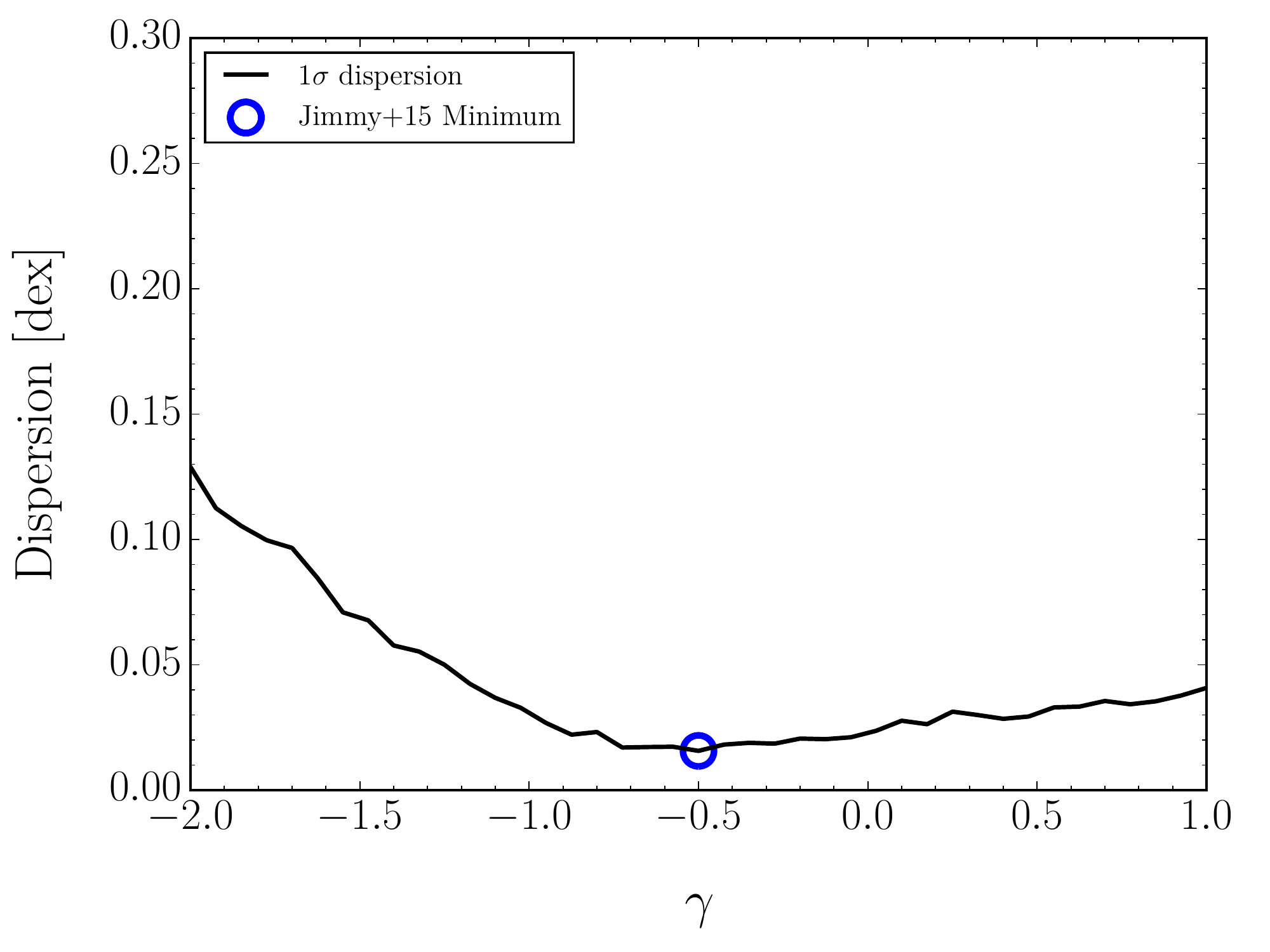, scale=0.45}
\caption{ Left: 1$\sigma$ scatter in the \fmlsfr\ relation as a function of $\zeta$.  We find the lowest scatter in the \fmlsfr\ relation when $\zeta = -0.35$  Right: 1$\sigma$ scatter in the means of the \fmlhi\ as a function of $\gamma$.   We find the lowest scatter in the \fmlhi\ relation when $\gamma$ = -0.50.}
\label{scatter_lzr}
\end{figure*}
We define a fundamental luminosity-metallicity relation (\fmlsfr ) using the star formation rate as: \begin{equation} \epsilon_{\zeta} = (\text{M}_B)-\zeta \text{log}(\text{SFR})\end{equation} 

To find the optimal value of $\zeta$, we project the ALFALFA/SDSS data onto the $\epsilon$-metallicity plane and bin our data using the same SFR bins used for Figure \ref{colorful_lzr_fmr}.  Each SFR binned dataset is binned a 2nd time in $\epsilon$ space similar to the luminosity binning performed in Section \ref{colorful} and we once again perform the same cut of 21 galaxies per $\epsilon$-SFR bin.  We take the mean metallicity of each $\epsilon$-SFR bin and fit a 4th degree polynomial to the binned  mean metallicity values.  This process is repeated for values of $\zeta$ ranging from -2 to 1.
 
\begin{figure}
\epsfig{ file=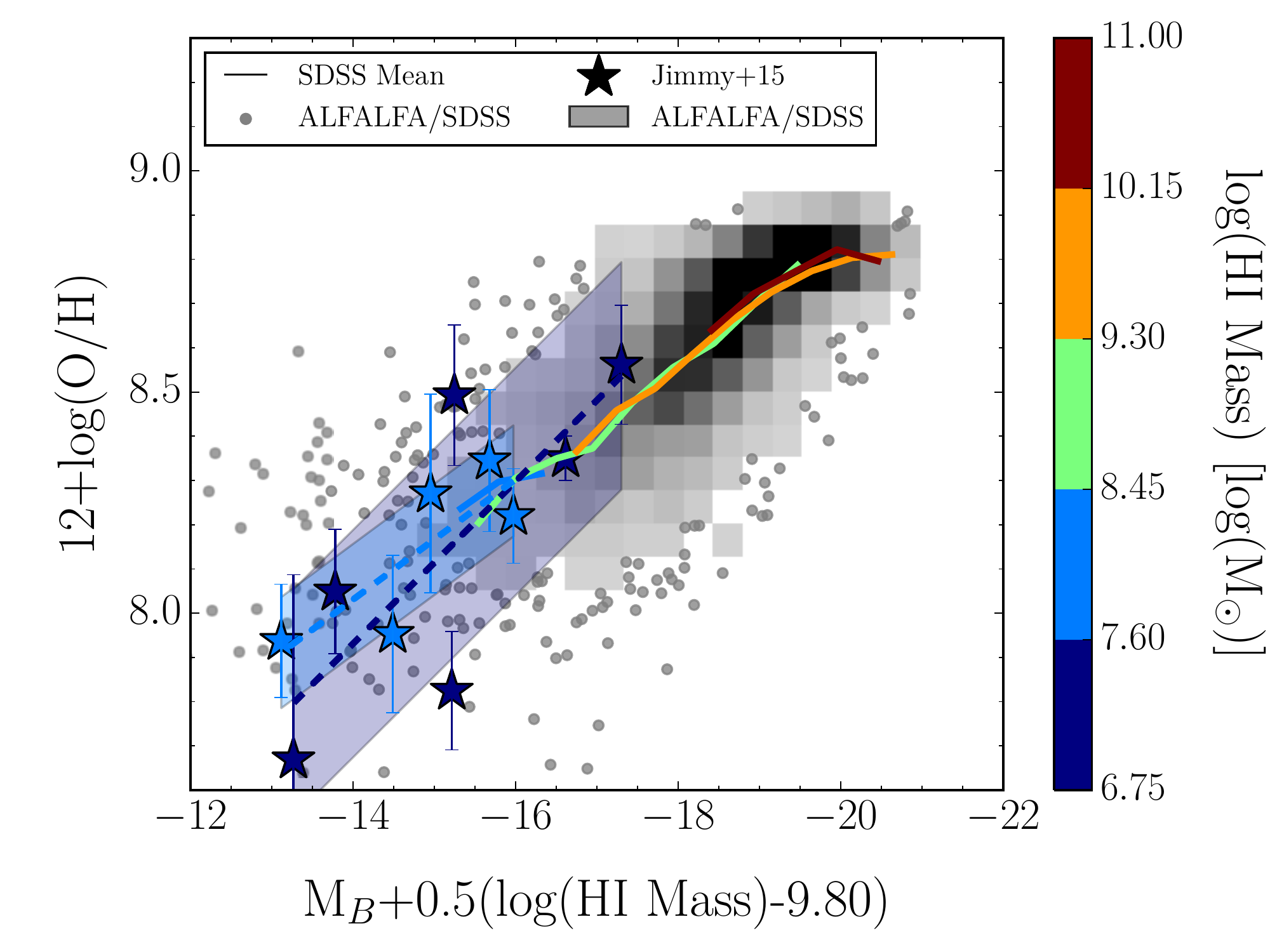, scale=0.45}
\epsfig{ file=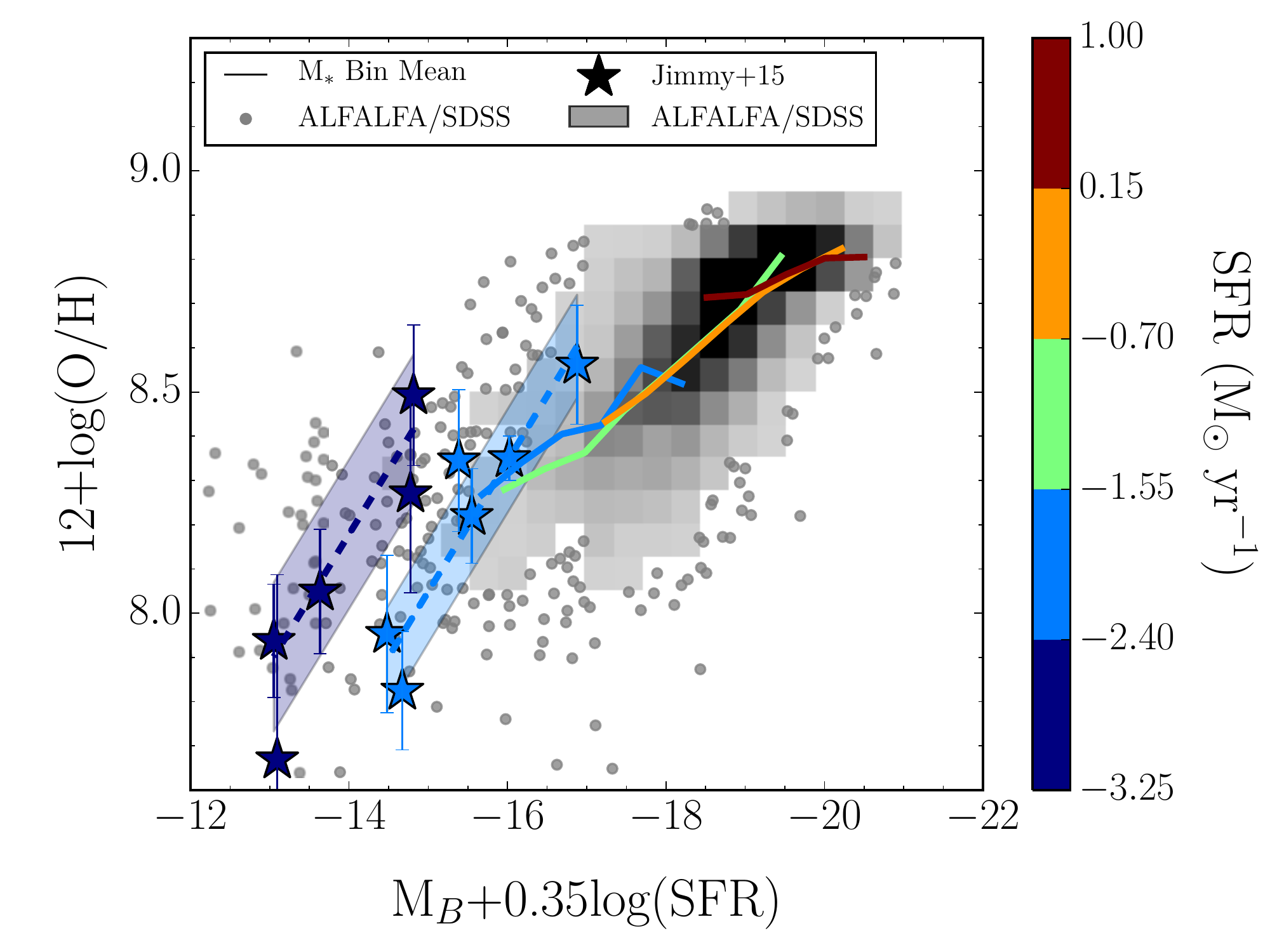, scale=0.45}
\caption{\fmlhi\ (top) and \fmlsfr\ (bottom) as calculated using the $\zeta$ and $\gamma$ values found to minimize scatter (Figure \ref{scatter_lzr}).  We find that the \fmlhi\ relation is consistent down to the lowest HI-mass bin, where as the lowest SFR bin in the \fmlsfr\ is offset 1$\sigma$ above the \fmlsfr . }
\label{fml}
\end{figure}

The scatter obtained for each value of $\zeta$ can be seen in the right hand side of Figure \ref{scatter_lzr}.  We find that a value of $\zeta$ = -0.35 provides the lowest scatter for the \fmlsfr\ (Table \ref{fmr_params}).  Using this $\zeta$ value we then produce the \fmlsfr\ plot using our optimal projections for Figure \ref{fml}.

For the \fmlsfr\ relation seen in Figure \ref{fml}, we find that although the 2nd lowest (light blue) SFR bin overlaps between the IFU observed dwarf galaxy sample and the larger crossmatched ALFALFA/SDSS sample, the lowest (dark blue) SFR bin is not consistent between the IFU observed dwarf galaxy sample and the crossmatched ALFALFA/SDSS sample. In fact the lowest SFR bin is offset 1$\sigma$ above the \fmlsfr\ derived from the ALFALFA/SDSS sample.  This is suggestive of a breakdown of the \fmlsfr\ relation for the lowest SFR galaxies.  However, as discussed in Section \ref{fmrsfr_sec}, it is possible that our SFRs are inaccurate in this range considering they are near the limit discussed in \citet{Kennicutt:12}.

\subsubsection{LZR HI-Mass Dependence (\fmlhi )}

We define a fundamental luminosity-metallicity relation (\fmlhi ) using the star formation rate as: : \begin{equation}\xi_{\gamma} = (\text{M}_B) - \gamma(\text{log}(M_{HI})-9.80). \end{equation} 

To find the optimal value of $\gamma$, we project the ALFALFA/SDSS data onto the $\xi$-metallicity plane and bin our data using the same HI mass bins used for Figure \ref{colorful_lzr_fmr}.  Each HI mass binned dataset is binned a 2nd time in $\xi$ space similar to the luminosity binning performed in Section \ref{colorful} and we once again perform the same cut of 21 galaxies per $\xi$-HI mass bin.  We take the mean metallicity of each $\xi$-HI mass bin and fit a 4th degree polynomial to the binned  mean metallicity values.  This process is repeated for values of $\gamma$ ranging from -2 to 1.  

The scatter obtained for each value of $\gamma$ can be seen in the right hand side of Figure \ref{scatter}.  We find that a value of $\gamma$ = -0.50  provides the lowest scatter for the \fmlhi\ (Table \ref{fmr_params}).  Using this $\gamma$ value we then produce the \fmlhi\ plot using our optimal projections for Figure \ref{fml}.

For the \fmlhi\ relation, we find that our linear fits to the IFU observed dwarf galaxy sample is, within the 1$\sigma$ uncertainties, consistent with the binned means of the larger ALFALFA/SDSS crossmatched survey, suggesting that the \fmlhi\ relation is likely to continue down to luminosities as low as M$_B = -12$.

\begin{deluxetable}{ l r r r }
\tablecolumns{4}
\tablecaption{Best Fitting Fundamental Relation Parameters}
\startdata

\hline
\hline
	  & & & \\
	$\mu_\alpha$ & $\alpha$ &  &  \\
	 $\text{log}(\text{M}_*)-\alpha \text{log}(\text{SFR})$ & 0.28 & & \\
	  & & & \\
	\hline
	  & & & \\
	 $\eta_\beta$ & $\beta$ &  &  \\
	 $\text{log}(\text{M}_*) - \beta (\text{log}(M_{HI})-9.80)$ & 0.32 & & \\
	  & & & \\

	\hline
	  & & & \\
	 $\epsilon_{\zeta}$& $\zeta$ &  &  \\
	   $(\text{M}_B)-\zeta \text{log}(\text{SFR})$ & -0.35 & & \\
	  & & & \\

	\hline
	  & & & \\
	$\xi_{\gamma}$ & $\gamma$ &  &  \\
	 $(\text{M}_B) - \gamma(\text{log}(M_{HI})-9.80)$ & -0.50 & & \\

\enddata
\tablecomments{ The values that produce the lowest scatter in Figures \ref{scatter} \& \ref{scatter_lzr}.}
\label{fmr_params}
\end{deluxetable}

\section{Discussion}

As can be seen in Figure \ref{MZRLZR}, the MZR holds across 5 orders of magnitude in stellar mass, with a 1$\sigma$ scatter of 0.05 dex.  There is an upturn in the MZR around M$_*$ = 10$^{8}$ M$_\sun$ which can be explained in the context of the FMR.  Galaxies with stellar masses below 10$^{8}$ M$_\sun$ are likely to have low star formation rates and low HI-gas masses which correlates with higher than average metallicity for a specific stellar mass.  For a galaxy with 10$^8$ M$_\sun$, for it to exist in the highest HI mass bin, it would have to have an HI-gas mass 2 orders of magnitude larger than the stellar-mass.

The LZR also holds over 7 magnitudes, with a 1$\sigma$ scatter of 0.03 dex.  A similar upturn can be seen in the LZR around M$_B$ = -15.  Again this can be explained by the expected low HI-mass and/or low SFR of low luminosity galaxies.  The lower scatter in the LZR supports the hypothesis of \citet{Berg:12} that accurately measured luminosities perform equally as well as stellar-masses.

To provide insight into the most physically motivated components of a fundamental metallicity relation, we examine the 1$\sigma$ scatter of the means in the larger ALFALFA/SDSS data set for each permutation of mass, luminosity, SFR, and HI mass used to form a possible fundamental relation.  Using the scatter of the means around each quartic fit as the evaluation criteria, we find that the lowest scatter of the means is in the combination of stellar mass and HI mass binning (\fmrhi ) for which we obtain a mean scatter of 0.01 dex, the other three relationships (\fmrsfr , \fmlhi , \fmlsfr ) have mean scatters of 0.02 dex.  This suggests that the \fmrhi\ may be the most physically significant fundamental relation.

\subsection{\fmrhi }

In examining the \fmrhi\ relation, seen in Figure \ref{fmr} we find that the dwarf IFU sample is consistent within 1 standard deviation with the \fmrhi\ relation observed within the larger ALFALFA/SDSS sample.  This suggests that the \fmrhi\ continues to hold down to stellar masses as low as M$_*$ = 10$^{6.5}$ M$_\sun$.  Our results suggest that the physical mechanism responsible for the \fmrhi\ must be active across the entire stellar mass range from 10$^{6.5}$ to 10$^{10}$ M$_\sun$ 

\citet{Bothwell:13} reach the conclusion that the \fmrhi\ is more physically motivated due to the reduced scatter and physical motivation of the \fmrhi .  They also noticed a lack of saturation in the high mass end of the \fmrhi .  In our analysis, which uses slightly different stellar-mass and HI-mass bins, we do observe saturation in stellar mass bins above log(M$_*$) = 10.0 in Figure \ref{colorful_fmr}.  We find the \fmrhi\ relation to be more physically motivated due to its behavior on the low mass end (log(M$_*$) $<$ 9.0) in which the IFU observed dwarf galaxy data is consistent with the larger ALFALFA/SDSS dataset.  The 1$\sigma$ scatter of the \fmrhi\ binned mean metallicities being the lowest of the 4 permutations supports this hypothesis.

Inflows of pristine gas diluting the metal content would explain why the FMR has an HI mass dependance.  \citet{Koppen:99} showed that the gas-phase metallicity of a galaxy can indeed be reduced if the gas infall rate is larger than the rate at which gas is converted into stars.  Our results suggest that inflows of pristine gas continue to drive the \fmrhi\ down to stellar masses as low as 10$^{6.5}$ M$_\sun$.  This agrees with recent models by \citet{Gronnow:15} which show that pristine gas rich mergers are partially responsible for the scatter in the FMR.

\subsection{\fmrsfr }

Unlike the \fmrhi , the \fmrsfr\ (Figure \ref{fmr}) does not appear to hold across our entire sample.  The lowest SFR bin in the IFU observed dwarf galaxy sample is offset 1$\sigma$ higher than the \fmrsfr\ as determined using the larger ALFALFA/SDSS sample.  If our results are accurate near the calibration limits of the SFR and oxygen abundance, they suggest a breakdown in the physical mechanism responsible for the \fmrsfr\ in very low SFR populations.

Possible explanations for the \fmrsfr\ have focused on outflows of metal-enriched gas caused by the intense winds of young star-forming regions within a galaxy.  Simulations by \citet{MacLow:99} have shown that supernova winds are capable of expelling metals efficiently enough to contribute to the MZR.  We find that this is likely to be true down to an SFR of -2.4 log(M$_*$ yr$^{-1}$), however, below that it is possible that stellar winds are not sufficiently strong enough to remove metal-enriched gas, and galaxies have a higher gas-phase metallicity than would be expected by the \fmrsfr .  Although limitations in our measurements could also explain this apparent breakdown.

\subsection{\fmlhi }
We find that the \fmrhi\ relation, seen in Figure \ref{fml}, is consistent within 1$\sigma$ between the IFU observed dwarf galaxy population and the larger ALFALFA/SDSS population.  These results are similar to those observed for \fmrhi\, however we find that the scatter of the binned $\xi _{-0.5}$ mean metallicities around the quartic fits is larger than the analogous fits for \fmrhi\ suggesting that stellar-mass is the more physically motivated parameter.

\subsection{\fmlsfr }
The \fmlsfr\ relation is inconsistent in both the high $\epsilon_{\zeta}$ and low $\epsilon_{\zeta}$ regions, as can be seen in Figure \ref{fmr}.  Similar to the \fmrsfr\ the lowest SFR bin is $>$ 1$\sigma$ offset from the \fmlsfr\ relation as determined by the larger ALFALFA/SDSS sample.  The highest SFR bin also deviates by as much as 1$\sigma$, suggesting that the high mass SFR bin is also poorly fit by the best fitting $\eta$ in the \fmlsfr .  The inconsistencies observed within the \fmlsfr\ relation provide further evidence that HI-gas mass may be the more physically motivated component in a fundamental metallicity relation.

\subsection{H$\alpha$ Line Flux Limitations}
\label{flux_limit}
\begin{figure}
\epsfig{ file=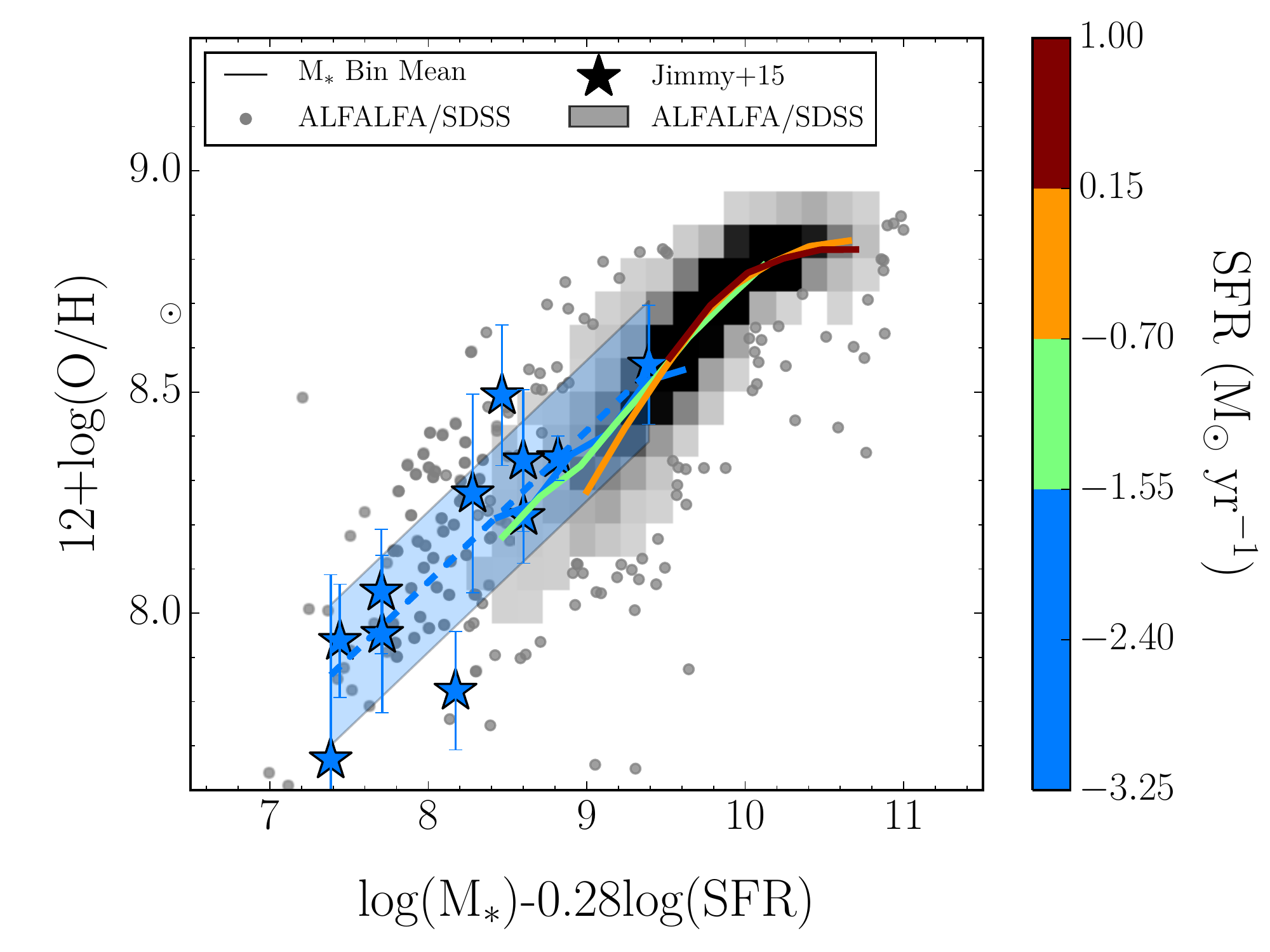, scale=0.45}
\caption{\fmrsfr\ with the two lowest SFR bins combined into one in order to compensate for the possible underestimation of SFR in the lowest SFR bin of Figure \ref{fmr}.  We see that in this scenario, the \fmrsfr\ is consistent across the stellar mass range observed.  }
\label{combined_bins}
\end{figure}

In order to explain the apparent breakdown in the \fmrsfr\ and \fmlsfr\ relations for the lowest SFR bins, we consider the possibility that our H$\alpha$ line flux limit may be causing inaccurate SFR estimations.  In comparing H$\alpha$ flux measurements between the AoN selected binned spectra and the segmentation map selected binned spectra, we find the segmentation map selection captures approximately 33\% more H$\alpha$ flux indicative of low SFR regions being missed.  It is also possible that metallicities in the fainter, low SFR, galaxies are over estimated due to limitations in measuring low NII $\lambda 6583$ fluxes, placing them preferentially above the \fmrsfr\ (See Appendix \ref{AppendixA}).

To test the hypothesis that our lowest SFR bins are underestimated, and our lowest oxygen abundances are overestimated, we re-examine the \fmrsfr\ considering the two lowest SFR bins as one bin.  Figure \ref{combined_bins} demonstrates that the combined bin is consistent with the \fmrsfr .  Therefore, it is possible that the \fmrsfr\ is indeed valid over the same stellar mass range as the \fmrhi\ however, care must be taken when obtaining IFU observations to ensure that the target depth is sufficient to accurately estimate the SFR and oxygen abundance.  To determine whether line flux measurement limitations or physical limitations are the cause of the 1$\sigma$ offset of the lowest SFR bin, we would require further observations with longer integration times.

\vspace{10 mm}

\section{Conclusion}

In order to extend the fundamental relations down to stellar masses as low as 10$^{6.5}$ and luminosities as low as -12 mag, we utilize a sample of IFU observed dwarf galaxies in combination with a larger ALFALFA/SDSS crossmatched population of galaxies.  Whenever possible, we perform the same analysis on both populations in order to ensure that results are consistent between the two data sets.

When comparing galaxies that overlap between the SDSS and IFU observed samples, we find that IFU observations are necessary to accurately measure the star formation rates of the dwarf irregular galaxies presented here due to the patchy nature of their star forming regions.  However, gas-phase metallicites appear to be adequately measured using a single SDSS fiber (Figure \ref{consistency}).

Using a sample of galaxies selected from the ALFALFA blind HI survey, in combination with galaxies observed with the VIMOS IFU spectrograph and a selection of long-slit observed galaxies, we investigate the mass-metallicity relation (MZR) with a particular focus on the low stellar mass/low luminosity regime.  We find that the MZR continues down to stellar masses as low as $10^{7.25}$ M$_\sun$ (Figure \ref{MZRLZR}).  We find that the MZR has a 1$\sigma$ scatter in the means around a quartic fit of 0.05 dex.

We find using the ALFALFA/SDSS crossmatched population that the fundamental metallicity relation as a function of SFR (\fmrsfr ) exhibits its lowest scatter in the means (1$\sigma$ = 0.02 dex) when $\alpha$ = 0.28.  We also find using the IFU observed dwarf galaxy population that the \fmrsfr\ relations appears to break down in the lowest SFR bins (Figure \ref{fmr}).   However limitations of strong-line metallicity estimations, and H$\alpha$ based SFRs in low stellar-mass dwarf galaxies may provide an alternative explanation for this apparent breakdown (Section \ref{flux_limit}).

We find using the ALFALFA/SDSS crossmatched sample that the fundamental metallicity relation (\fmrhi ) exhibits the lowest scatter in the means (1$\sigma$ = 0.01 dex) when $\beta$ = 0.32.  We also find that the \fmrhi\ relations are both consistent between the IFU dwarf population and the ALFALFA/SDSS population across the entire HI-gas mass range.

We find that the luminosity-metallicity relation (LZR) continues down to B-band luminosities as low as -14 mag.  We find that the LZR has a 1$\sigma$ scatter in the means around the quartic fit of 0.03. The lowest scatter in the \fmlhi\ it is 0.02 dex (for $\zeta$ = -0.35).  Similarly we find that the lowest scatter in the \fmlsfr\ is 0.02 dex for $\gamma$=-0.50. 

When comparing our sample of dwarf galaxies to the \citet{Mannucci:10} and \citet{Bothwell:13} analysis, we find that our results are consistent with theirs in that our derived $\alpha$ and $\beta$ values agree $\pm$ 0.04.  In agreement with the conclusions of \citet{Bothwell:13}, we find that the \fmrhi\ is more physically motivated than \fmrsfr\ based on the most significant reduction of scatter.

In summary, the \fmrhi\ appears to be the most physically significant driver of the fundamental metallicity relation, suggesting that inflows of pristine gas would be a possible explanation for the fundamental relations observed.

\vspace{10 mm}

\acknowledgments
\section*{Acknowledgments}
We would like to thank the referee for the careful review and the valuable comments, which provided insights that helped improve the paper.

We are very grateful to the entire ALFALFA team for the extensive work involved in observing, processing, flagging, and cataloguing the ALFALFA data, which this paper is based on. We also wish to acknowledge the important contributions of Martha Haynes, Riccardo Giovanelli, and Marco Scodeggio in devising the VIMOS-IFU follow-up project based on ALFALFA-detected dwarf galaxies.

SB acknowledges the funding support from the Australia Research Council through aFuture Fellowship (FT140101166).

J would like to thank Dustin Lorshbough for his valuable discussions.

Funding for SDSS-III has been provided by the Alfred P. Sloan Foundation, the Participating Institutions, the National Science Foundation, and the U.S. Department of Energy Office of Science. The SDSS-III web site is http://www.sdss3.org/.

SDSS-III is managed by the Astrophysical Research Consortium for the Participating Institutions of the SDSS-III Collaboration including the University of Arizona, the Brazilian Participation Group, Brookhaven National Laboratory, Carnegie Mellon University, University of Florida, the French Participation Group, the German Participation Group, Harvard University, the Instituto de Astrofisica de Canarias, the Michigan State/Notre Dame/JINA Participation Group, Johns Hopkins University, Lawrence Berkeley National Laboratory, Max Planck Institute for Astrophysics, Max Planck Institute for Extraterrestrial Physics, New Mexico State University, New York University, Ohio State University, Pennsylvania State University, University of Portsmouth, Princeton University, the Spanish Participation Group, University of Tokyo, University of Utah, Vanderbilt University, University of Virginia, University of Washington, and Yale University.

\newpage
\appendix
\section{\\Appendix A - Ability to Recover Metallicity} \label{AppendixA}

Due to the spectral resolution and instrumental dispersion of the VIMOS LR Blue grism, some emission lines are severely blended together, as can be seen if Figure \ref{line_fit_output}.  Our science results depend on the [NII] and H$\alpha$ emission lines which are separated by $\sim$20 \AA .  To test our ability to recover the emission line fluxes of the blended [NII] and H$\alpha$ emission lines within our spectra, we create a simulated galaxy with known emission line ratios and test our ability to recover them using our Python-based Gaussian fitting routines.

To create the simulated galaxy, we first take a pipeline reduced data cube from one of the galaxies in our sample that was too faint to be detected by the IFU spectrograph (AGC220261) and find a region of the field of view which contains only sky with little instrumental artifacts to interfere.  We also check to ensure that none of the galaxy that was originally intended to be observed is located near this region.  We then generate a simulated galaxy and place it within this region.  This simulated galaxy is then run through our full data reduction pipeline to measure the recovered emission line ratios.  This process is repeated several times to test for metallicity dependance of our ability to recover emission line ratios.

To ensure that we generate physically realistic simulated galaxies, we draw our input emission line ratios from the \citet{Berg:12} survey of low-luminosity galaxies.  For every galaxy in the \citet{Berg:12} sample, we use the long slit measured emission line ratios to create a simulated galaxy with the same emission line ratios.  We scale the brightest central spaxel in the simulated galaxy to have an H$\beta$ emission line flux of $3.0 \times 10^{-16} \text{ erg cm}^{-2} \text{ s}^{-1} \text{ \AA} ^{-1}$, which corresponds to the average H$\beta$ flux in the brightest spaxel of each galaxy in the dwarf IFU sample.  The flux of the simulated galaxy decreases exponentially with radius.  Each emission line is given a dispersion equal to the average instrumental dispersion of 7.5 \AA .

\begin{figure}[h!]
\centering
\epsfig{ file=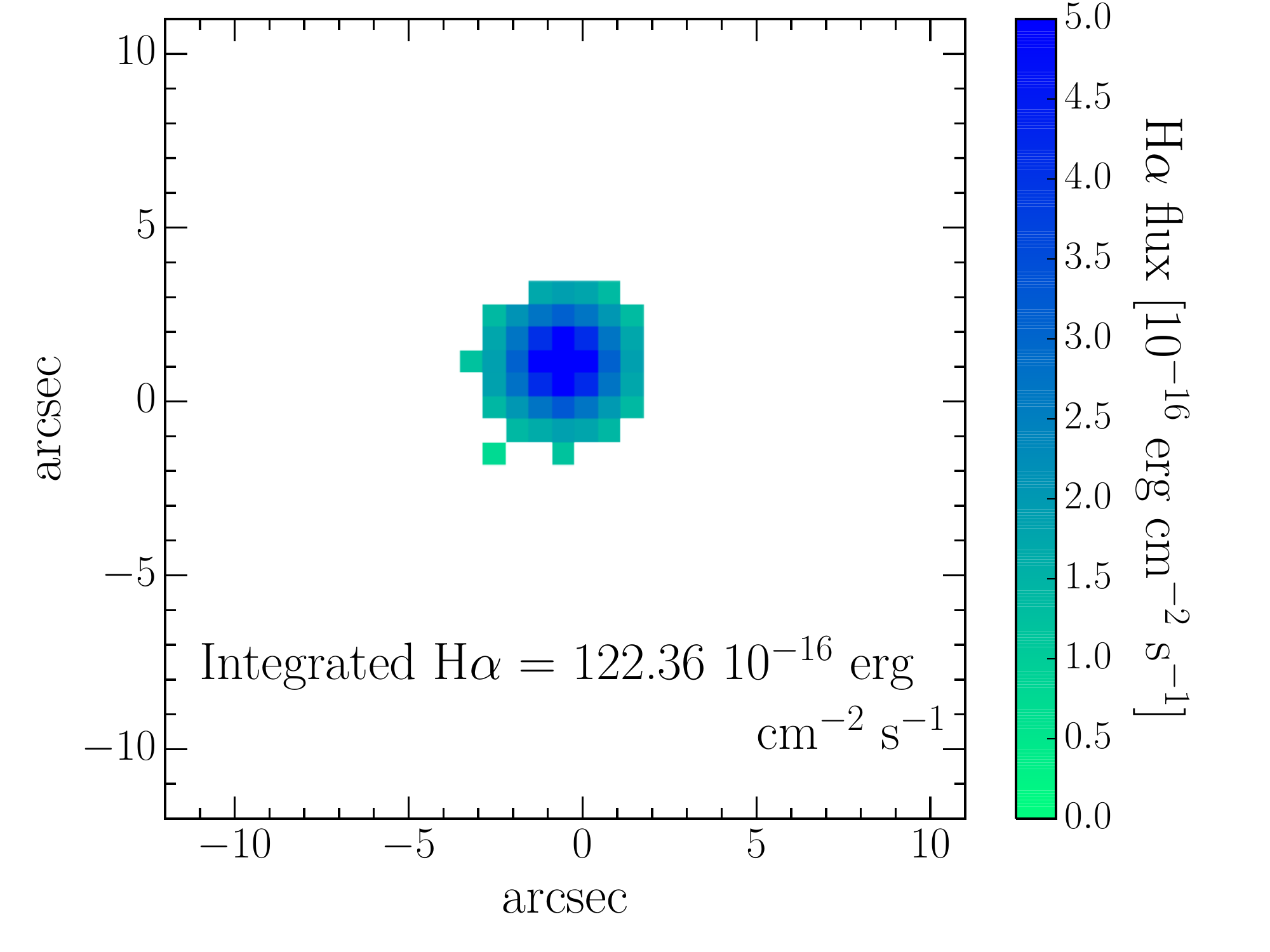, scale=0.5}
\caption{Example of the H$\alpha$ emission observed in a simulated galaxy as recovered by our Python-based Gaussian fitting routines.  Also shown is the recovered H$\alpha$ flux after integrating the spectra of each spaxel shown.}
\label{fake_h_alpha_map}
\end{figure}

\begin{figure}[h!]
\hspace{-5mm}\epsfig{ file=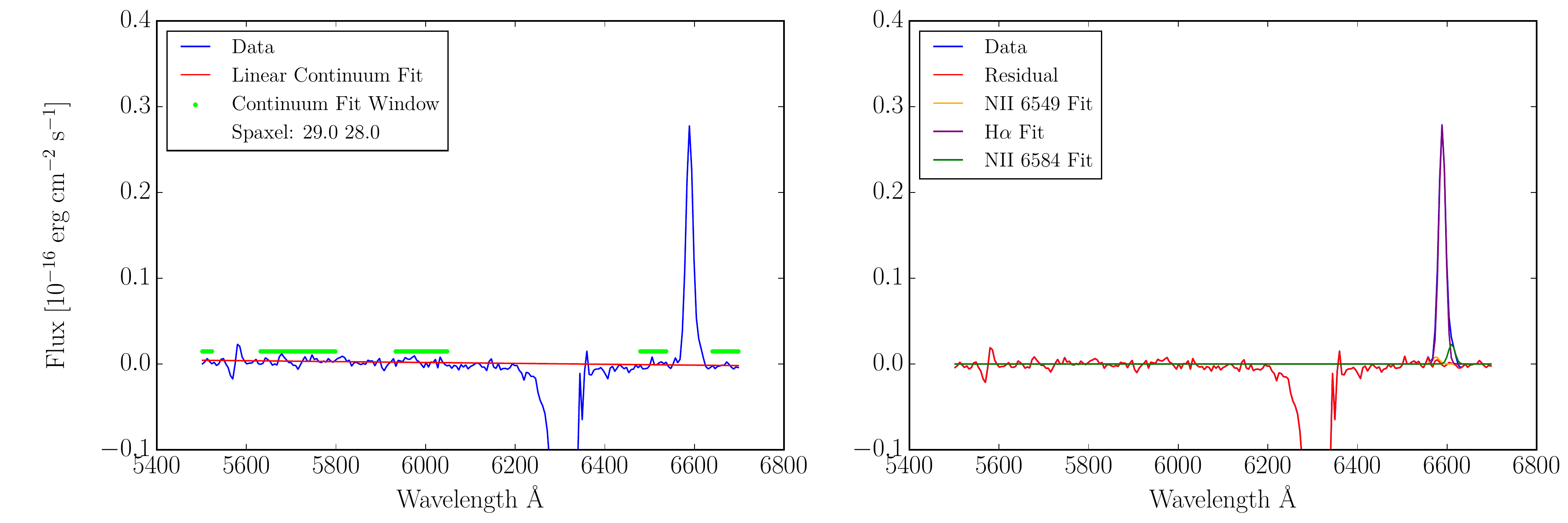, scale=0.5}
\caption{Left: Input simulated spaxel for our simulated galaxy.  Blue curves indicate the simulated spectral data, red lines indicates the linear continuum fit done in the first stage of our Python-based Gaussian fitting routines, and green points indicate the fitting window that was used to determine the linear fit.  The fitting window excludes residuals from skylines and the large internal reflection feature at $\sim 6300\AA$.  Right: Three Gaussian fits of the H$\alpha$ and [NII] emission lines as found by our routines.  The routines are able to successfully resolve the 3 blended emission lines.}
\label{FakeData}
\end{figure}

An example of the H$\alpha$ emission map for one of our simulated galaxies can be seen in Figure \ref{fake_h_alpha_map}.  Emission lines for a sample fiber of our simulated galaxy can be seen in Figure \ref{FakeData} which can be compared to Figure \ref{line_fit_output} visually to confirm that the simulated galaxies being created are physically realistic.

\newpage
\clearpage

\section{\\Appendix B - Comparison between N2 and O3N2 Based Calibrations} \label{AppendixB}

\begin{figure*}[h!]
\hspace{-5mm}\epsfig{ file=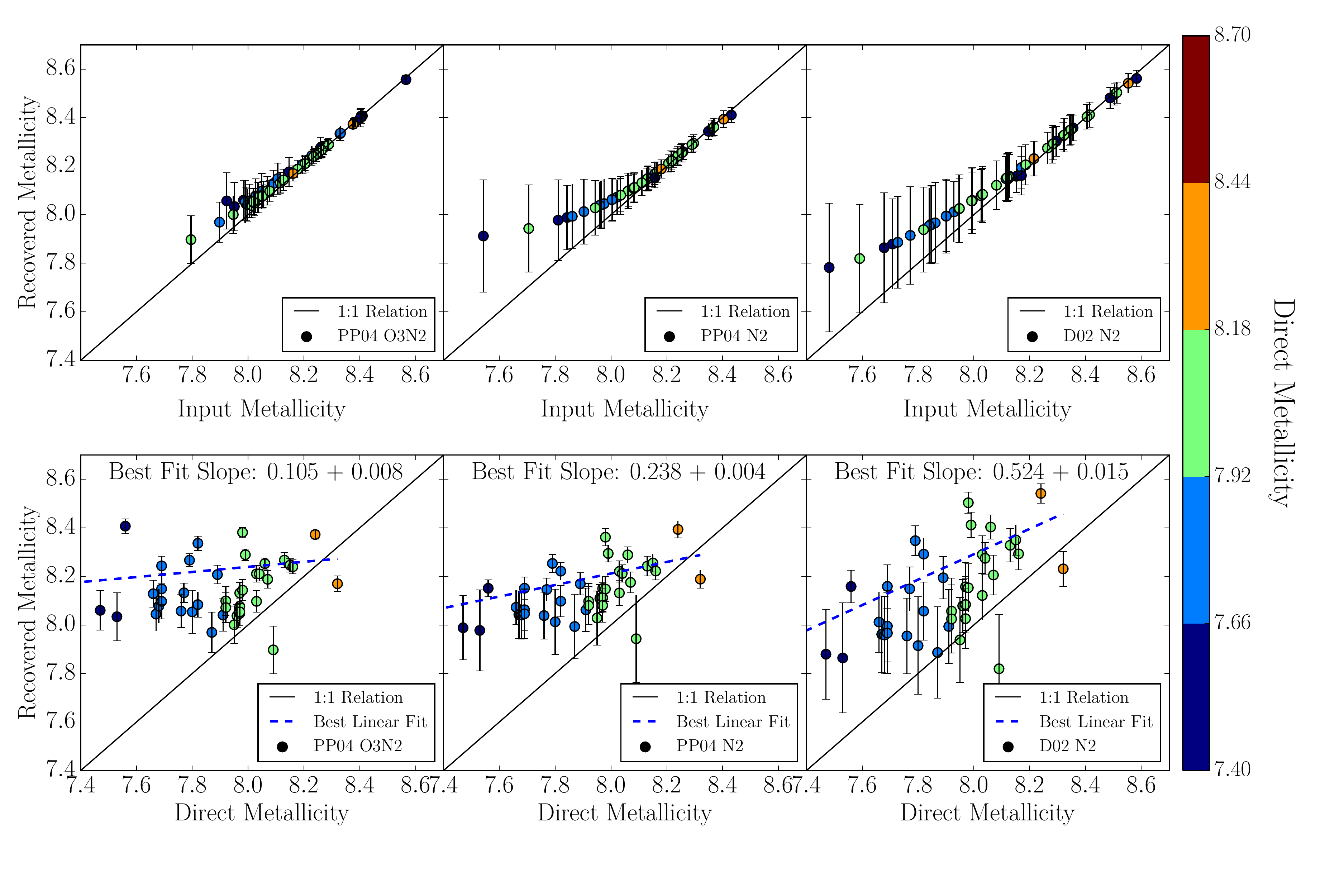, scale=0.5}
\caption{Showing the ability of our Python-based Gaussian fitting routines to recover metallicities in various systems after creating a simulated galaxy.  In the top row, the x-axis shows the input gas-phase metallicity based on the input line flux ratios, and the y-axis shows the result measured using our routines.  In the bottom row, the x-axis shows the direct metallicity that was used to create the simulated galaxy, and the y-axis shows the gas-phase metallicity measured using our routines.  We find that the D02 oxygen abundance calibration is more accurate than the PP04 based systems, however PP04 O3N2 is more precise than the D02 based systems.}
\label{metallicity_comparison}
\end{figure*}

We simulate every galaxy in the \citet{Berg:12} sample using the method outlined in Appendix \ref{AppendixA} to test both our ability to recover input strong-line emission ratios, and to compare our recovered metallicity estimations to direct metallicity measurements.  The top row of Figure \ref{metallicity_comparison} shows the comparison between input and recovered metallicities in each of the three calibrations: PP04 O3N2, PP04 N2, and D02 N2.  We find that our Python-based Gaussian fitting routines are able to successfully recover line fluxes within 1 standard deviation down to the lowest metallicity simulated.  The larger error bars in N2 line ratios of the low-metallicity simulated galaxies measured are due to the amplitude of the [NII] lines being comparable to the noise of the spectra.

The bottom row of Figure \ref{metallicity_comparison} shows the gas-phase metallicity measured directly along with the recovered gas-phase metallicity.  Although the O3N2 based estimation has smaller error bars due to the increased information from the [OIII] and H$\beta$ emission line information, it is less accurate than the N2 based estimations due to the saturation of the [OIII]/H$\beta$ ratio known to occur in O3N2 based oxygen abundance estimations at low gas-phase metallicity \citep{Pettini:04}.

We find based on the bottom row of Figure \ref{metallicity_comparison} that D02 N2 oxygen abundance estimations are better correlated with direct metallicity measurements than PP04 O3N2 estimations for low-metallicity galaxies such as those presented in this study.  We find that PP04 O3N2 provides more precise, but less accurate observations, whereas D02 N2 provides more accurate, but less precise observations for the lowest metallicity galaxies.

For the sake of completeness, we have reproduced all of the figures from the main text which contain oxygen abundance estimations using the O3N2 based oxygen abundance estimation from \citet{Pettini:04}: \begin{equation} 12+\text{log(O/H)} = 8.73 - 0.32 \times \text{O3N2} \label{PP04O3N2}\end{equation} where \begin{equation}O3N2 = \text{log} \left\{  \frac { [OIII] \lambda 5007/\text{H}\beta }{[NII] \lambda 6583/\text{H}\alpha} \right\}. \label{O3N2} \end{equation}  We find that in general, the trends observed in the paper and the conclusions of the paper continue to hold, especially on the high-metallicity end, however the oxygen abundance floor of the O3N2 based estimations cause the conclusions to be less solid than as observed in the N2 based results of the main text.

\FloatBarrier
\begin{figure}[h!]
\epsfig{ file=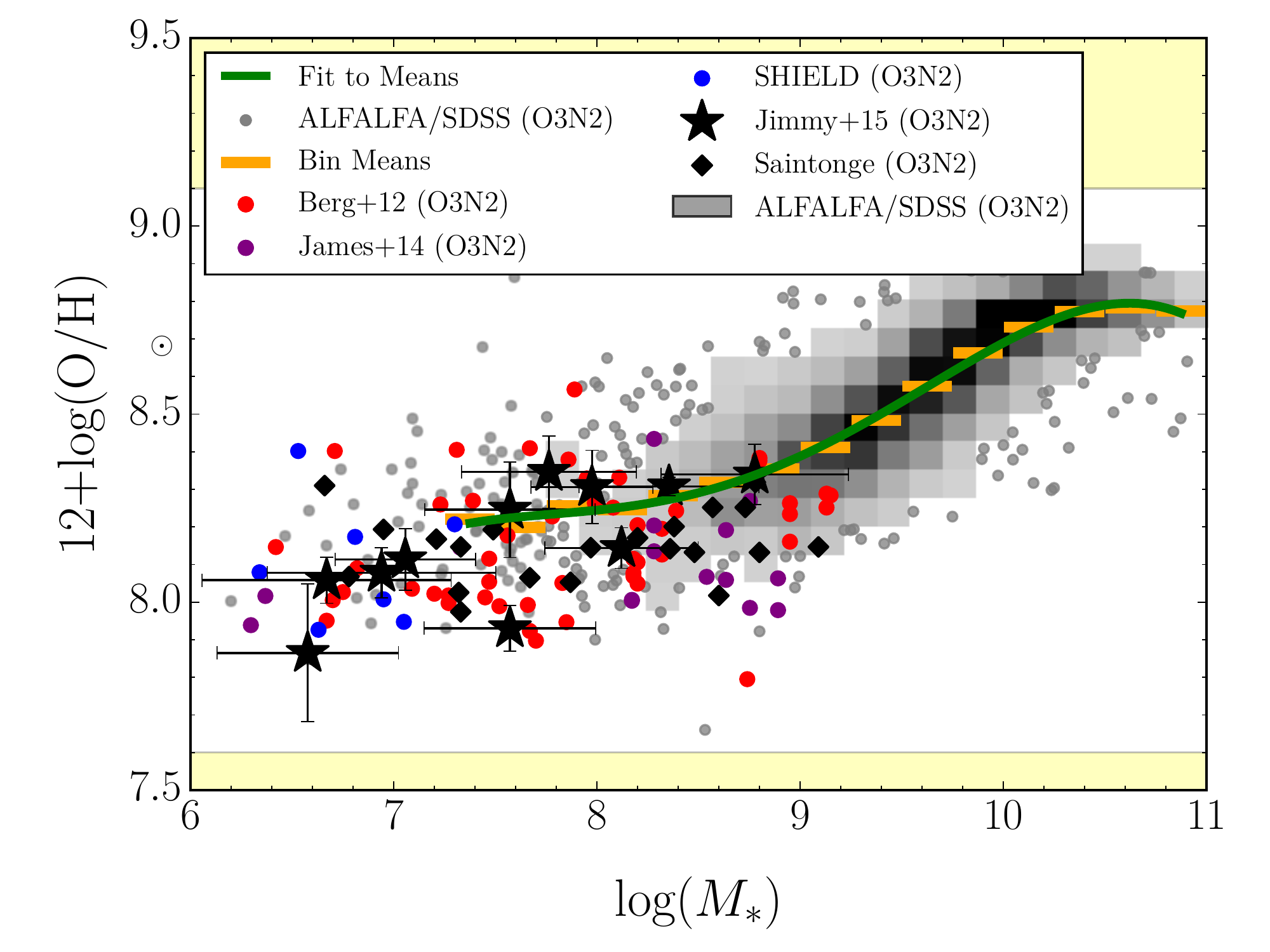, scale=0.45}
\epsfig{ file=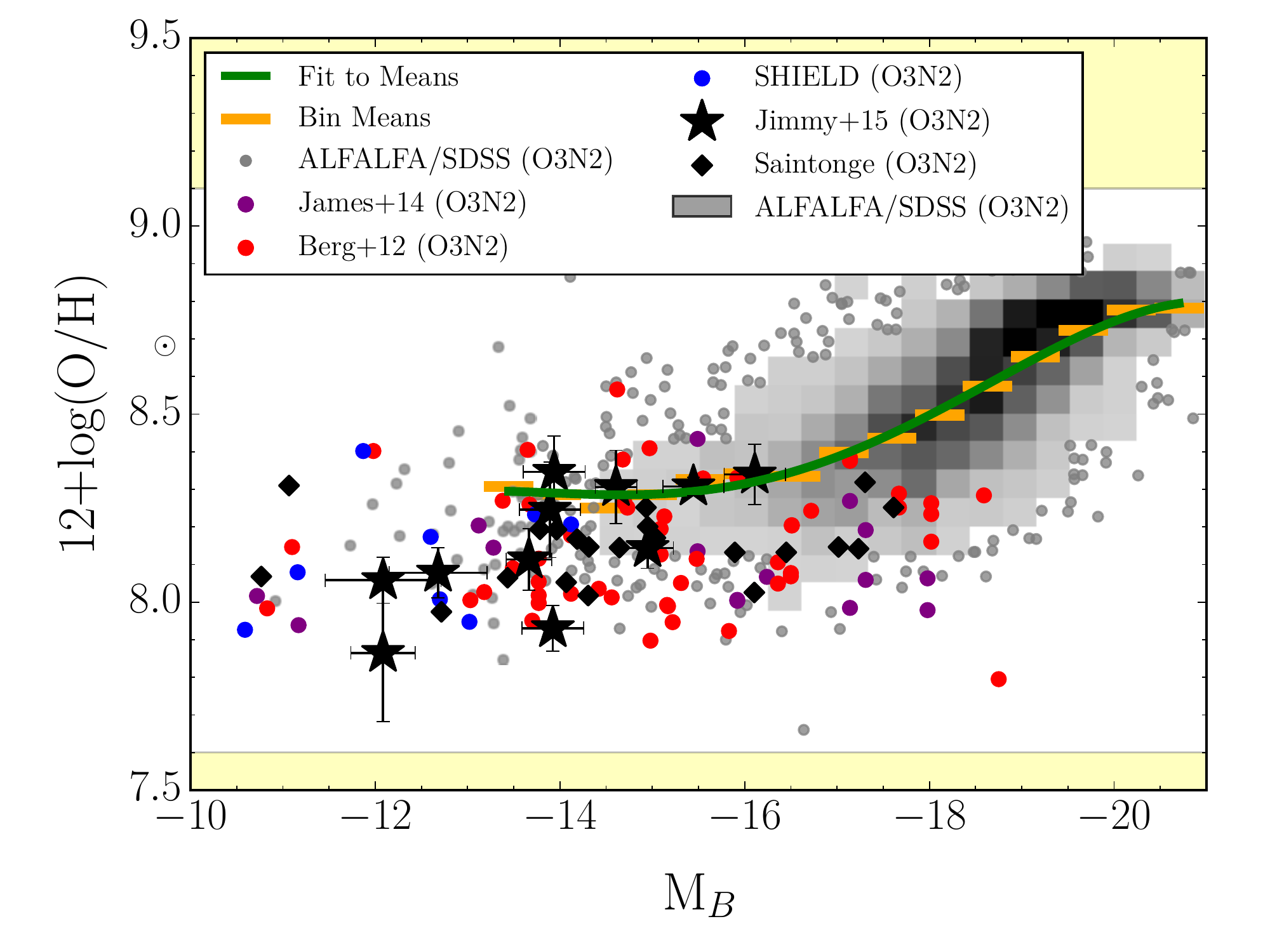, scale=0.45}
\caption{Same as Figure \ref{MZRLZR} however using the PP04 O3N2 oxygen abundance calibration in place of the D02 N2 calibration.  Left: The mass-metallicity relation for the IFU observed dwarf galaxy sample, along with several low stellar mass galaxy observations and the ALFALFA/SDSS sample.  Right: The luminosity-metallicity relation for the IFU observed dwarf galaxy sample, along with several low stellar mass galaxy observations and the ALFALFA/SDSS sample.  The scatter in both the MZR and the LZR fit around the bin means is 0.05 dex.  For reference we plot a sun symbol at solar metallicity, showing that all of the IFU observed dwarf galaxy oxygen abundances are below solar.  The yellow regions indicate the calibration limits of the D02 N2 oxygen abundance estimation.  We have used the published emission line fluxes from other authors to estimate the oxygen abundance in the same PP04 O3N2 system.  The mass-metallicity relation appears to flatten out in the low stellar-mass/low luminosity regime. }
\label{mass_metallicity_O3N2}
\end{figure}

\begin{figure}[h!]
\epsfig{ file=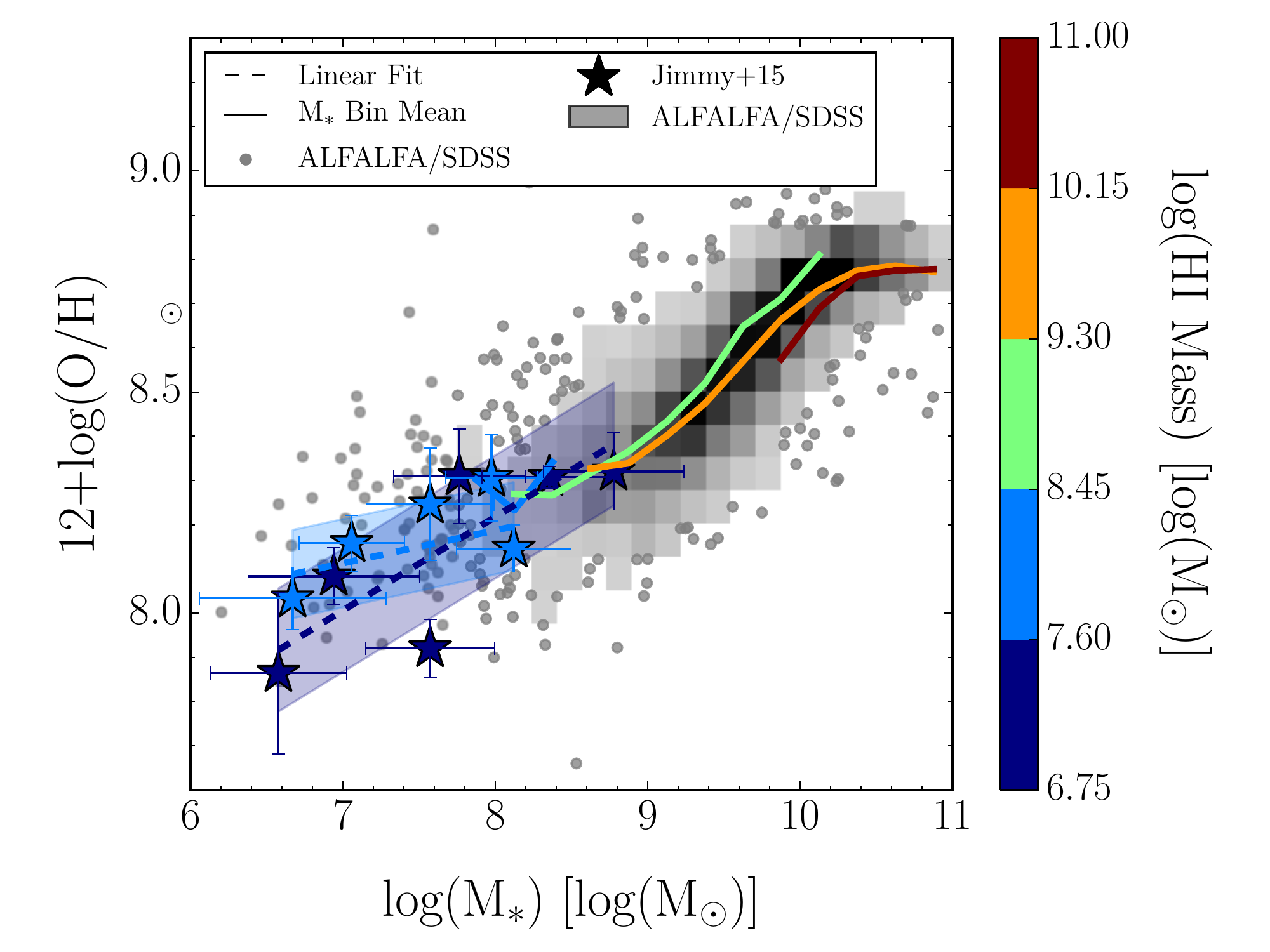, scale=0.45}
\epsfig{ file=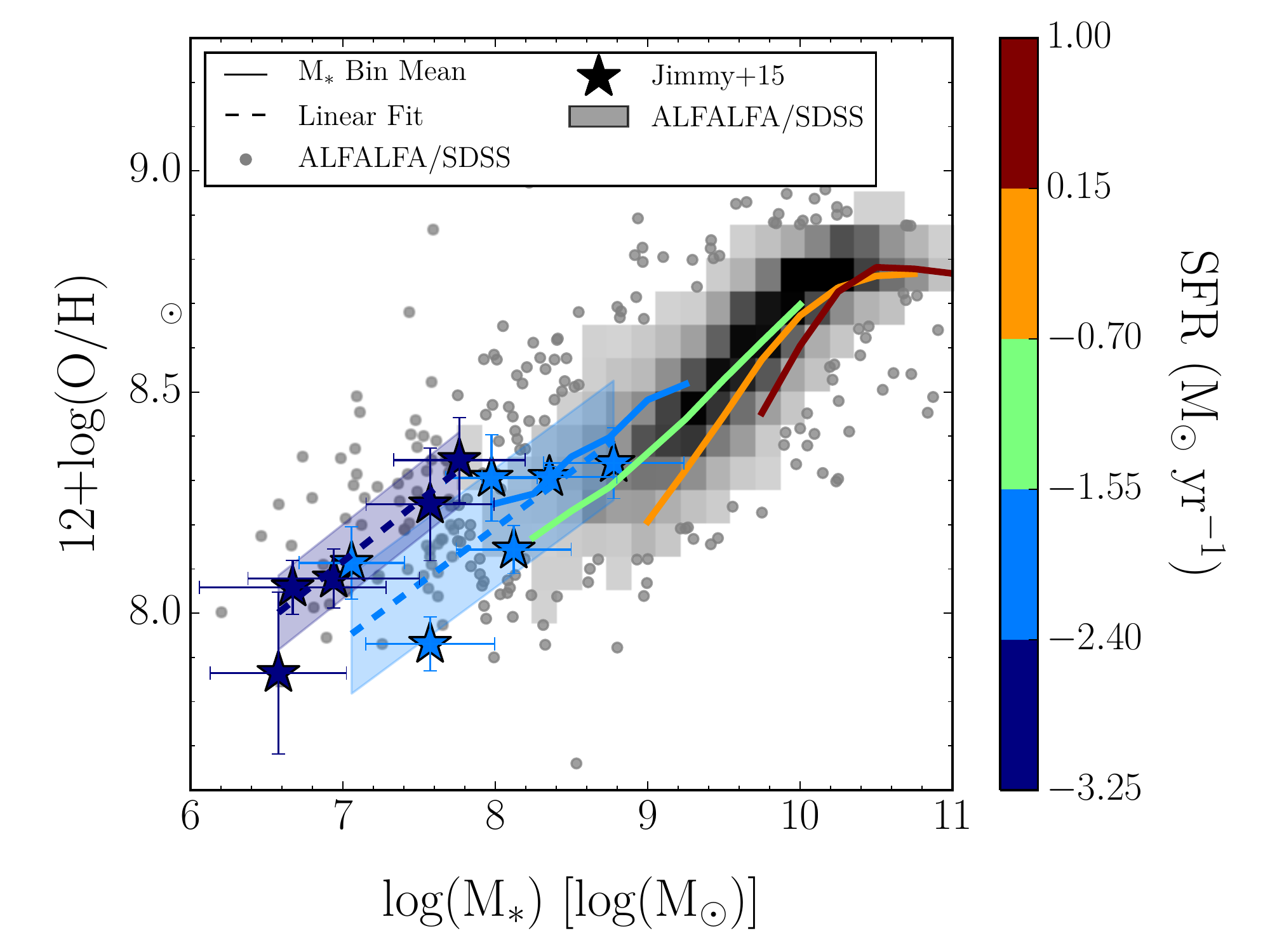, scale=0.45}
\caption{Same as Figure \ref{colorful_fmr} however using the PP04 O3N2 oxygen abundance calibration in place of the D02 N2 calibration.  The mass-metallicity relation as seen in Figure \ref{mass_metallicity_O3N2} color-coded by HI mass (left) and SFR (right).  The stars indicate individual observations within our IFU observed sample of dwarf galaxies.  The dashed lines indicate linear least squares fits to these points, and the shaded regions indicate the 1$\sigma$ standard deviations to these fits.  The solid curves indicate the mean values of the ALFALFA/SDSS sample, separated into HI mass bins.  Shown in the background are the ALFALFA/SDSS points which are binned to produce the color-coded means.  We find little overlap between the SFR bins (right).}
\label{colorful_hi_mass_metallicity_O3N2}
\end{figure}

\begin{figure}[h!]
\epsfig{ file=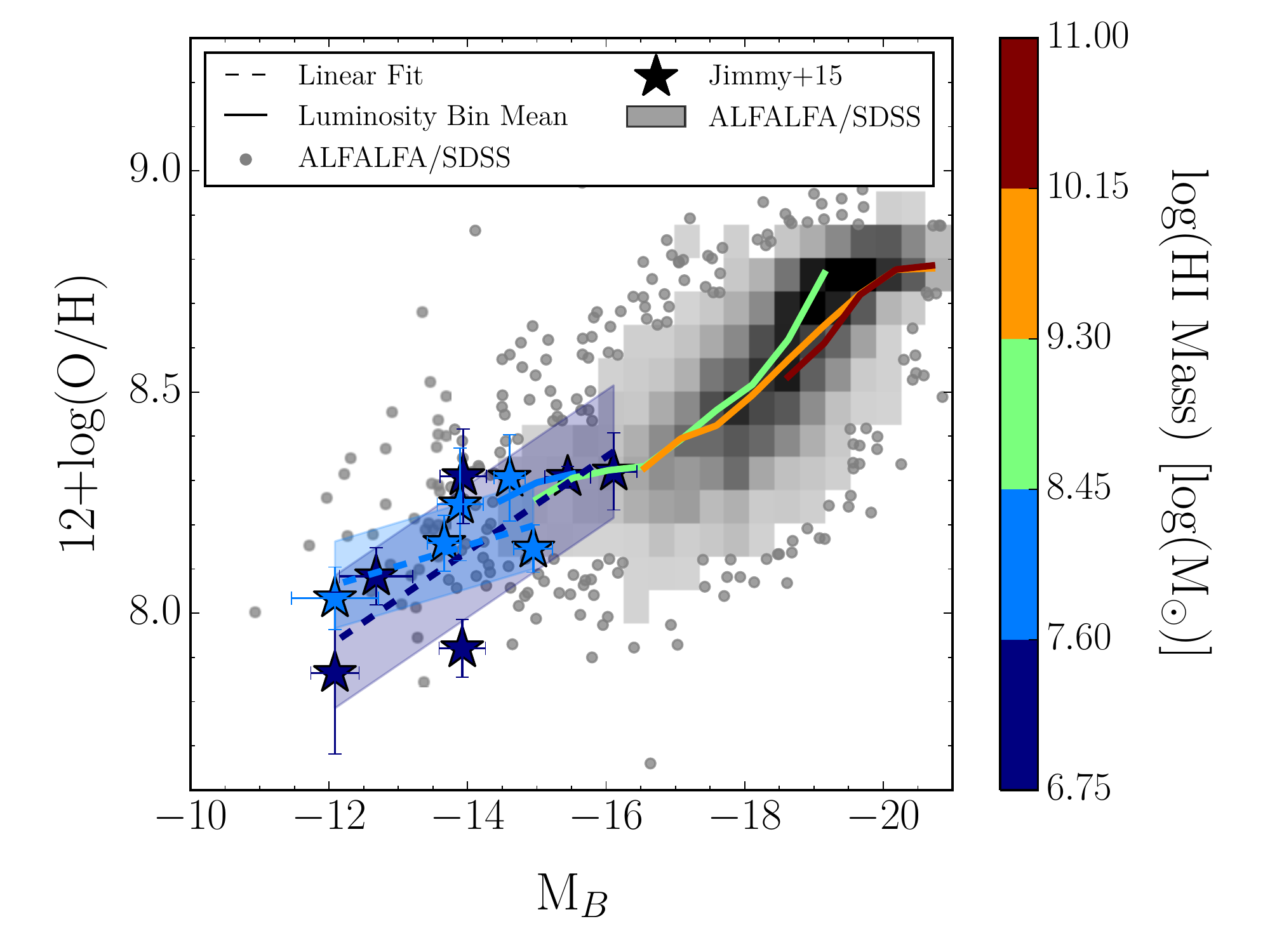, scale=0.45}
\epsfig{ file=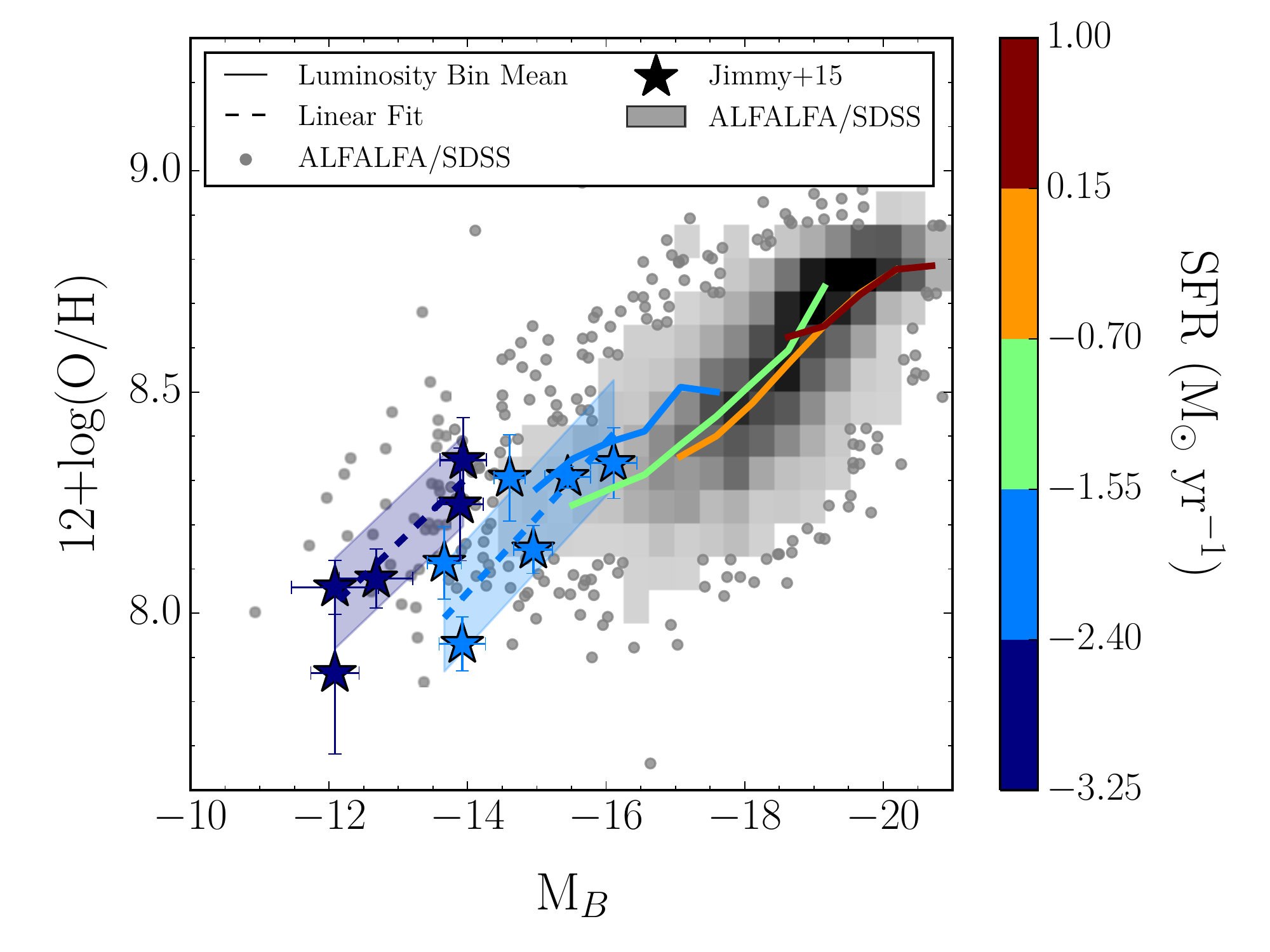, scale=0.45}
\caption{Same as Figure \ref{colorful_lzr_fmr} however using the PP04 O3N2 oxygen abundance calibration in place of the D02 N2 calibration.  The luminosity-metallicity relation as seen in Figure \ref{mass_metallicity_O3N2} color-coded by HI mass (left) and SFR (right).  The stars indicate individual observations within our IFU observed sample of dwarf galaxies.  The solid curves indicate the mean values of the ALFALFA/SDSS sample, separated into HI mass bins.  Shown in the background are the ALFALFA/SDSS points which are binned to produce the color-coded means.  We find little overlap between the SFR bins (right). }
\label{colorful_hi_mass_luminosity_metallicity_O3N2}
\end{figure}

\begin{figure}[h!]
\epsfig{ file=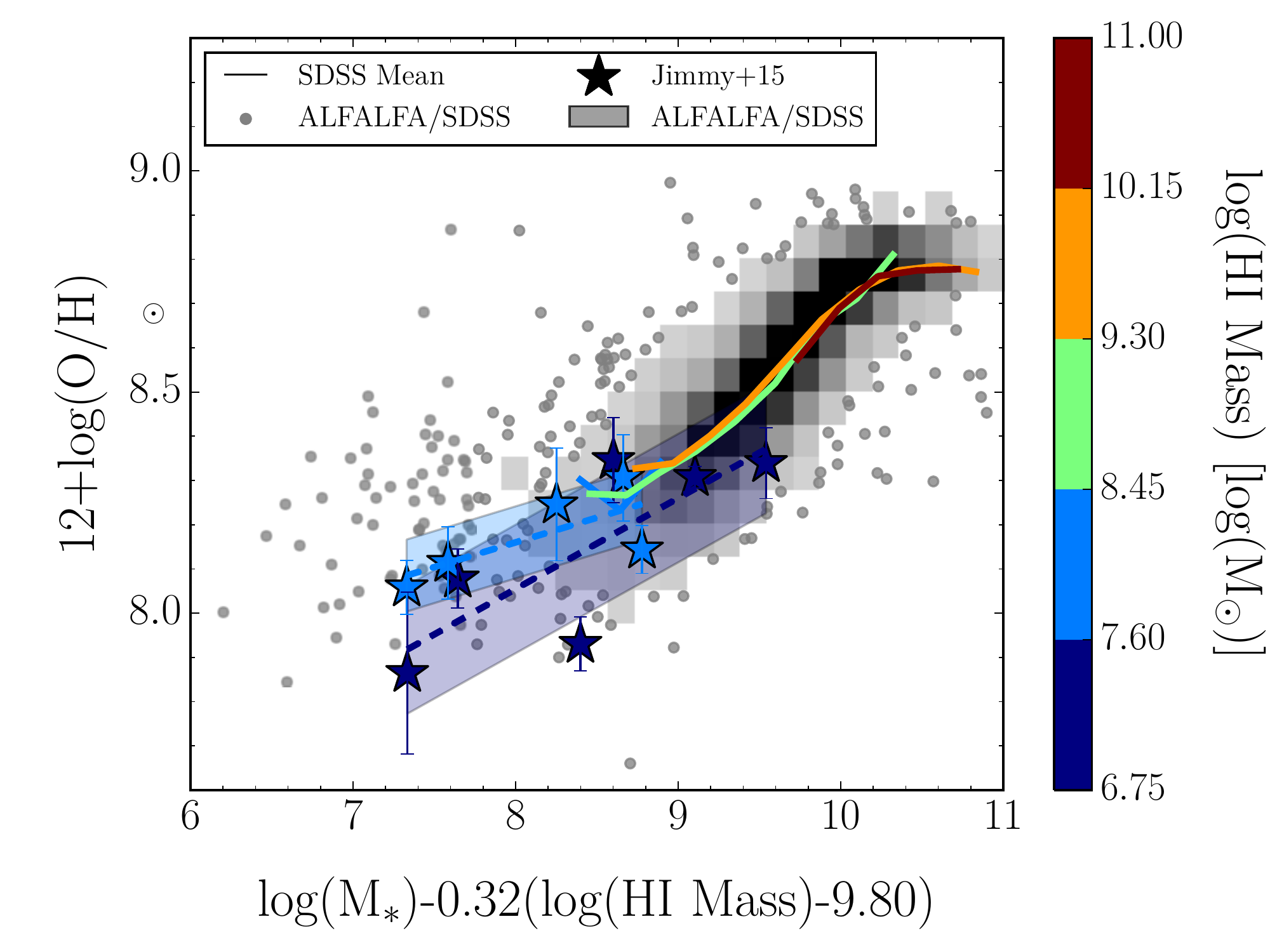, scale=0.45}
\epsfig{ file=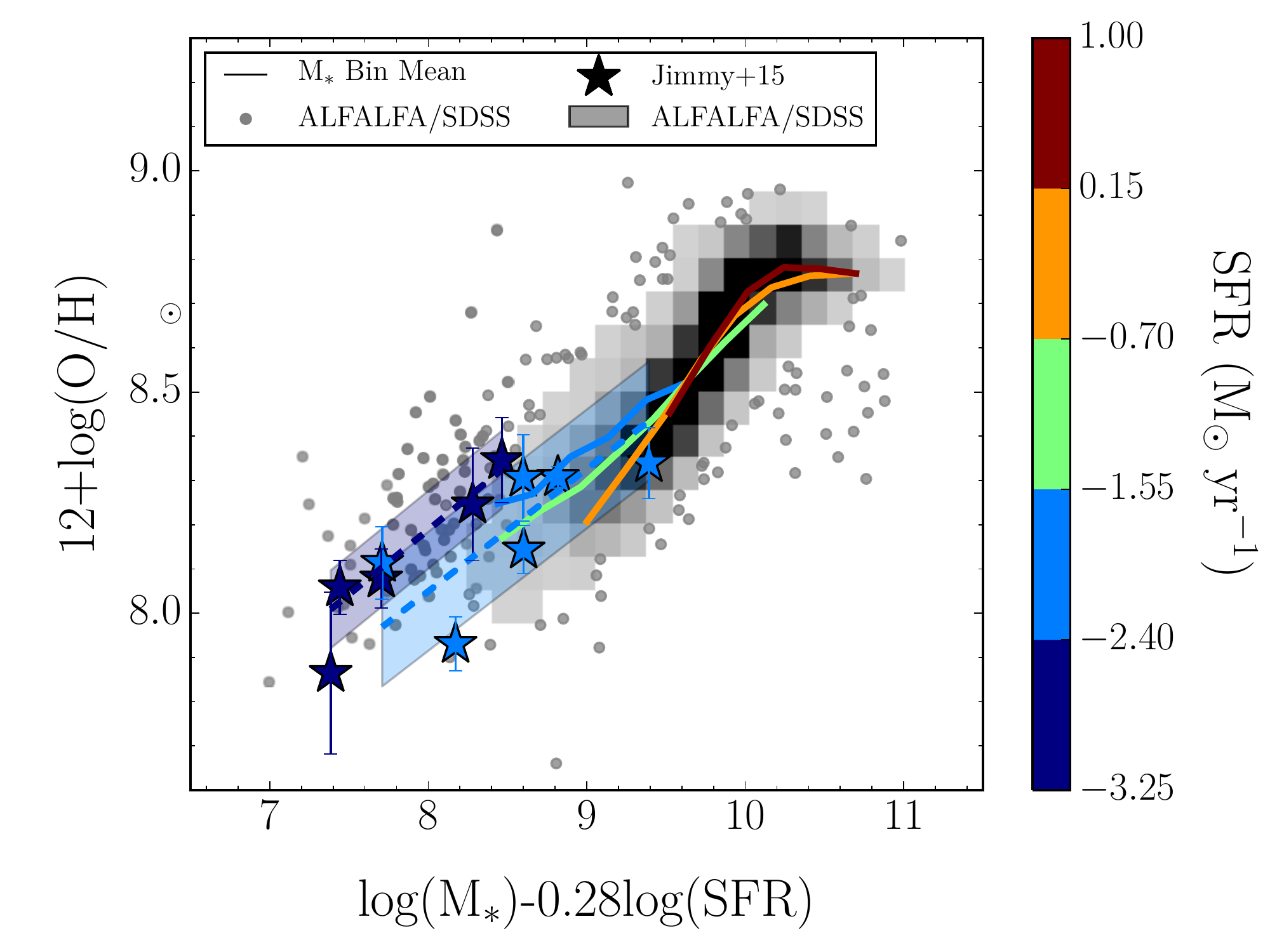, scale=0.45}
\caption{Same as Figure \ref{fmr} however using the PP04 O3N2 oxygen abundance calibration in place of the D02 N2 calibration.  \fmrhi\ (left) and \fmrsfr\ (right) as calculated using the $\alpha$ and $\beta$ values found to minimize scatter (Figure \ref{scatter}).  We find that the \fmrhi\ relation is consistent down to the lowest HI-mass bin, where as the lowest SFR bin in the \fmrsfr\ is offset 1$\sigma$ above the ALFALFA/SDSS derived \fmrsfr .}
\label{fmr_O3N2}
\end{figure}

\begin{figure}[h!]
\epsfig{ file=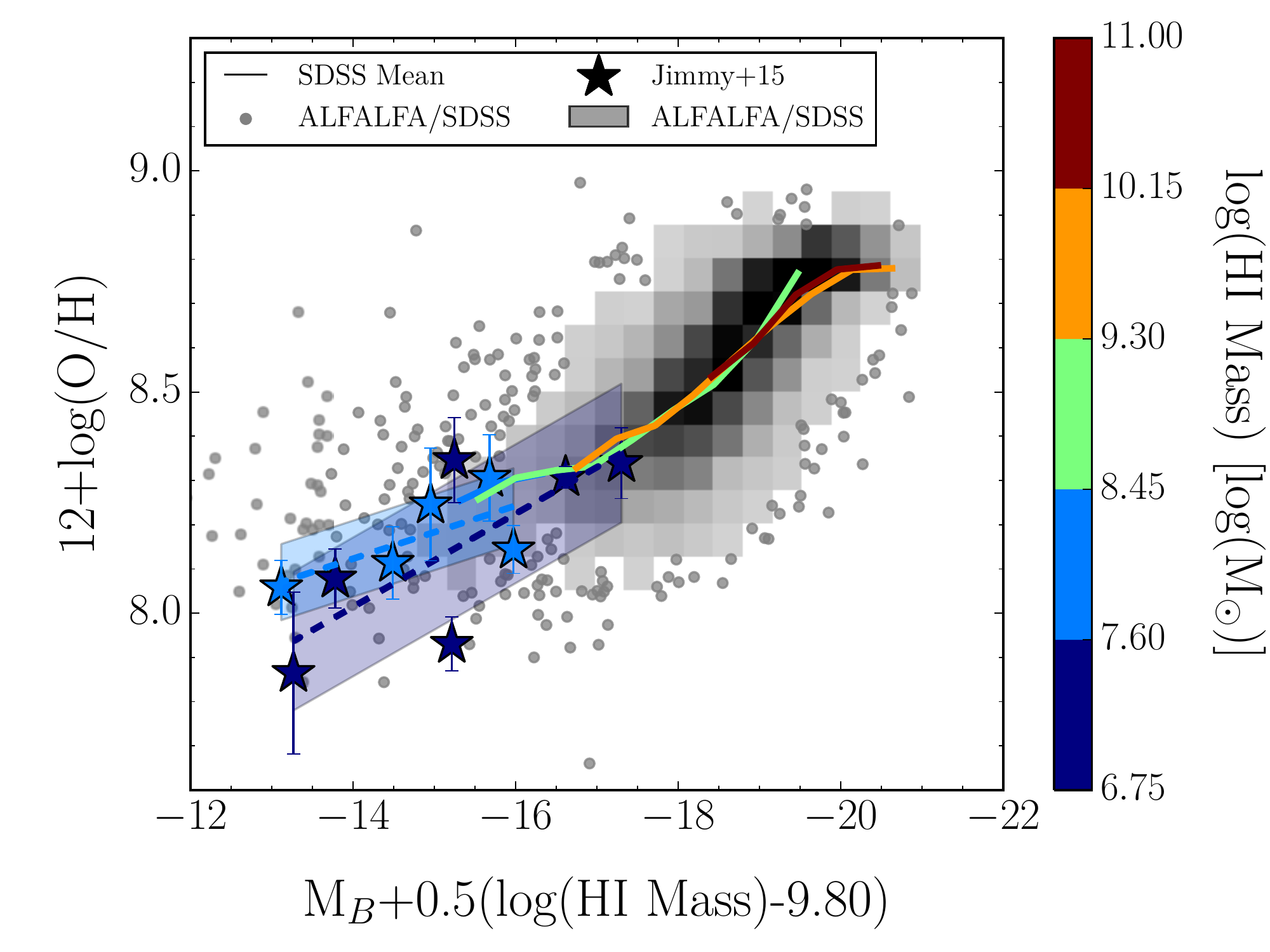, scale=0.45}
\epsfig{ file=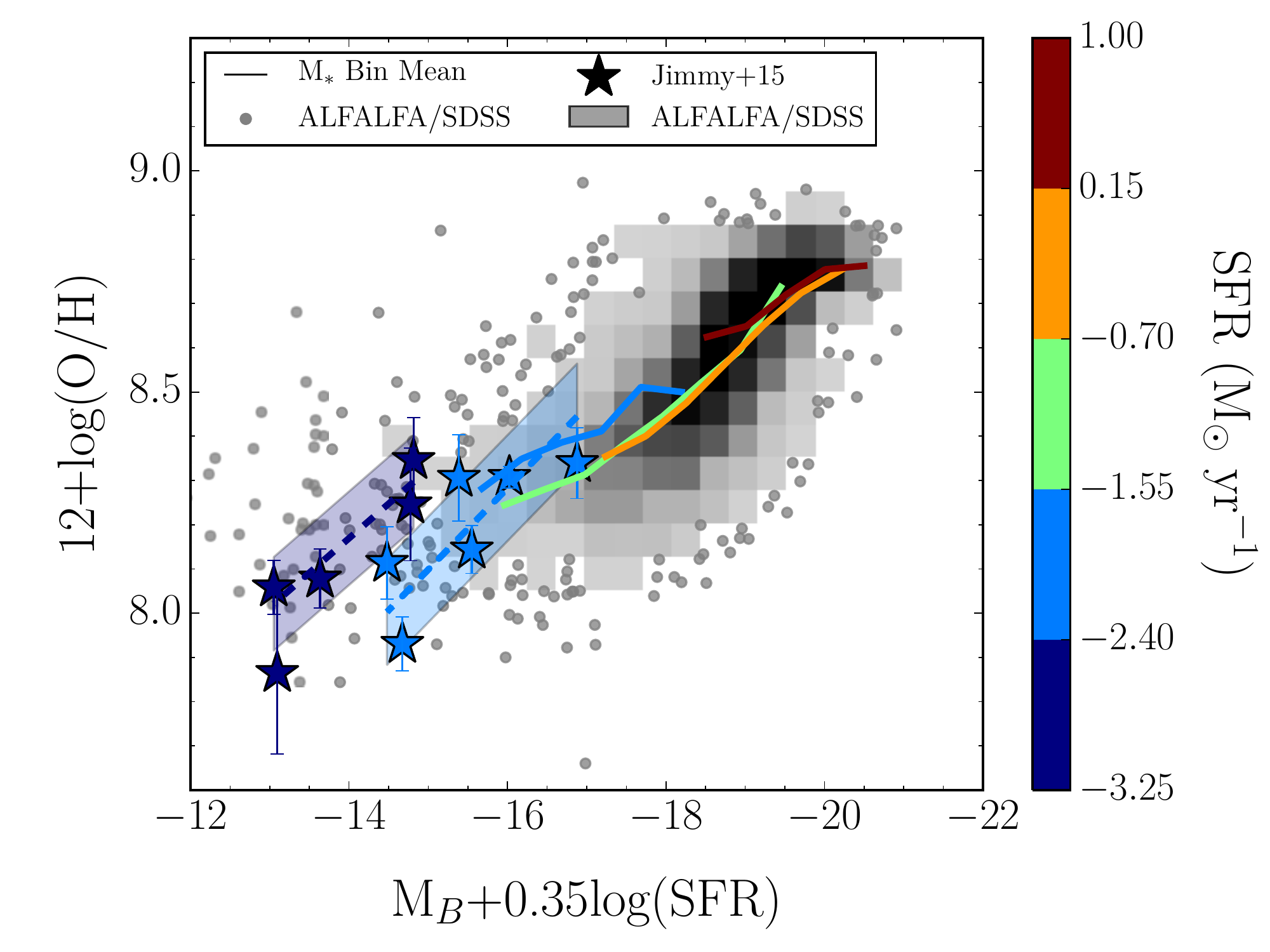, scale=0.45}
\caption{Same as Figure \ref{fml} however using the PP04 O3N2 oxygen abundance calibration in place of the D02 N2 calibration.  \fmlhi\ (left) and \fmlsfr\ (right) as calculated using the $\zeta$ and $\gamma$ values found to minimize scatter (Figure \ref{scatter_lzr}).  We find that the \fmlhi\ relation is consistent down to the lowest HI-mass bin, where as the lowest SFR bin in the \fmlsfr\ is offset 1$\sigma$ above the \fmlsfr .}
\label{fml_O3N2}
\end{figure}
\FloatBarrier

\section{\\Appendix C - SFR and HI Mass Mean Bins} \label{AppendixC}
To correspond to the linear fits, we provide a set of tables showing the mean metallicity values in each one of the SFR and HI mass binned data for both the MZR and LZR.  These values could be used to reproduce the solid curves of the SFR or HI-mass bins in Figures \ref{colorful_fmr} \& \ref{colorful_lzr_fmr}.  

\begin{deluxetable*}{ l r r r r r}
\tablecolumns{6}
\tablecaption{ALFALFA/SDSS SFR Binned Mean Metallicity Values}
\startdata
\hline
\hline
	 Stellar Mass & 12+log(O/H) & 12+log(O/H)  & 12+log(O/H)  &  12+log(O/H) &  12+log(O/H) \\
	 Bin Range &  Unbinned SFR Range & SFR -2.6 to -1.7  & SFR -1.7 to -0.8 & SFR -0.8 to 0.1 & SFR 0.1 to 1.0 \\
	 
\hline
\\
7.75 to 8.0 & 8.22 & 8.22 & - & - & -\\
8.0 to 8.25 & 8.23 & 8.25 & 8.17 & - & -\\
8.25 to 8.5 & 8.29 & 8.36 & 8.26 & - & -\\
8.5 to 8.75 & 8.35 & 8.41 & 8.33 & - & -\\
8.75 to 9.0 & 8.42 & 8.52 & 8.43 & 8.28 & -\\
9.0 to 9.25 & 8.5 &         - & 8.53 & 8.42 & -\\
9.25 to 9.5 & 8.58 &       - & 8.62 & 8.55 & -\\
9.5 to 9.75 & 8.67 &      - & 8.70 & 8.67 & 8.58\\
9.75 to 10.0 & 8.74 &   - & 8.79 & 8.74 & 8.70\\
10.0 to 10.25 & 8.78 & - & -       & 8.79 & 8.77\\
10.25 to 10.5 & 8.81 & - & -       & 8.83 & 8.80\\
10.5 to 10.75 & 8.83 & - & -       & 8.84 & 8.82\\
10.75 to 11.0 & 8.82 & - & -       & -       & 8.82\\

 & & & & & \\
 \hline
\hline

	 Stellar Mass & 12+log(O/H) & 12+log(O/H)  & 12+log(O/H)  &  12+log(O/H) &  12+log(O/H) \\
	 Bin Range &  Unbinned HI Mass Range &  HI Mass 7.6 to 8.45  & HI Mass 8.45 to 9.3 & HI Mass 9.3 to 10.15 & HI Mass 10.15 to 11.0 \\
	 
\hline
\\

7.75 to 8.0 & 8.22 & 8.30 & - & - & -\\
8.0 to 8.25 & 8.23 & 8.22 & 8.25 & - & -\\
8.25 to 8.5 & 8.29 & 8.37 & 8.27 & - & -\\
8.5 to 8.75 & 8.35 & -      & 8.35 & 8.35 & -\\
8.75 to 9.0 & 8.42 & -      & 8.44 & 8.40 & -\\
9.0 to 9.25 & 8.5 & -       & 8.53 & 8.48 & -\\
9.25 to 9.5 & 8.58 & -     & 8.63 & 8.57 & -\\
9.5 to 9.75 & 8.67 & -     & 8.72 & 8.67 & -\\
9.75 to 10.0 & 8.74 & -   & 8.78 & 8.74 & 8.67\\
10.0 to 10.25 & 8.78 & - & 8.78 & 8.78 & 8.75\\
10.25 to 10.5 & 8.81 & - & -       & 8.81 & 8.80\\
10.5 to 10.75 & 8.83 & - & -      & 8.83 & 8.82\\
10.75 to 11.0 & 8.82 & - & -       & 8.82 & 8.82\\
 & & & & & \\

\hline
\hline

	 Luminosity & 12+log(O/H) & 12+log(O/H)  & 12+log(O/H)  &  12+log(O/H) &  12+log(O/H) \\
	 Bin Range &  Unbinned SFR Range & SFR -2.6 to -1.7  & SFR -1.7 to -0.8 & SFR -0.8 to 0.1 & SFR 0.1 to 1.0 \\
	 
\hline
\\

-15.26 to -14.74 & 8.29 & 8.27 & - & - & -\\
-15.78 to -15.26 & 8.33 & 8.34 & 8.28 & - & -\\
-16.3 to -15.78 & 8.37   & 8.40 & 8.33 & - & -\\
-16.81 to -16.3 & 8.38 & 8.43   & 8.36 & - & -\\
-17.33 to -16.81 & 8.47 & 8.56 & 8.46 & 8.43 & -\\
-17.85 to -17.33 & 8.52 & 8.52 & 8.53 & 8.50 & -\\
-18.37 to -17.85 & 8.59 & -      & 8.61 & 8.57 & -\\
-18.89 to -18.37 & 8.66 & -      & 8.69 & 8.65 & 8.71\\
-19.41 to -18.89 & 8.73 & -      & 8.81 & 8.73 & 8.72\\
-19.93 to -19.41 & 8.77 & -      & -       & 8.78 & 8.77\\
-20.44 to -19.93 & 8.81 & -      & -       & 8.82 & 8.80\\
-20.96 to -20.44 & 8.81 & -      & -       & -       & 8.80\\

 & & & & & \\

\hline
\hline

	 Luminosity & 12+log(O/H) & 12+log(O/H)  & 12+log(O/H)  &  12+log(O/H) &  12+log(O/H) \\
	 Bin Range  &  Unbinned HI Mass Range &  HI Mass 7.6 to 8.45  & HI Mass 8.45 to 9.3 & HI Mass 9.3 to 10.15 & HI Mass 10.15 to 11.0 \\

\hline
\\

-14.74 to -14.22 & 8.24 & 8.24  & - & - & -\\
-15.26 to -14.74 & 8.29  & 8.30 & 8.20 & - & -\\
-15.78 to -15.26 & 8.33  & 8.32 & 8.31 & - & -\\
-16.3 to -15.78   & 8.37  & -       & 8.35 & - & -\\
-16.81 to -16.3   & 8.38 & -        & 8.37 & 8.37 & -\\
-17.33 to -16.81 & 8.47 & -        & 8.48 & 8.46 & -\\
-17.85 to -17.33 & 8.52 & -        & 8.56 & 8.51 & -\\
-18.37 to -17.85 & 8.59 & -        & 8.61 & 8.59 & -\\
-18.89 to -18.37 & 8.66  & -       & 8.70 & 8.66 & 8.64\\
-19.41 to -18.89 & 8.73 & -        & 8.79 & 8.73 & 8.72\\ 
-19.93 to -19.41 & 8.77 & -        & -       & 8.77 & 8.77\\
-20.44 to -19.93 & 8.81 & -        & -       & 8.80 & 8.82\\
-20.96 to -20.44 & 8.81 & -        & -       & 8.81 & 8.80\\

\enddata
\tablecomments{ Binned means used to produce the solid lines observed in Figures \ref{colorful_fmr} \& \ref{colorful_lzr_fmr}. }
\label{sfr_means_table}
\end{deluxetable*}

\FloatBarrier
\section{\\Appendix D - Table of Flux Values} \label{AppendixD}
We provide our emission line measurements for the integrated spectra from each galaxy's IFU observations so that others may use them for their own analysis.

\FloatBarrier
\begin{deluxetable*}{ l r r r r r r}
\tablecolumns{7}
\tablecaption{Observed Emission Line Fluxes of Dwarf Galaxies}
\startdata
\hline
\hline
	 Galaxy AGC\# & H$\beta$ 4861 \AA\ & [OIII] 5007 \AA\  &  H$\alpha$ 6563 \AA\ & [NII] 6583 \AA\ \\
\hline
\\

AGC191702 & 82.1 $\pm$ 11.3 & 247.4 $\pm$ 25.5 & 249.6 $\pm$ 31.8 & 6.0 $\pm$ 2.3 \\
AGC202218 & 56.3 $\pm$ 8.3 & 223.4 $\pm$ 28.0 & 223.7 $\pm$ 31.1 & 13.1 $\pm$ 4.0 \\
AGC212838 & 47.0 $\pm$ 6.3 & 174.3 $\pm$ 21.3 & 151.1 $\pm$ 20.5 & 5.2 $\pm$ 2.2 \\
AGC220755 & 10.7 $\pm$ 3.7 & 23.4 $\pm$ 7.7 & 43.0 $\pm$ 13.9 & 5.9 $\pm$ 2.3 \\
AGC220837 & 12.3 $\pm$ 3.3 & 35.1 $\pm$ 9.8 & 65.1 $\pm$ 15.9 & 11.2 $\pm$ 3.9 \\
AGC220860 & 60.7 $\pm$ 5.1 & 321.2 $\pm$ 23.3 & 212.2 $\pm$ 14.9 & 3.6 $\pm$ 1.5 \\
AGC221000 & 125.1 $\pm$ 7.2 & 230.6 $\pm$ 12.5 & 492.5 $\pm$ 29.3 & 43.5 $\pm$ 6.3 \\
AGC221004 & 38.1 $\pm$ 13.5 & 69.9 $\pm$ 23.4 & 126.2 $\pm$ 43.0 & 11.0 $\pm$ 4.1 \\
AGC225852 & 21.8 $\pm$ 9.4 & 48.8 $\pm$ 19.1 & 66.5 $\pm$ 28.3 & 4.6 $\pm$ 2.6 \\
AGC225882 & 52.4 $\pm$ 7.5 & 111.2 $\pm$ 12.8 & 142.0 $\pm$ 21.4 & 3.6 $\pm$ 1.9 \\
AGC227897 & 14.3 $\pm$ 0.6 & 74.2 $\pm$ 1.3 & 41.7 $\pm$ 1.0 & 0.4 $\pm$ 0.6 \\
\enddata
\tablecomments{ All line fluxes measurements are in units of $10^{-16} \text{ erg cm}^{-2} \text{ s}^{-1} \text{ \AA} ^{-1}$.  These are the line fluxes used to calculate oxygen abundance estimations throughout this work.}
\label{line_flux_table}
\end{deluxetable*}
\FloatBarrier

\newpage
\clearpage

\bibliographystyle{apj}
\bibliography{Jimmy2015}

\end{document}